\documentclass[aps,prd,twocolumn,showpacs,floatfix,preprintnumbers,amsfont,amsmath,amssymb,nofootinbib, superscriptaddress]{revtex4-1}

\usepackage{color}
\usepackage{hyperref}
\usepackage[normalem]{ulem}
\usepackage{lipsum}
\usepackage{url}
\usepackage{float}
\usepackage{ctable}
\usepackage{graphicx}
\usepackage[font=small, justification=raggedright, format=plain
  ]{caption} 
\usepackage{subcaption}
\usepackage{mathtools}
\usepackage{bm}
\usepackage{siunitx}

\colorlet{RED}{red}

\newcommand{\teja}[1]{ {\color{olive} #1} }
\newcommand{\jr}[1]{ {\color{violet} #1} }

\newcommand{\change}[2]{{#1}}

\newcommand{\sk}[1]{}

\newcommand{\tabline}{\specialrule{.1em}{.05em}{.05em}}

\newcommand{\snr}{{\rm SNR}}

\def\chieff{\chi_{\rm eff}}

\newcommand{\be}{\begin{equation}}
\newcommand{\ee}{\end{equation}}
\newcommand{\ba}{\begin{eqnarray}}
\newcommand{\ea}{\end{eqnarray}}

\allowdisplaybreaks

\begin{document}

\title{New binary black hole mergers in the second observing run of Advanced LIGO and Advanced Virgo}
\author{Tejaswi Venumadhav}
\email{tejaswi@ias.edu}
\affiliation{\mbox{School of Natural Sciences, Institute for Advanced Study, 1 Einstein Drive, Princeton, NJ 08540, USA}}
\author{Barak Zackay}
\affiliation{\mbox{School of Natural Sciences, Institute for Advanced Study, 1 Einstein Drive, Princeton, NJ 08540, USA}}
\author{Javier Roulet}
\affiliation{\mbox{Department of Physics, Princeton University, Princeton, NJ, 08540, USA}}
\author{Liang Dai}
\affiliation{\mbox{School of Natural Sciences, Institute for Advanced Study, 1 Einstein Drive, Princeton, NJ 08540, USA}}
\author{Matias Zaldarriaga}
\affiliation{\mbox{School of Natural Sciences, Institute for Advanced Study, 1 Einstein Drive, Princeton, NJ 08540, USA}}

\date{\today}


\begin{abstract}
We report the detection of new binary black hole merger events in the publicly available data from the second observing run of advanced LIGO and advanced Virgo (O2).
The mergers were discovered using the new search pipeline described in \change{\citet{2019PhRvD.100b3011V}}{\citet{2019arXiv190210341V}}, and are above the detection thresholds as defined in \citet{LIGOScientific:2018mvr}. Three of the mergers (GW170121, GW170304, GW170727) have inferred probabilities of being of astrophysical origin $p_{\rm astro} > 0.98$. The remaining three (GW170425, GW170202, GW170403) are less certain, with $p_{\rm astro}$ ranging from $0.5$ to $0.8$.
The newly found mergers largely share the statistical properties of previously reported events, with the exception of GW170403, the least secure event, which has a highly negative effective spin parameter $\chi_{\rm eff}$.
The most secure new event, GW170121 ($p_{\rm astro} > 0.99$), is also notable due to its inferred negative value of $\chi_{\rm eff}$, which is inconsistent with being positive at the $\approx 95.8\%$ confidence level.\sk{---95.8\% of the support of the $\chi_{\rm eff}$ posterior has negative values.}
The new mergers nearly double the sample of gravitational wave events reported from O2, and present a substantial opportunity to explore the statistics of the binary black hole population in the Universe.\sk{intrinsic statistical quantities and the formation rate of binary black holes.}
The number of detected events is not surprising since we estimate that the detection volume of our pipeline \change{may be larger than that of other pipelines by as much as a factor of two (with significant uncertainties in the estimate).}{is nearly twice that of other pipelines.} 
The increase in volume is larger when the constituent detectors of the network have very different sensitivities, as is likely to be the case in current and future runs.

\end{abstract}

\maketitle

\section{Introduction}

The LIGO--Virgo Collaboration (LVC) reported the detection of gravitational waves from ten binary black hole (BBH) and one binary neutron star mergers in their two latest observing runs, O1 and O2~\cite{LIGOScientific:2018mvr, GW150914, GW151226, O1catalog, GW170104, GW170608, GW170814, GW170817}. The intrinsic properties of the mergers, namely the masses, the mass ratio and the spins of the black holes, are important observables that can inform us about how and where the binaries were assembled. All the BBH events are consistent with mergers of black holes with comparable masses (notably, GW170729 is mildly inconsistent with equal component masses \cite{GW170729HigherModes}); in two of the events, at least one of the components had a nonzero spin. \sk{(in the range $10\,M_\odot$ to $60\,M_\odot$). A number of binary formation channels have been proposed in the literature.} 

Currently, inference about the origin of the BBHs is limited by the small number of detected events. 
Future observations with improved sensitivity will enlarge the sample and map out the parameter space of the BBH population;
meanwhile, it is important to adopt analysis techniques that maximize the yield of existing data.

 \sk{As more observing time with increasing sensitivity is accumulated the number of events will increase and inferences about the formation channels will become more secure. In the mean time one can increase the sensitive volume by doing a more optimal analysis of the data, in particular by making improvements to more properly deal with the non-Gaussian nature of the detector noise.} 

A number of data analysis pipelines have been developed to search for transient events in LIGO--Virgo data.
The two\change{}{standard} modeled searches used by the LVC are \texttt{PyCBC}~\cite{PYCBCPipeline} and \texttt{GstLAL}~\cite{gstlal}; these pipelines use matched filtering with a template bank of target compact binary coalescence signals.
In addition, the LVC runs an unmodeled transient search with the coherent Wave Burst pipeline~\cite{cWB}. 
There are also groups external to the LVC running independent pipelines~\cite{NitzCatalog, 2019arXiv190210341V} on the public data released by the LVC, which now includes the O1 and O2 observing runs \cite{gwosc_url, GWOSC}.
In Ref.~\change{\cite{2019PhRvD.100b3011V}, henceforth TV19,}{\cite{2019arXiv190210341V}} we presented a new and independent pipeline to analyze public LIGO data, which we applied to the public data from the O1 observing run. The cumulative improvements significantly increased the sensitive volume (at the same detection thresholds as those of Ref.~\cite{LIGOScientific:2018mvr}), and led to the detection of a new event, which is consistent with the merger of rapidly spinning and heavy black holes~\cite{GW151216}. 
In this paper, we present results from our search of coincident triggers in the Hanford (H1) and Livingston (L1) public data from the second observing run of advanced LIGO (O2) \cite{GWOSC, gwosc_url}.

 \sk{which is significantly more sensitive than the LIGO one. The  pipeline  incorporates  different  techniques  and  makes  independent  implementation  choices  in all its stages including the search design, the method to construct template banks, the automatic routines to detect bad data segments (“glitches”) and to insulate good data from them, the procedure to account for the non-stationary nature of the detector noise, the signal-quality vetoes at the single-detector level and the methods to combine results from multiple detectors. Using this pipeline we found a new candidate event with a very high spin in O1 \cite{GW151216}.}  

The paper is organized as follows: \change{in Section~\ref{sec:preview} we begin with a preview of our final results to motivate what follows. We then discuss the details of the changes in the pipeline used in this paper compared to the one used to analyze the O1 data in Section~\ref{sec:changes}.}{in Section~\ref{sec:changes} we describe the changes in the pipeline used in this paper compared the one used to analyze the O1 data.} Section~\ref{sec:previous_events} summarizes the results of our analysis on the events that were previously detected by the LVC analysis pipelines in Ref.~\cite{LIGOScientific:2018mvr}. We describe the new events we found in Section~\ref{sec:events}, estimate our improvement in sensitivity in Section~\ref{sec:sensitivity}, and conclude with some remarks in Section~\ref{sec:conclusions}. In Appendices A--C, we present the posterior distributions for the parameters of the new events, as well as some technical details.

\section{Preview of results}
\label{sec:preview}

\begin{figure*}
    \centering
    \includegraphics[width=\linewidth]{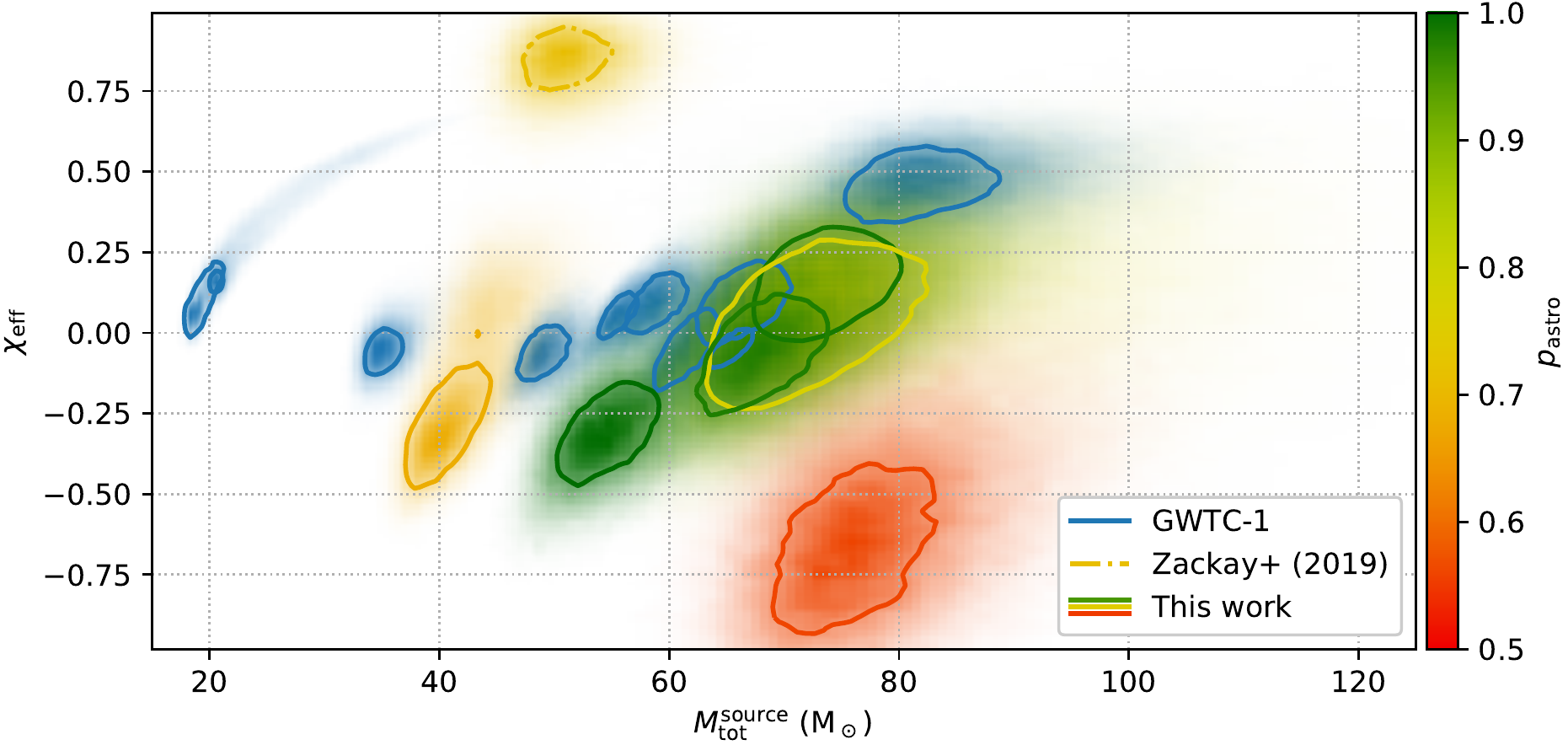}
    \caption{Source-frame total mass and effective spin for the BBH events found in Hanford--Livingston coincidence, over O1 and O2. We recovered all the previously reported events \change{that were in the data we searched in }{}with high confidence, \change{i.e., the probability of astrophysical origin}{} $p_{\rm astro} \approx 1$, except for \change{}{GW170608 and }GW170818, see Sec.~\ref{sec:previous_events}. We found seven additional events ranging from marginal triggers to confident detections: one in O1 \cite{GW151216} and six in O2 (this work)\change{; we color-code the posteriors according to the value of $p_{\rm astro}$}{}. \change{The densities use samples from the posterior distribution of each event, and the curves are $1\sigma$ contours (i.e., they enclose $1-e^{-1/2} \approx 39\%$ of the posterior probability). We used a prior that is uniform in detector-frame $m_1$, $m_2$, $\chi_{\rm eff}$, and luminosity volume. Appendix~\ref{Appendix:posteriors} shows other projections of the posteriors.}{The dots and error bars show median and 90\% confidence intervals, respectively. The spin $\chi_{\rm eff}$ and the mass can be correlated (not shown). The full posteriors can be found in Appendix~\ref{Appendix:posteriors}. The prior used was uniform in $m_1$, $m_2$, $\chi_{\rm eff}$, and luminosity volume.}}
    \label{fig:events}
\end{figure*}

\change{We improved upon the methods we previously developed to analyze the O1 data (see TV19 and Sec.~\ref{sec:changes} for details), and applied them to look for massive binary black hole mergers in the H1 and L1 data from the O2 data released by the LVC. 
We recover with high confidence all the BBH events that the LVC searches previously found in the bulk data release, with the notable exception of GW170818, which is one of our highest ranked candidates, but is effectively close to a single-detector trigger that would be hard to confirm using the methods adopted in this search (see Sec.~\ref{sec:previous_events} for details). 

This serves as an independent validation of the events presented in GWTC-1 \cite{LIGOScientific:2018mvr}, and a confirmation of the sensitivity of our pipeline. 
The fact that we can only place upper bounds on the false alarm rates (FARs) for these events, in particular, even for the least significant event in GWTC-1, GW170729, is consistent with an improvement in sensitivity over the range of parameters included in our search. 
Also consistent with this is the fact that we find six new events with (a) false alarm rates below the threshold of 1 in 30 days defined in GWTC-1 (even after applying appropriate trials factors), and (b) probabilities of astrophysical origin $p_{\rm astro} > 0.5$, as effectively quantified by their consistency with the rates of occurrence of louder events.

Figure \ref{fig:events} places these new events in context of the population of events in the GWTC-1 catalog by showing their marginalized posterior probability distributions in the plane of the effective spin-parameter, $\chi_{\rm eff} = (m_1 \chi_{1, z} + m_2 \chi_{2, z}) / (m_1 + m_2)$ (where $m_i$ and $\chi_{i, z}$, respectively, are the masses and projections of the dimensionless spin onto the orbital angular momentum), and the total source-frame mass. 
We chose to present the points for the new events colored according to their estimated values of $p_{\rm astro}$ to emphasize that these events come as a population, with these numbers being an integral part of their interpretation. 
In the rest of the paper, we will present the changes to our analysis methods, the procedure we use to assign values of FAR and $p_{\rm astro}$ to events, and a deeper view into the results.}{}

\section{Changes to the O1 analysis pipeline}
\label{sec:changes}

Our analysis pipeline is similar in overall structure to the one we used in the O1 analysis~\change{}{\cite{2019arXiv190210341V}}. \change{
The overall flow of the search is described in Section II of TV19: for completeness, we briefly repeat the description here. We divide our search space into a set of banks, and within each bank, construct a set of templates that approximate the waveforms generated using the \texttt{IMRPhenomD} approximant \cite{2016PhRvD..93d4007K} to a desired degree of fidelity in the presence of characteristic detector noise. 
We then generate a time-series of matched-filtering scores for each template against the data streams from the Hanford and Livingston detectors. 
When these scores cross a threshold value in a detector, we call this a {\em trigger} in that detector: we collect coincident triggers between H1 and L1 (with templates indexed by the same coefficients, and within \SI{10}{\milli\second} of each other). 
Subsequently we locally refine the triggers in each detector onto a finer template bank, apply vetoes based on signal quality, and pick the best coincident trigger from the subsets of refined H1 and L1 triggers. 
Finally we assign triggers (and hence the underlying signal/background event) to banks, compute a ranking statistic for the triggers in a bank, and assign a false alarm rate to triggers according to the empirically measured distribution of the ranking statistic.}{} 

\change{The interested reader can find a full description of the details of the methods in TV19. In this section, we focus on the differences between the current and previous analysis.}{} In detail, the pipeline for the O2 analysis differs in the following aspects:
\begin{enumerate}
    \item \textit{Construction of the template bank:} 
    \change{Our search covers the parameter space of compact binary mergers with component masses between 3 and 100 $M_\odot$, with mass-ratios q $\in [1/18, 1]$ and the projections of the dimensionless spins of the constituents on the orbital angular momentum satisfying $\vert \chi_i \vert < 0.99$. 
    As in the O1 analysis, we partition this search space into five banks (\texttt{BBH 0-4}) based on the detector-frame chirp masses, and divide each bank into sub-banks based on the shape of the frequency-domain amplitude profile.
    Within each sub-bank, we use the procedure described in Ref.~\cite{templatebankpaper} to construct grids of templates.}{We use the template bank described in \cite{templatebankpaper}. 
    As in the O1 analysis, we divide our BBH search into five banks (\texttt{BBH 0-4}) based on the chirp masses of the templates, and divide each bank into sub-banks based on the shape of the frequency-domain amplitude profile.} We now use the $k$-means algorithm to automatically divide each bank into sub-banks, each of which has a frequency-domain amplitude, $\overline{A}(f)$, that is the root-mean-square average of the amplitudes of the constituent astrophysical waveforms.
    In Fig.~\ref{fig:amplitudes sub-banks} we show $\overline{A}(f)$ for all the sub-banks in our banks covering detector chirp masses 20--40 ${\rm M}_\odot$ (bank \texttt{BBH 3}) and above $40\,{\rm M}_\odot$ (\texttt{BBH 4}). \change{We show the profiles only for these banks to avoid overcrowding the figure, and since all the events we report in this paper fall in these banks. The curves for the other banks (\texttt{BBH 2}, \texttt{BBH 1}, and \texttt{BBH 0}) continue the trend that is visible in Fig.~\ref{fig:amplitudes sub-banks}, in which the cutoff frequency shifts rightward as we go to lower masses.}{All the events we report in this paper fall in these banks. }
 
\begin{figure}
    \centering
    \includegraphics[width=\linewidth]{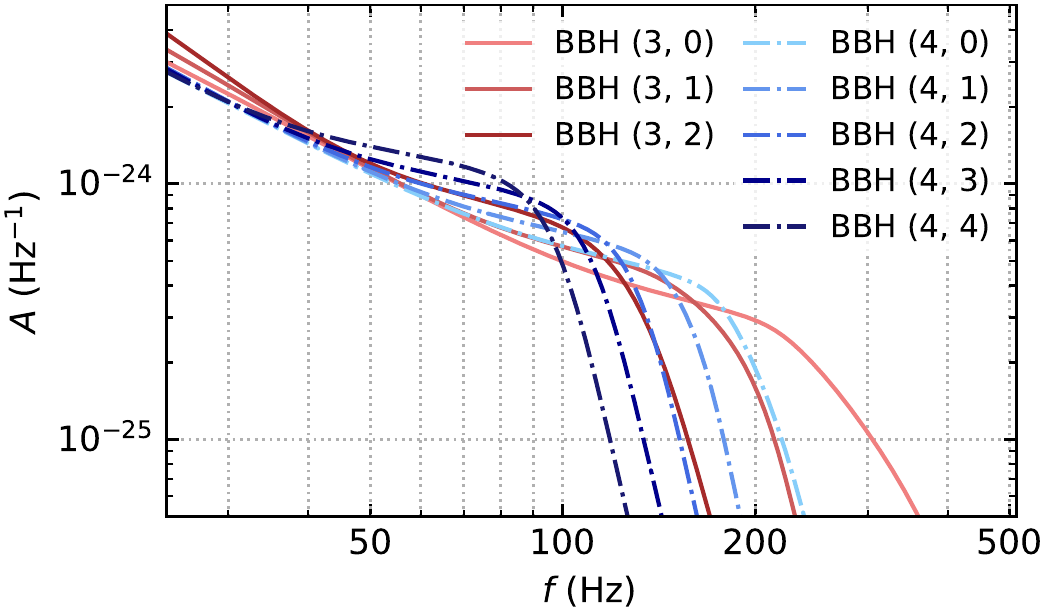}
    \caption{Sub-bank amplitude profiles as a function of frequency $f$ for banks \texttt{BBH 3} and \texttt{BBH 4}.}
    \label{fig:amplitudes sub-banks}
\end{figure}

    The phases of the templates are linear combinations of basis functions, $\psi_\alpha(f)$, whose form depends on the noise power spectral density (PSD); we use the PSD estimated from a representative set of files from the run (instead of a model PSD as was done in the O1 analysis). 
    We also restrict the templates to frequencies between $24$ and \SI{512}{\hertz}. Cumulatively, these changes enable us to cover the same astrophysical parameter space using $\sim 30\%$ fewer templates, and achieve slightly better effectualness~\cite{templatebankpaper}.
    
    \item \textit{Preprocessing and flagging the data:}
    Our analysis pipeline produces a stream of whitened data per \SI{4096}{\second} file. 
    \change{As in our O1 analysis, we}{We} perform several tests on the data to identify prolonged or transient disturbances in the detector that appear as segments with excess power, and discard these segments to avoid polluting our search. \change{The overall nature and number of these tests is unchanged from the O1 analysis; a full description of these tests can be found in Section III C of TV19. What is different, however, is how we set thresholds to trigger these tests. If the thresholds are too low, they can be triggered by real astrophysical events. Conversely, if they are too high, the tests are not easily triggered by bad data segments. To mitigate the former effect, in the O1 analysis, we set the threshold for each test to the power achieved by signals with a fiducial signal-to-noise ratio $(\snr)_{\max}$ in the absence of noise. This method of setting thresholds is problematic, since in the presence of additive Gaussian random noise, real signals with $\snr < (\snr)_{\max}$ can have excess power above these thresholds due to upward fluctuations.}{We would like these tests not to trigger on astrophysical signals of interest. Towards this end, in our previous analysis we set the threshold for each test to the power achieved in the absence of noise by signals with a fiducial signal-to-noise ratio $(\snr)_{\max}$. In the presence of noise, real signals with $\snr < (\snr)_{\max}$ can have excess power above these thresholds due to upward fluctuations.} Hence, in our O1 analysis we used a relatively high value of $(\snr)_{\max} = 30$ to ensure completeness at lower values~\cite{2019arXiv190210341V}. 
    
    \change{In the analysis described in this paper, we instead set the thresholds for tests by demanding a given false-positive probability for signals with a desired signal-to-noise ratio.}{A better design for tests would be to ensure a given false-positive probability for signals with a desired signal-to-noise ratio.}
    For whitened data, power in a given band, and on a given timescale, is distributed according to a non-central chi-squared distribution in the presence of a signal; hence, we can set thresholds such that signals with a given value of $(\snr)_{\max}$ are flagged with a probability $< 10^{-4}$. 
    \change{Given this promise, we are able to set thresholds with a lower target value $(\snr)_{\max}=20$, without losing relevant signals in this search. By nature, the thresholds resulting from this criterion, and hence the amount of data flagged as bad, are dependent on the bank (i.e., at the same value of \snr, the thresholds on excess power are higher for a signal in BBH 4 compared to those in BBH 0). Regardless of bank, the total length of bad data never exceeds 2 \% of the overall runtime of O2, and hence, the effective reduction in `runtime' is not an important factor to take into account in spacetime volume estimates for the search.}{Given this promise, we lower the target value to $(\snr)_{\max}=20$.}
    
    \item \textit{Refining coincident triggers:} 
    The phases of our templates belong to a vector space, $\mathcal{V}$, spanned by the basis functions $\{ \psi_\alpha(f): \alpha = 1, 2 \dots n \}$; the template bank is a discrete subset of this space with basis coefficients that live on an $n$-dimensional grid. 
    We determine the extent of the grid in all dimensions by projecting a large random sample of astrophysical waveforms into $\mathcal{V}$ and ensuring that every waveform has a nearby grid point. We allow comparatively large mismatches ($\lesssim 10\%$) between astrophysical waveforms and the best template in the bank, which enables us to work with coarser grids on $\mathcal{V}$ when generating triggers. We then reduce the mismatch for significant triggers by refining their coefficients on a finer local grid. In the O1 analysis, we chose this grid to be a uniform regular grid centered on a trigger of interest~\cite{2019arXiv190210341V}.
    
    For heavy BBH waveforms, the set $\mathcal{S}$ of projected astrophysical waveforms is typically thin and mildly curved in higher dimensions ($\alpha \gtrsim 2$)~\cite{templatebankpaper}. The strategy used in the O1 analysis can cause us to step outside $\mathcal{S}$ in these dimensions, and introduce unphysical degrees of freedom that pick up noise but no signal. 
    In this analysis, we change the spacing of the finer grid and excise unnecessary elements to ensure that we enumerate over templates within $\mathcal{S}$. 
    Note that different choices of how this refinement is done can lead to different quoted SNRs for the same astrophysical signal depending on how closely the finer grid approaches it. Hence, we apply the same strategy to the background triggers (found via time slides) to avoid biasing the calculation of false-alarm rates.
    
    \item \textit{Reducing cross-contamination between banks:} High $\snr$ triggers tend to appear in several of our chirp-mass banks, both in the time slides used to estimate our background, as well as in the set of coincident triggers. In the O1 analysis, we assigned triggers to the best sub-bank in a given chirp-mass bank (as determined by the incoherent network $\snr^2 = \rho_{\rm H}^2 + \rho_{\rm L}^2$, where $\rho_{\rm H}^2$ and  $\rho_{\rm L}^2$ are the incoherent squared $\snr$s in Hanford and Livingston, respectively), but allowed them to appear in multiple banks~\change{\cite{2019PhRvD.100b3011V}}{\cite{2019arXiv190210341V}}.
    This choice was conservative, in that it caused us to overestimate the FAR for real events. In this analysis, we improve upon this in two ways: we assign both background and coincident triggers to a unique bank (and sub-bank within), and instead of the incoherent network $\snr^2$, we use a discriminator that better accounts for the different structures of the sub-banks. Appendix~\ref{Appendix:collection} contains the expression (see Eq.~\eqref{eq:likebank}), and outlines a derivation. \sk{. and do it in a way  the bank in which they had the highest coherent network $\snr$ minus the log of the number of templates in the bank.
    
    , but since we eventually assign real coincident triggers events to only one bank, it is consistent to treat the background in the same way. }
    
    \item \textit{Computing the false-alarm rate:}
    \begin{figure}
        \centering
        \includegraphics[width=\linewidth]{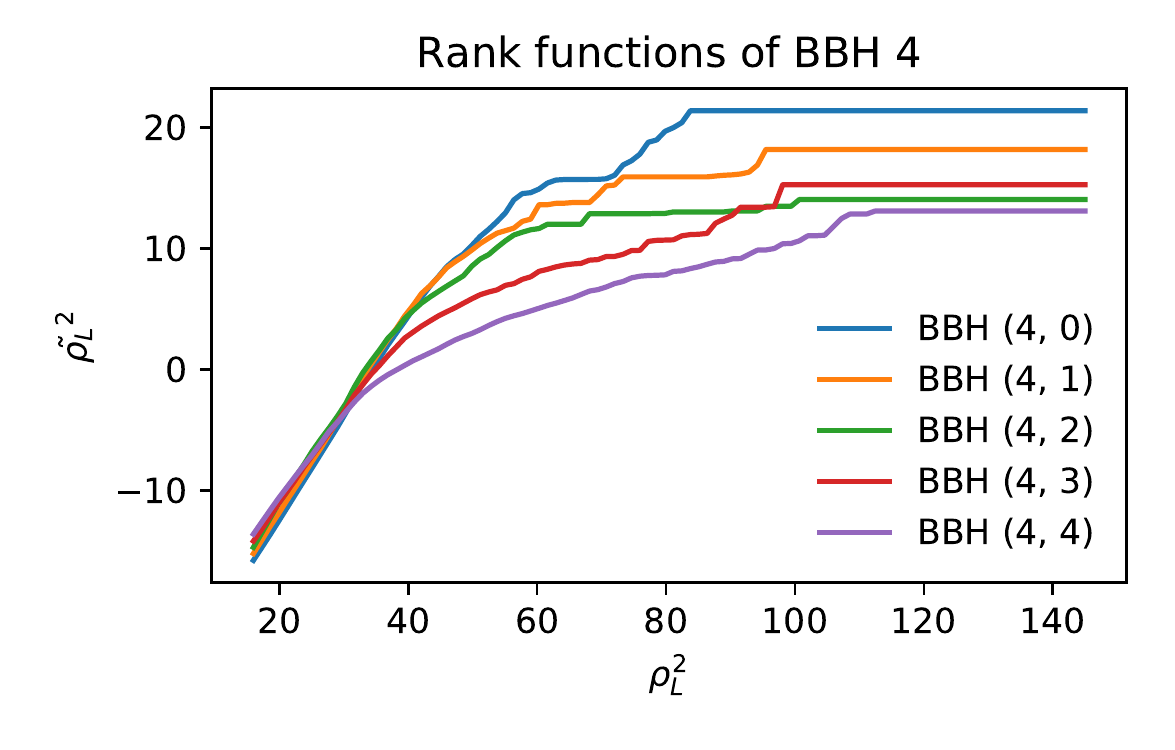}
        \caption{Rank functions for all sub-banks in bank \texttt{BBH 4}\change{, as a function of the $\snr^2$ in the L1 detector. The quantity on the y-axis is twice the logarithm of the number of triggers above a particular value of $\rho_{\rm L}^2$, with an offset.}{.}
        The different sub-banks have substantially different background rates: glitches prefer the higher-mass sub-banks and, even in the Gaussian part, the slopes may differ due to the different number of degrees of freedom.} 
        \label{fig:RankFunctions}
    \end{figure}
    After collecting triggers and assigning them to banks, we estimate the FAR for a given coincident trigger by comparing it to the background triggers (collected using time slides) within the same chirp-mass bank. 
    Our statistic for comparing triggers is the coherent score \cite{CoherentScore}, which for a given trigger $t$ is an approximation of the likelihood ratio under the signal ($\mathcal{L}(t \mid \mathcal{S})$) and noise ($\mathcal{L}(t\mid \mathcal{N})$) hypotheses. 
    \change{In order to estimate $\mathcal{L}(t\mid \mathcal{N})$, we approximate the distribution of background triggers in each detector (as a function of $\snr^2$) by its survival function. 
    The survival function (or more correctly, the complementary cumulative distribution function) has the virtue that we can estimate it empirically by just ranking triggers as a function of the value of $\snr^2$ in that detector (see Section III J of TV19).}{}
    
    In the O1 analysis, when estimating $\mathcal{L}(t\mid \mathcal{N})$, we assumed that the background rate per template is flat over all sub-banks in a given bank~\change{\cite{2019PhRvD.100b3011V}}{\cite{2019arXiv190210341V}}. \change{We revisited the validity of this assumption while developing the analysis in this paper. Figure \ref{fig:RankFunctions} shows the rank functions for the five sub-banks of the bank \texttt{BBH 4} as a function of the $\snr^2$ in L1, $\rho_{\rm L}^2$ (these curves are completely empirically determined and not fits; the quantity on the y-axis is twice the logarithm of the number of triggers above a particular value of $\rho_{\rm L}^2$, with an offset. See discussion in Section III J of TV19, and below.). The figure clearly shows that the sub-banks have substantially different background-rates per template. The distributions of $\rho_{\rm L}^2$ transition from exponential (chi-squared-like) to power-laws when glitches become important, and hence the rank functions flatten: this happens at progressively lower values of $\rho_{\rm L}^2$ for higher-mass sub-banks.}{However, we empirically see that the various sub-banks have substantially different background-rates per template.} If we do not account for this, a sub-bank with more glitches can disproportionately influence the background in the search. 
    
    \change{In this search, when computing the false alarm rates of a trigger, we use the rank function of the sub-bank it is drawn from. This enables us to give a more local estimate of the false-alarm rate, which better accounts for the fact that templates in some regions of the search space are more prone to triggering on glitches (this effect has been noted before~\cite{2014CQGra..31a5016D, 2016PhRvD..94l2004B}). A technical point is that we compute FARs over all templates in a bank (which is the union of several sub-banks), and hence we need to properly fix the normalization of the rank functions when comparing triggers from different sub-banks. Appendix~\ref{Appendix:normrank} describes the procedure; the rank functions shown in Fig.~\ref{fig:RankFunctions} were normalized in this way. Curiously, we see that the rank functions have different slopes in different sub-banks even at low values of $\rho_{\rm L}^2$ (in the Gaussian-noise-dominated regime); this is due to the maximization over templates when we collect triggers (see Appendix~\ref{Appendix:collection} for a derivation in a different context).}{}
    
    \change{}{In order to estimate $\mathcal{L}(t\mid \mathcal{N})$, we approximated the distribution of background triggers in each detector (as a function of $\snr^2$) by its survival function, or complementary cumulative distribution function, which in turn we obtained empirically by ranking triggers as a function of $\snr^2$ in that detector. 
    In this analysis, we separately rank triggers in each sub-bank. Figure \ref{fig:RankFunctions} shows the rank functions for the five sub-banks of the bank \texttt{BBH 4} as a function of the $\snr^2$ in L1, $\rho_{\rm L}^2$. 
    The distributions of $\rho_{\rm L}^2$ transition from exponential (chi-squared-like) to power-laws when glitches become important, and hence the rank functions flatten: this happens at progressively lower values of $\rho_{\rm L}^2$ for higher-mass sub-banks. 
    Moreover, even at low values of $\rho_{\rm L}^2$ (in the Gaussian-noise-dominated regime), the rank functions have different slopes in different sub-banks; this is due to the maximization over templates when we collect triggers (see Appendix~\ref{Appendix:collection} for a derivation in a different context). 
    
    We compute FARs over all templates in a chirp-mass bank (which is the union of several sub-banks), and hence we need to fix the normalization of the rank functions when comparing triggers from different sub-banks. Appendix~\ref{Appendix:normrank} describes the procedure; the rank functions shown in Fig.~\ref{fig:RankFunctions} were normalized in this way.}
    
    \item \textit{Computing $p_{\rm astro}$:} We define $\mathcal{R}{({\rm event} \mid \mathcal{N})}$ and $\mathcal{R}{({\rm event} \mid \mathcal{S})}$ to be the rates of a given event under the noise ($\mathcal{N}$) and signal ($\mathcal{S}$) hypotheses. The probability that an event is astrophysical is 
\begin{align}
~~~ & \!\!\!\!
    p_{\rm astro}({\rm event}) \nonumber \\
    & = \frac{\mathcal{R}{({\rm event} \mid \mathcal{S})}}
    {\mathcal{R}{({\rm event} \mid \mathcal{N})} + \mathcal{R}{({\rm event} \mid \mathcal{S})}}. \label{eq:pastrodef}
\end{align}
We define a rate $\mathcal{R}$ for each bank to be the overall number of astrophysical events satisfying $\rho_{\rm H}^2,\rho_{\rm L}^2 > 16$ and $\rho_{\rm H}^2 + \rho_{\rm L}^2 > \rho^2_{\rm th}$, where \change{\begin{align} \nonumber
    \rho^2_{\rm th\,\texttt{BBH 0}}&=67 \\ \nonumber
    \rho^2_{\rm th\,\texttt{BBH 1}}&=65.5 \\ \nonumber
    \rho^2_{\rm th\,\texttt{BBH 2}}&=63.5 \\ \nonumber
    \rho^2_{\rm th\,\texttt{BBH 3}}&=59.5 \\ \nonumber
    \rho^2_{\rm th\,\texttt{BBH 4}}&=56.5 \,.
\end{align}}

The rate $\mathcal{R}$ is for a hypothetical network consisting of two identical detectors, each having a sensitivity equal to the median Hanford sensitivity in the O2 run, which observe in coincidence for $118$ days.
$\mathcal{R}$ is assumed to be uniform across templates within the bank, regardless of which sub-bank they might fall in. For a given event, we have
\begin{equation}
     \mathcal{R}{({\rm event} \mid \mathcal{S})}=W ({\rm event}) \mathcal{R}, \label{eq:wdef}
\end{equation}
where the factor $W$ depends on the instantaneous sensitivities of the detectors, as well as the extrinsic parameters of the event. Note that $W$ does not depend on the (unknown) astrophysical rate. \sk{\teja{$w$ is not a rate} \jr{which defines the normalized rate $W$.} \teja{H1 is also hanford, better to use S and N. I think S is consistent in this paper, where is it that it is inconsistent?} \jr{[Should we go back to $H_0, H_1$ instead of $\mathcal{S, N}$? $\mathcal S$ was something else in the previous section and also we are inconsistent with the other paper.]}}We determine the terms in Eqs.~\eqref{eq:pastrodef} and \eqref{eq:wdef} in a similar manner to that of our O1 analysis~\cite{2019arXiv190210341V}, but with two changes. Firstly, we estimate the rate of producing triggers under the noise hypothesis, $\mathcal{R}{({\rm event} \mid \mathcal{N})}$, using only the background triggers in the respective sub-banks that the candidates belong to. Secondly, we determine the rate of astrophysical events, $\mathcal{R}$, from the data itself rather than assuming it \change{from the loudest events detected in the search.}{.} 

\sk{We determine $\mathcal{R}({\rm event} \mid \mathcal{N})$ by estimating the local density of background triggers at the location of the event by using time slides. We do this estimate on a sub-bank per sub-bank basis. 

Rather than quoting $p_{\rm astro}({\rm event})$ for a fiducial value of $\mathcal{R}$ as we did in \cite{2019arXiv190210341V} we will determine it from the data itself.} 
Given a particular value of the rate $\mathcal{R}$, the likelihood of the data is
\begin{align}
    ~~~~~ \mathcal{L}({\rm data} \mid \mathcal{R}) & \propto e^{-\mathcal{R}} \times \nonumber \\
    & \!\!\!\!\! \prod_{\rm triggers} [{\mathcal{R}({\rm event} \mid \mathcal{N})}+ W ({\rm event}) \mathcal{R}],
\end{align}
where the product is over all the triggers in the bank, including those detected originally by the LVC. 
Using this likelihood, we compute a posterior on $\mathcal{R}$, assuming a uniform prior $P(\mathcal{R})$ between 0 and 50. \change{As earlier, this number is the overall number of triggers of astrophysical origin in a region in the $\rho_H^2$--$\rho_L^2$ plane for a hypothetical two--detector network with each detector having the median H1 sensitivity over the O2 run, i.e., a period of 118 days. 
We expect this number to be of the order of the number of detections, and determine it from the data itself; hence the most important requirement is that the prior be non-informative over the relevant range.}{} For any value of the overall rate $\mathcal{R}$, we can calculate the probability that an event \change{is of}{has} astrophysical origin; our final quoted values were obtained by marginalizing over $\mathcal{R}$:
\begin{align}
    p_{\rm astro}({\rm event}) = \int_0^\infty {p_{\rm astro}({\rm event} \mid \mathcal{R})P(\mathcal{R})d\mathcal{R}}.
\end{align}
    
\end{enumerate}

\section{Results on the previously reported events}
\label{sec:previous_events}

\begin{table*}
    \centering
    \caption{Events already reported by the LIGO--Virgo Collaboration \cite{LIGOScientific:2018mvr} as detected with our pipeline. The rate distributions used to compute $p_{\rm astro}$ are shown in Fig.~\ref{fig:post rates}. The maximum likelihood rates are $\mathcal{R}_{\rm max} = 8/{\rm O2}$ and $5/{\rm O2}$ in banks \texttt{BBH 3} and \texttt{BBH 4}, respectively.}
    \begin{tabular}{|c|c|c|cc|ccc|}
        \tabline
        \tabline
         Name & Bank & GPS time\footnote{The times given are the `linear-free' times of the best fit templates in our bank \change{in the Hanford detector}{}; with this time as the origin, the phase of the template is orthogonal to shifts in time, given the fiducial PSD.} & $\rho_{\rm H}^2$ & $\rho_{\rm L}^2$ & \change{IFAR}{${\rm FAR}^{-1}$} (O2)\footnote{The \change{inverse false alarm rates, or IFARs,}{FARs given} are computed within each bank\change{, and do not include any additional trials factors}{}; our BBH analysis has 5 chirp-mass banks. The \change{IFAR}{inverse FAR} is given in terms of ``O2" \change{instead of physical time, since the ranking statistic includes a time-dependent volumetric correction factor to account for the significant and systematic changes in the network's sensitivity over the run. If the network's sensitivity were constant during the observing run, the unit ${\rm ``O2"} \approx 118$ days.}{to reflect the volumetric weighting of events. Under the approximation of constant sensitivity of the detectors during the observing run, the unit ``O2" corresponds to $\approx 118$ days.}} & $\frac{W(\text{event})}{\mathcal{R}(\text{event}\mid \mathcal{N})}$ (O2) & $p_{\rm astro}$  \\
         \tabline
         
         GW170104 & \texttt{BBH (3,0)} & $1167559936.582$ & $85.1$ & $104.3$ & $>\num{2e4}$ & $>100$ & $>0.99$\\
         GW170809 & \texttt{BBH (3,0)} & $1186302519.740$ & $40.5$ & $113$ & $>\num{2e4}$ & $>100$ & $>0.99$\\
         GW170814 & \texttt{BBH (3,0)} & $1186741861.519$ & $90.2$ & $170$ & $>\num{2e4}$ & $>100$ & $>0.99$\\
         GW170818 & \texttt{BBH (3,0)} & $1187058327.075$ & $19.4$ & $95.1$ & 1.7\footnote{See discussion in Sec.~\ref{sec:previous_events}. \label{Foot:SeeLIGODiscussion}} & --- & ---\textsuperscript{\ref{Foot:SeeLIGODiscussion}}\\
         GW170729 & \texttt{BBH (3,1)} & $1185389807.311$ & $62.1$ & $53.6$ & $>\num{2e4}$ & $>100$ & $>0.99$\\
         GW170823 & \texttt{BBH (3,1)} & $1187529256.500$ & $46.0$ & $90.7$ & $>\num{2e4}$ & $>100$ & $>0.99$\\
         \tabline
         \tabline
    \end{tabular}
    \label{tab:lvc events}
\end{table*}

Table~\ref{tab:lvc events} summarizes our pipeline's results for the O2 events published by the LVC~\cite{LIGOScientific:2018mvr}. We detect all previously reported BBH events except for GW170608, for which the LVC did not provide the Hanford data in their bulk data release, and thus that time was not part of our coincidence search. We only report results from our BBH search in this paper, so we exclude the binary neutron star GW170817 from our results. Nearly all of the LVC events have only an upper limit for the FAR of $1/(\num{20000}\, {\rm O2})$.
All of these events are certainly astrophysical sources with lower limit on $p_{\rm astro} \geqslant 0.99$.

An interesting special case is GW170818, which was not found by the \texttt{PyCBC} pipeline, and deemed potentially interesting but not confirmed by the \texttt{GstLAL} pipeline using Hanford and Livingston alone~\cite{LIGOScientific:2018mvr}. 
It was subsequently detected with high confidence by \texttt{GstLAL} when Virgo data was included; we did not analyze Virgo for the search reported in this paper. 
\sk{Empirically, our background distributions do not have any louder L1 triggers than this event (except for other certain BBH mergers).}
\change{Due to the low score in H1, this is close to a single-detector trigger, and hence the detection heavily relies on our understanding of the background distribution in the vicinity of the L1 $\snr^2$. 
Empirically, our background distributions do not have any L1 triggers that are louder than this event. 
In the regime relevant to triggers like GW170818, the background is dominated by only a few loud glitches, and hence the L1 ranking score, $\tilde{\rho}_{\rm L}^2$, that we use to define our test statistic saturates (see Fig.~\ref{fig:RankFunctions}) and comes with significant error bars. 
Moreover, we cannot compute the {\em local} probability density of the background in the $\rho_{\rm H}^2$--$\rho_{\rm L}^2$ plane without extrapolating from lower values, and this is a vital ingredient in the calculation of $p_{\rm astro}$ (see Eq.~\eqref{eq:pastrodef}). 
Hence we need a more careful analysis to reliably assess the FAR and probabilities of astrophysical origin of such events.}{The low score in H1, combined with the saturation of the L1 ranking scores $\tilde{\rho}_{\rm L}^2$ at high SNR (see Figure \ref{fig:RankFunctions}), together imply that we need a more careful analysis to reliably assess the FAR and probability of astrophysical origin of such events.}
We defer this analysis to a follow-up paper, in which we will show that such events can be detected using the Livingston and Hanford detectors alone. 
In this subsequent work, we will also report the results of a search for similar events, i.e., events which are loud in one detector (and saturate the rank score), and which either have low scores in the other detectors, or have no coincident data.

\change{The following technical details about the quantities we report in Table~\ref{tab:lvc events} (and subsequent tables) are worth noting for clarity:
\begin{itemize}
    \item Our false alarm rates (FARs) are calculated per bank, and do not include any additional trials factors. Hence, the reader should interpret these numbers in the context of the set of banks used in the search. We adopt this convention since (a) FARs reported this way are invariant even if we subsequently expand the search space (i.e., add extra banks), and (b) it makes it convenient for readers to apply their own priors when computing the look-elsewhere correction, or the trials factor (i.e., the interpretation of a particular FAR is different in regions of parameter space known to have larger astrophysical rates).
    \item We report the `linear-free' times of the best fit templates in the template bank in the Hanford detector. The `linear-free' time is defined as follows: suppose that for a particular set of intrinsic parameters, we have a frequency domain template $h(f) = A(f) \exp{\left[ i \psi(f) \right]}$, where the phase $\psi(f)$ is unwrapped as a function of frequency. Suppose through the inverse Fourier transform, we have a time-domain template $h(t) = \int {\rm d} f \, h(f) \exp(i 2 \pi f t)$, defined on the domain $[0, T]$. The `linear-free' time of the template $h(t)$ occurs at $t = T - t_0$, where $t_0$ satisfies the equations
    \begin{align}
        \begin{bmatrix}
          \langle f, f \rangle & \langle f, 1 \rangle \\
          \langle f, 1 \rangle & \langle 1, 1 \rangle
        \end{bmatrix} 
        \begin{bmatrix}
          2 \pi t_0 \\
          \psi_0
        \end{bmatrix}
        & =
        -\begin{bmatrix}
          \langle \psi, f \rangle \\
          \langle \psi, 1 \rangle
        \end{bmatrix}, {\rm with} \label{eq:linearfree} \\
        \langle a(f), b(f) \rangle & = \int {\rm d} f \frac{A^2(f)}{S(f)} a(f) b(f). \label{eq:innerproduct}
    \end{align}
    In the above equation, $S(f)$ is the PSD of the detector noise. 
    Intuitively, after applying the parameter-dependent shift $t_0$ to the template, the measurements of arrival time and intrinsic parameters are uncorrelated in a single detector (see related discussion in Ref.~\cite{templatebankpaper}). When using Eqs.~\eqref{eq:linearfree}-\eqref{eq:innerproduct}, i.e., only to define $t_0$, we used the same fiducial PSD that we used to construct the template bank as described in Sec.~\ref{sec:changes}. 
    For the short waveforms in the heavier binary black hole banks, the linear-free time is practically close to the merger time.
\end{itemize}
}{}

\section{Overview of the newly discovered BBH mergers}
\label{sec:events}

\begin{table*}
    \centering
    \caption{New events with astrophysical probability $> 50\%$ in all of the BBH banks. The rate distributions used to compute $p_{\rm astro}$ are shown in Fig.~\ref{fig:post rates}, the maximum-likelihood rates in banks \texttt{BBH 3} and \texttt{BBH 4} are $\mathcal{R}_{\rm max} = 8/{\rm O2}$ and $5/{\rm O2}$, respectively.}
    \begin{tabular}{|c|c|ccc|c|cc|ccc|}
        \tabline
        \tabline
         Name & Bank & $\mathcal M^{\rm det} (\rm M_\odot)$ & $\chi_{\rm eff}$ & $z$ & GPS time\footnote{The times given are the `linear-free' times of the best fit templates in our bank \change{in the Hanford detector}{}; with this time as the origin, the phase of the template is orthogonal to shifts in time, given the fiducial PSD.} & $\rho_{\rm H}^2$ & $\rho_{\rm L}^2$ & \change{IFAR}{${\rm FAR}^{-1}$} (O2)\footnote{The \change{inverse false alarm rates, or IFARs,}{FARs given} are computed within each bank\change{, and do not include any additional trials factors}{}; our BBH analysis has 5 chirp-mass banks. The \change{IFAR}{inverse FAR} is given in terms of ``O2" \change{instead of physical time, since the ranking statistic includes a time-dependent volumetric correction factor to account for the significant and systematic changes in the network's sensitivity over the run. If the network's sensitivity were constant during the observing run, the unit ${\rm ``O2"} \approx 118$ days.}{to reflect the volumetric weighting of events. Under the approximation of constant sensitivity of the detectors during the observing run, the unit ``O2" corresponds to $\approx 118$ days.}} & $\frac{W(\text{event})}{\mathcal{R}(\text{event}\mid \mathcal{N})}$ (O2) & $p_{\rm astro}$  \\
         \tabline
         GW170121 & \texttt{BBH (3,0)} & $29 ^{+4}_{-3}$ & $-0.3 ^{+0.3}_{-0.3}$ &  $0.24 ^{+0.14}_{-0.13}$ & $1169069154.565$ & $29.4$ & $89.7$ & $2.8\times 10^3$ & $>30$ & $>0.99$ \\
         GW170304 & \texttt{BBH (4,0)} & $47 ^{+8}_{-7}$ & $0.2 ^{+0.3}_{-0.3}$ &  $0.5 ^{+0.2}_{-0.2}$ & $1172680691.356$ & $24.9$ & $55.9$ & 377 & $13.6$ & $0.985$ \\
         GW170727 & \texttt{BBH (4,0)} & $42^{+6}_{-6}$     & $-0.1 ^{+0.3}_{-0.3}$ &  $0.43 ^{+0.18}_{-0.17}$ & $1185152688.019$ & $25.4$ & $53.5$ & 370 & $11.8$ & $0.98$ \\
         GW170425 & \texttt{BBH (4,0)} & $47^{+26}_{-10}$ & $0.0 ^{+0.4}_{-0.5}$ &  $0.5 ^{+0.4}_{-0.3}$ & $1177134832.178$ & $28.6$ & $37.5$ & 15 & $0.65$ & $0.77$ \\
         GW170202 & \texttt{BBH (3,0)} & $21.6_ {-1.4} ^ {+4.2} $ & $-0.2 ^{+0.4}_{-0.3}$ &  $0.27 ^{+0.13}_{-0.12}$ & $1170079035.715$ & $26.5$ & $41.7$ & 6.3 & $0.25$ & $0.68$ \\
         GW170403 & \texttt{BBH (4,1)} & $48 ^{+9}_{-7}$    & $-0.7 ^{+0.5}_{-0.3}$ &  $0.45 ^{+0.22}_{-0.19}$ & $1175295989.221$ & $31.3$ & $31.0$   & 4.7   & $0.23$ & $0.56$ \\
         \tabline
         \tabline
    \end{tabular}
    \label{tab:signalsFound}
\end{table*}

Table~\ref{tab:signalsFound} summarizes the basic properties of the newly discovered events: their parameters, \change{inverse FAR, or IFAR}{FAR}, and estimated probabilities of being of astrophysical origin, $p_{\rm astro}$. 
We need the rate of astrophysical events in the detector, $\mathcal{R}$, to calculate \change{these values of $p_{\rm astro}$, and hence, we also report the values of the factor $W$ (from Eq.~\eqref{eq:wdef}), so that external sources of information about the rate (e.g., future runs) can be easily incorporated if needed}{these probabilities}.

\begin{figure}
    \centering
    \includegraphics[width=\linewidth]{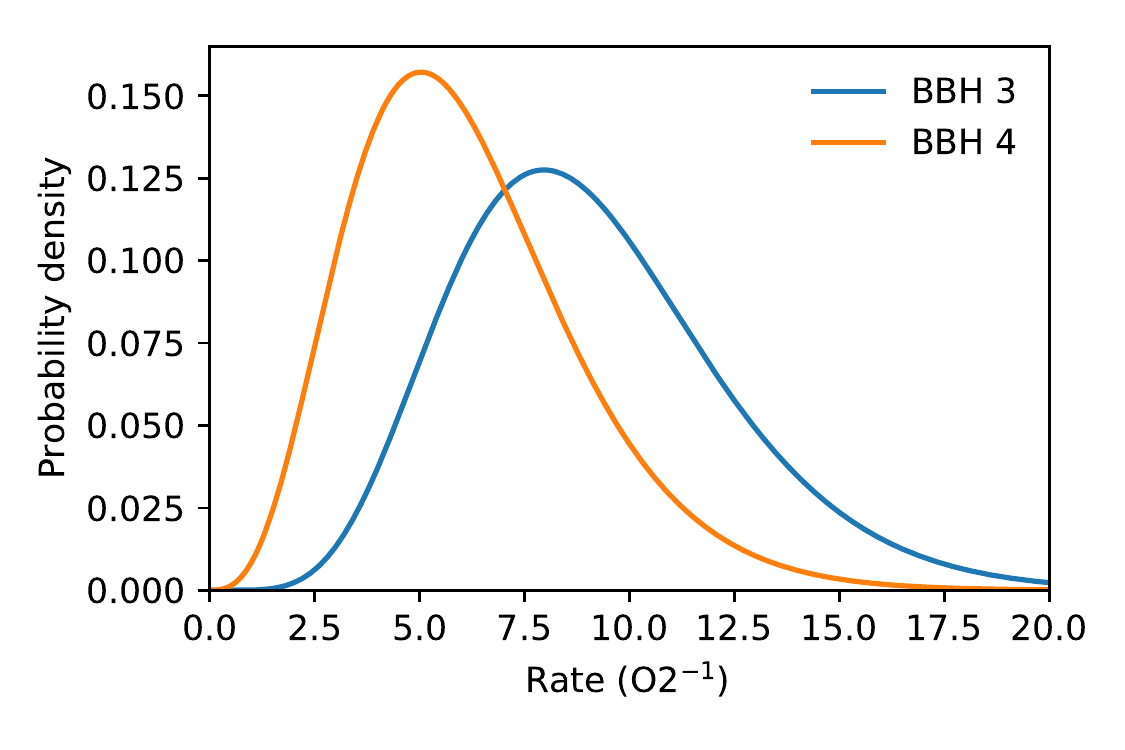}
    \caption{Rates inferred from all the events. The rates are simplistic and bank-specific observed rates for black hole mergers, which we only use to determine the $p_{\rm astro}$ of the events. The rates are defined as the astrophysical occurence rate of signals satisfying $\rho_{\rm H}^2,\rho_{\rm L}^2 > 16$ and $\rho_{\rm H}^2 + \rho_{\rm L}^2 > \rho^2_{\rm th}$, where $\rho^2_{\rm th}=60$ for \texttt{BBH 3} and $\rho^2_{\rm th}=57$ for \texttt{BBH 4}.
    The assumed network consists of two identical detectors with median Hanford sensitivity, observing in coincidence for $118$ days. More careful analysis is required in order to infer astrophysically meaningful volumetric rates.}
    \label{fig:post rates}
\end{figure}

We derive a distribution for $\mathcal{R}$ from all the events we detected (including those already detected by the LVC analyses), and at the same time estimate the $p_{\rm astro}$ of each event, using the procedure described in Section~\ref{sec:changes}. 
Figure~\ref{fig:post rates} shows the posteriors on the rates for chirp-mass banks \texttt{BBH 3} and \texttt{BBH 4}; the values of $p_{\rm astro}({\rm event})$ quoted in Tables~\ref{tab:lvc events} and \ref{tab:signalsFound} were marginalized over these posteriors. \change{As is apparent from Table~\ref{tab:signalsFound}, all of our detections are in the high-mass region of parameter space, which is covered by the banks \texttt{BBH 3} and \texttt{BBH 4}. Hence, curves analogous to those in Fig.~\ref{fig:post rates} for the other banks only yield upper bounds on the respective rates (of the mergers detected in previous runs, GW151012 and GW151226 lie in banks \texttt{BBH 2} and \texttt{BBH 1}, respectively).}{}

\change{As in Table~\ref{tab:lvc events}, the IFARs in Table~\ref{tab:signalsFound} are calculated per bank, and do not include any additional trials factors. The criterion in Table~\ref{tab:signalsFound} is based on the value of $p_{\rm astro}$, but all the events have significantly higher values of the `per-bank' IFAR than the threshold of 1 in 30 days, or $\sim 0.25$ O2, adopted in GWTC-1 \cite{LIGOScientific:2018mvr} (i.e., their IFARs exceed 0.25 O2 even if divided by a trials factor of five).}{}

\begin{figure*}
     \centering
     \includegraphics[width=.3\linewidth]{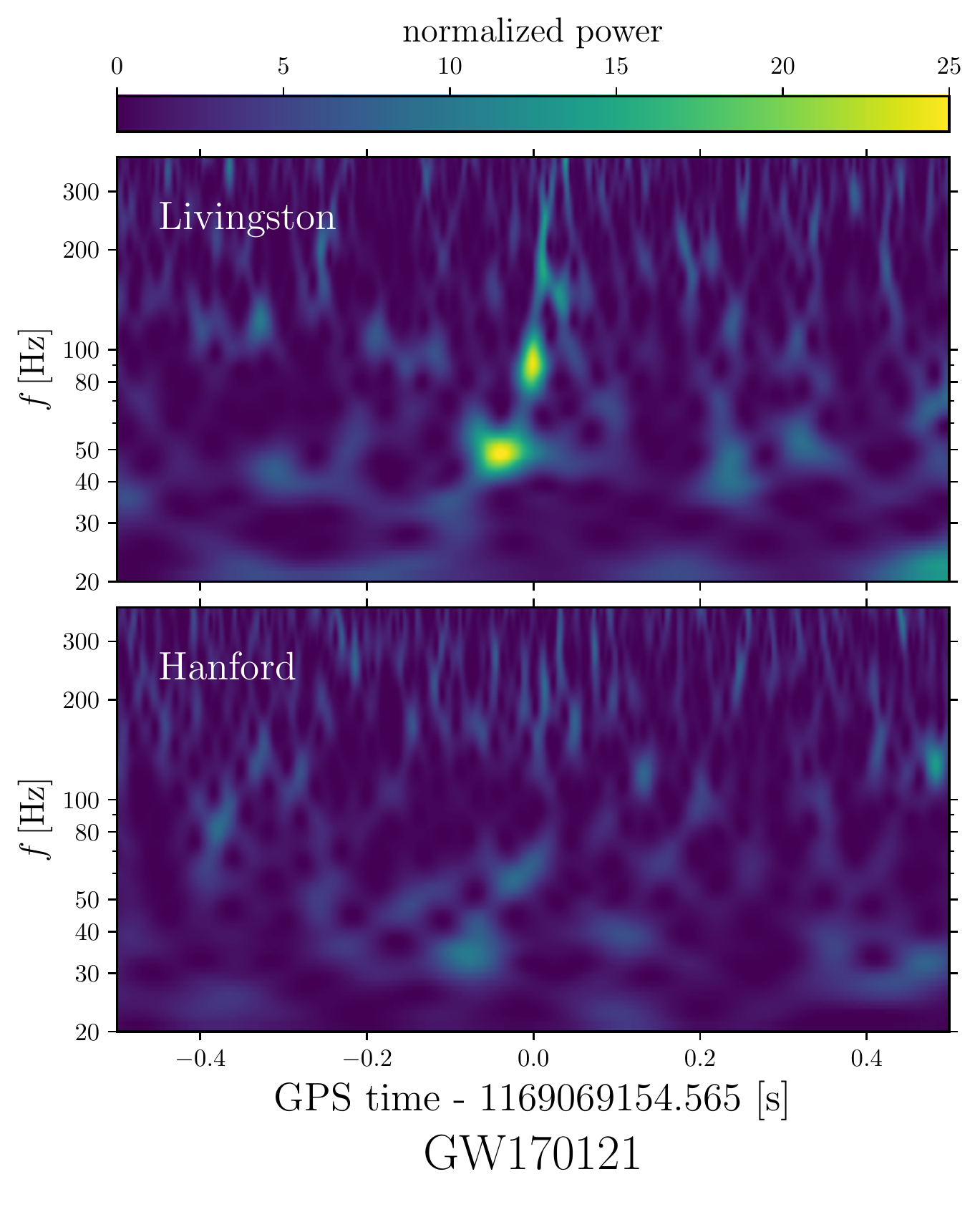}
     \includegraphics[width=.3\linewidth]{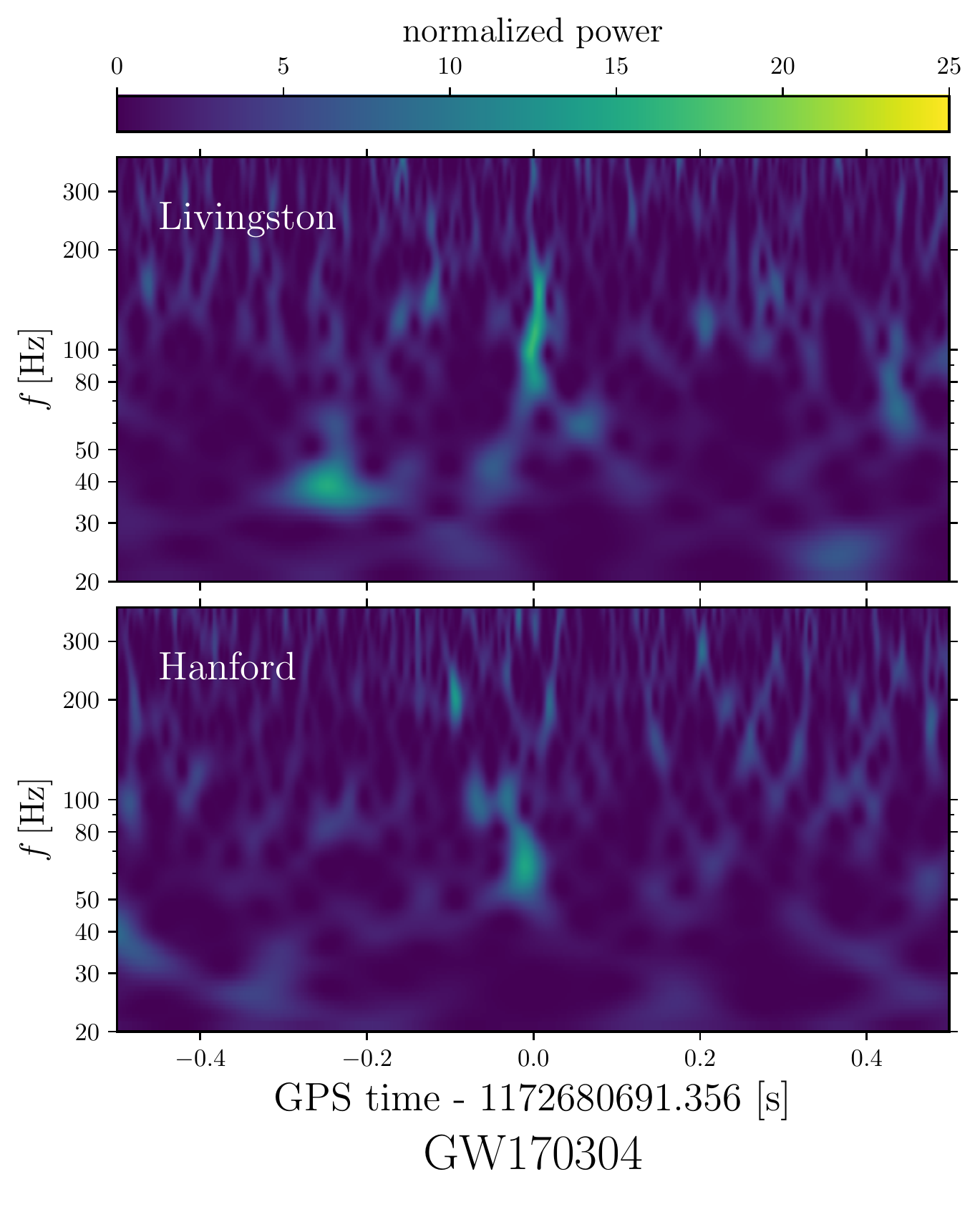}
     \includegraphics[width=.3\linewidth]{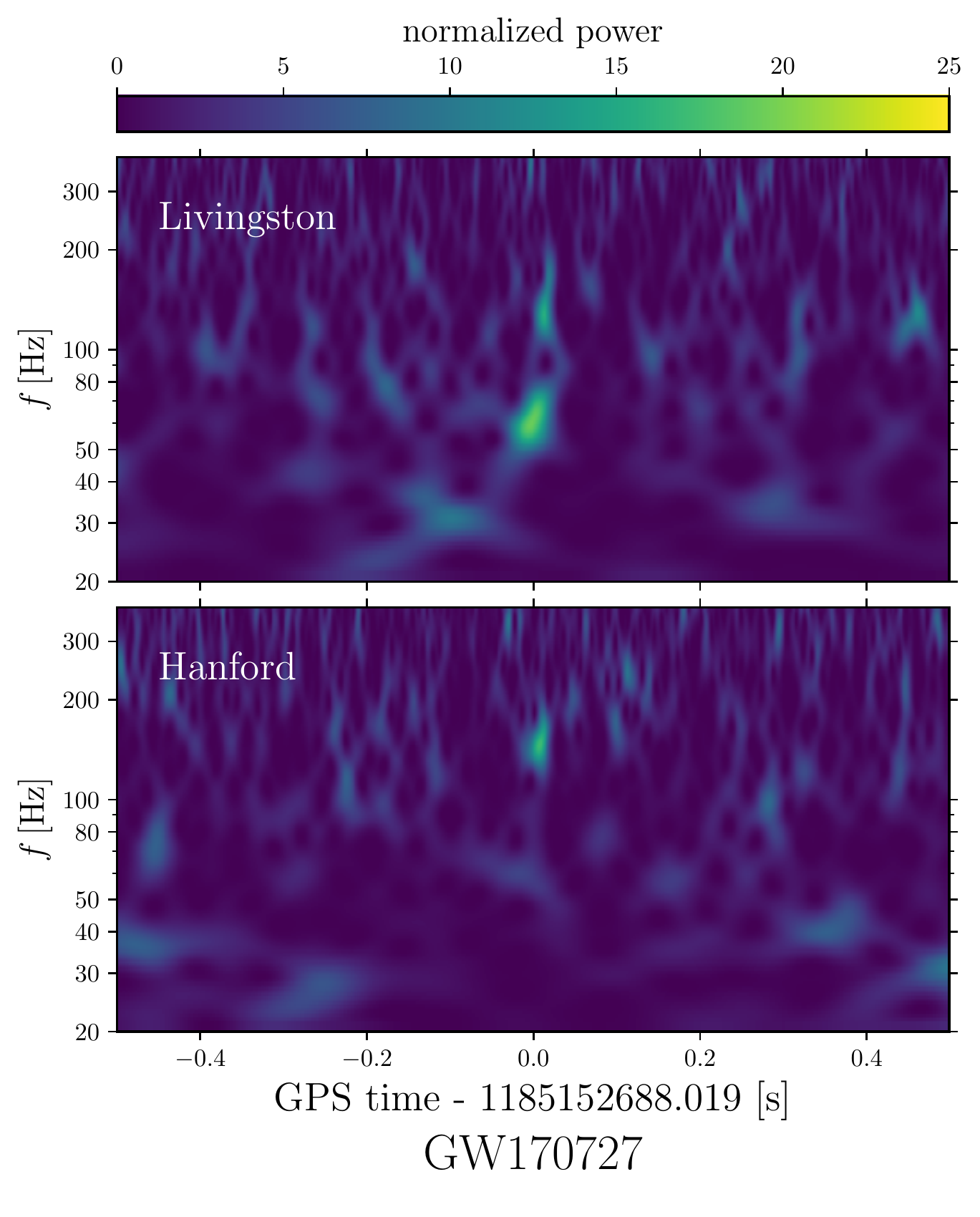}\\
     \includegraphics[width=.3\linewidth]{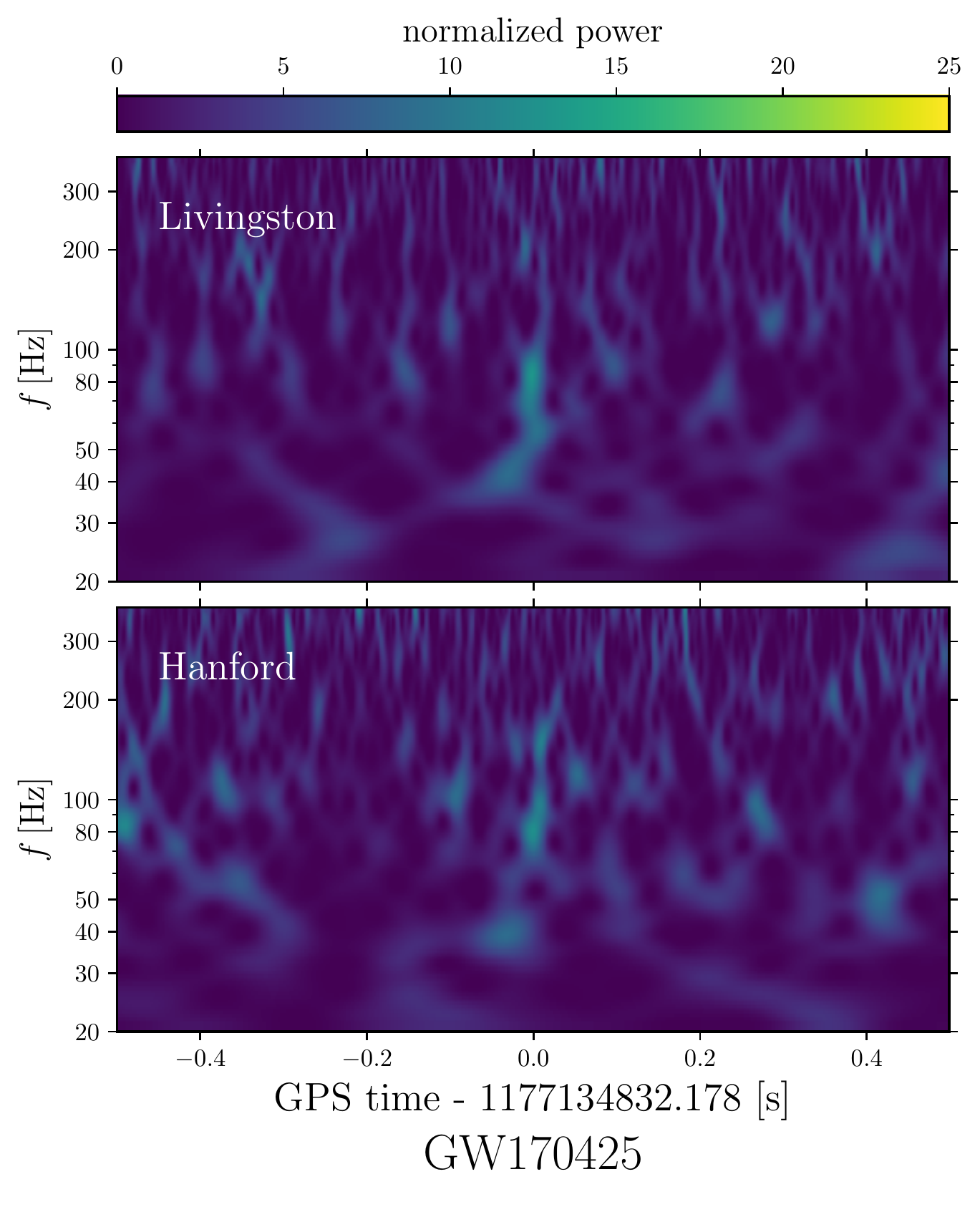}
     \includegraphics[width=.3\linewidth]{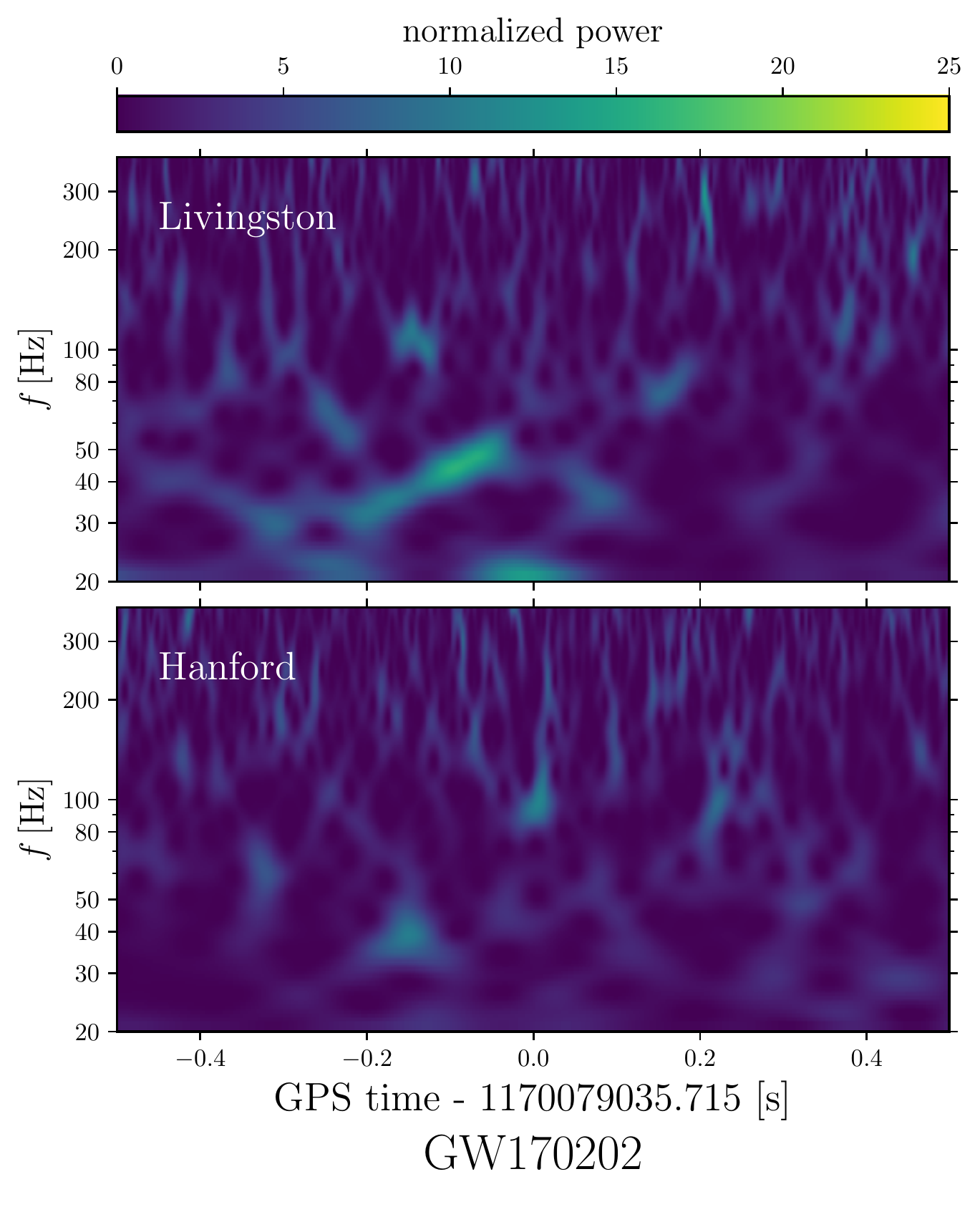}
     \includegraphics[width=.3\linewidth]{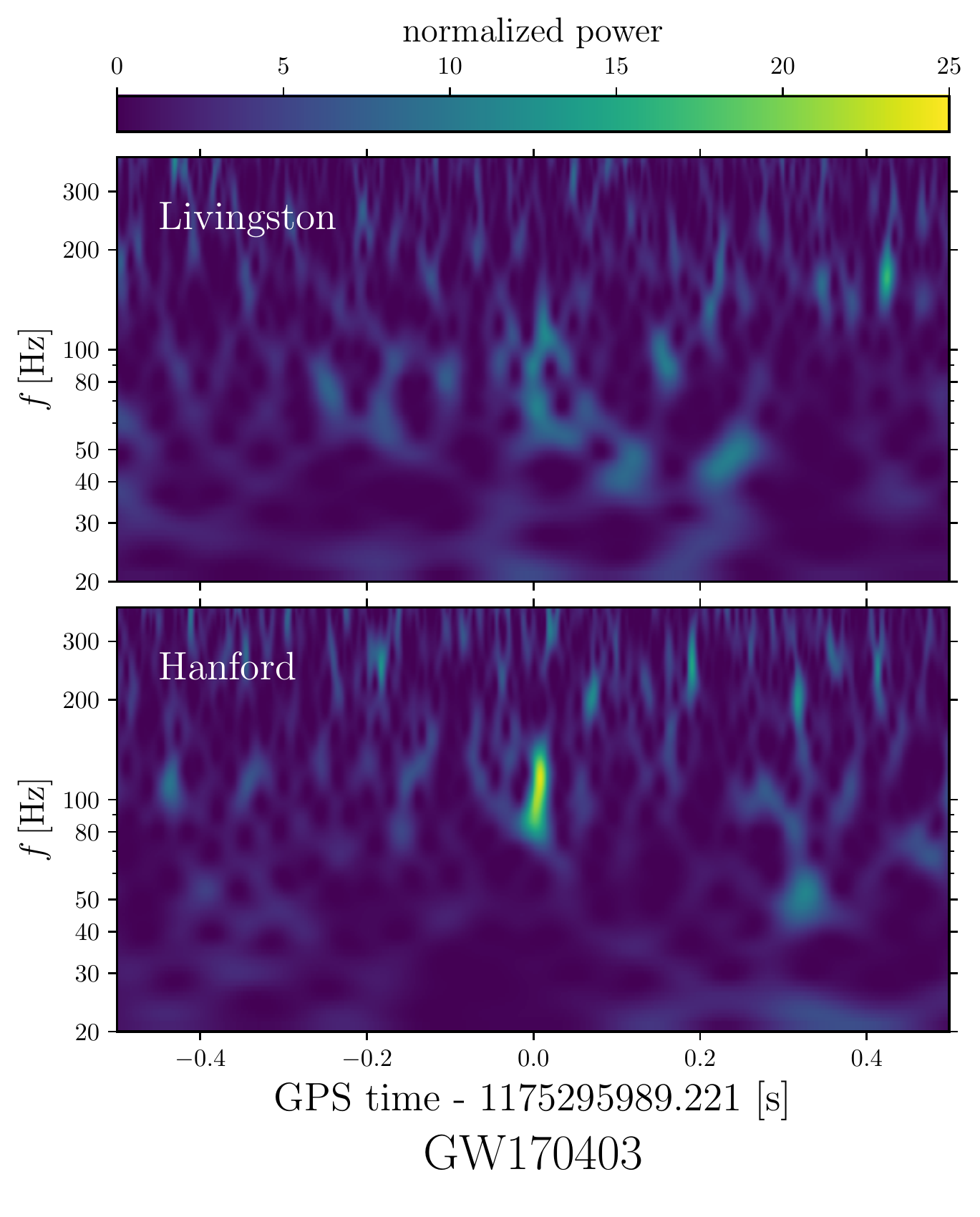}
     \caption{\label{fig:specgrambbh} \change{The various panels show spectrograms for 1 s segments of data around the GPS times of the events in the Livingston and Hanford detectors. In each figure, the data shown was whitened with a filter constructed using the PSD measured from a 4096 s segment of data containing the event but not corrected for any short-term PSD variations within.}{}}
 \end{figure*}

\change{Figure \ref{fig:specgrambbh} shows the spectrograms for segments of data of length 1 s around the GPS times of the events in the Hanford and Livingston detectors. 
These spectrograms show no obvious evidence of glitches in the immediate vicinity of the events, and are visually consistent with how binary black hole merger signals with the reported values of SNR would look like in the presence of additive Gaussian noise. 
Note that visual inspection of spectrograms is not a particularly good, or even well-defined way to identify either glitches or signals in the data: our pipeline automatically flags bad segments of data using mathematically well-defined test statistics (see TV19 for details), and looks for signals using matched filtering.}{}

Appendix~\ref{Appendix:posteriors} includes the posteriors for the parameters of all the new events. Figure~\ref{fig:events} places these events in the context of the previous LVC events, as well as the one we reported in Ref.~\cite{GW151216}, by showing their distribution in the plane of source-frame total mass and the effective spin parameter $\chieff$. In the remainder of this section, we briefly comment on the properties of each of the newly found events. 

\paragraph{GW170121:} This event has the lowest FAR among those not reported by the LVC ($\change{\rm IFAR}{{\rm FAR}^{-1}} \approx 2.8 \times 10^3\,\rm O2$), and $p_{\rm astro} > 0.99$. The posterior distribution of $\chieff$ has most of its support at negative values, $\chieff > 0$ being ruled out at the $95\%$ confidence level. The chirp mass, mass-ratio and redshift of this event are similar to those of the heavy BBHs reported by the LVC. Its redshift is $z \sim 0.24$.

\paragraph{GW170304 \& GW170727:} 
These two events have $\change{\rm IFAR}{{\rm FAR}^{-1}} \sim 370\,\rm O2$ and
$p_{\rm astro}\approx 0.98$. 
Their masses and spins are similar to those of the heavy BBHs detected by the LVC. Both events are consistent with zero $\chieff$, and are on the massive end of the population. They have relatively high redshifts $z \sim 0.5$ and $0.43$, respectively.

\paragraph{GW170425:} This candidate has $p_{\rm astro}\approx 0.77$ and $\change{\rm IFAR}{{\rm FAR}^{-1}} \approx 29\,\rm O2$. Its inferred parameters are similar to those of the heavy BBHs reported by the LVC; the effective spin $\chieff$ is consistent with zero. The posterior distribution has a tail extending to large values for the masses. Its inferred redshift is large, $z\sim 0.5$.

\paragraph{GW170202:} This candidate has $p_{\rm astro}\approx 0.7$ and $\change{\rm IFAR}{{\rm FAR}^{-1}} \approx 6\,\rm O2$. The masses and the spins are similar to those of the heavy LVC BBHs. It is found in the bank with the largest number of secure detections (\texttt{BBH 3}). It has a bimodal posterior, in which the solution with lower masses has a more negative spin, and is located closer. The inferred redshift is $z \sim 0.27$.

\paragraph{GW170403:} This candidate has $\change{\rm IFAR}{{\rm FAR}^{-1}} \approx 5\,\rm O2$ and $p_{\rm astro}\approx 0.55$; this is close to the threshold $p_{\rm astro} = 0.5$ to make it into a list of detections (as defined in Ref.~\cite{LIGOScientific:2018mvr}). The inferred redshift is $z \sim 0.45$.
Interestingly, the posterior for $\chieff$ is inconsistent with positive values.

\begin{table*}
    \centering
    \caption{Sub-threshold candidates with astrophysical probability above 10\% in all of the BBH banks. The rate distributions used to compute $p_{\rm astro}$ are shown in Fig.~\ref{fig:post rates}, the maximum-likelihood rates in banks \texttt{BBH 3} and \texttt{BBH 4} are $\mathcal{R}_{\rm max} = 8/{\rm O2}$ and $5/{\rm O2}$, respectively.}
    \begin{tabular}{|c|c|cc|ccc|}
        \tabline
        \tabline
         Bank & GPS time\footnote{The times given are the `linear-free' times of the best fit templates in our bank \change{in the Hanford detector}{}; with this time as the origin, the phase of the template is orthogonal to shifts in time, given the fiducial PSD.} & $\rho_{\rm H}^2$ & $\rho_{\rm L}^2$ & \change{IFAR}{${\rm FAR}^{-1}$} (O2)\footnote{The \change{inverse false alarm rates, or IFARs,}{FARs given} are computed within each bank\change{, and do not include any additional trials factors}{}; our BBH analysis has 5 chirp-mass banks. The \change{IFAR}{inverse FAR} is given in terms of ``O2" \change{instead of physical time, since the ranking statistic includes a time-dependent volumetric correction factor to account for the significant and systematic changes in the network's sensitivity over the run. If the network's sensitivity were constant during the observing run, the unit ${\rm ``O2"} \approx 118$ days.}{to reflect the volumetric weighting of events. Under the approximation of constant sensitivity of the detectors during the observing run, the unit ``O2" corresponds to $\approx 118$ days.}} & $\frac{W(\text{event})}{\mathcal{R}(\text{event}\mid \mathcal{N})}$ (O2) & $p_{\rm astro}$ \\
         \tabline
         \texttt{BBH (4,1)} & $ 1172487817.477$ & $48.6$ & $19.1$ & $0.82$ & $0.147$ & $0.45$\\
         \texttt{BBH (3,0)} & $ 1170914187.455$ & $20.4$ & $41.4$ & $0.43$ & $0.044$ & $0.28$ \\
         \texttt{BBH (3,1)} & $ 1172449151.468$ & $29.5$ & $32.4$ & $0.31$ & $0.025$ & $0.18$ \\
         \texttt{BBH (4,0)} & $ 1174138338.385$ & $37.1$ & $28.4$ & $0.62$ & $0.034$ & $0.17$ \\
         \texttt{BBH (3,0)} & $ 1171863216.108$ & $46.5$ & $21.6$ & $0.27$ & $0.016$ & $0.125$ \\
         \texttt{BBH (3,1)} & $ 1187176593.222$ & $20.3$ & $42.0$ & $0.2$ & $0.014$ & $0.12$ \\
         \texttt{BBH (3,0)} & $ 1182674889.044$ & $34.1$ & $28.7$ & $0.23$ & $0.016$ & $0.12$ \\
         \texttt{BBH (3,1)} & $ 1171410777.200$ & $40.8$ & $21.0$ & $0.18$ & $0.014$ & $0.11$ \\
         \tabline
         \tabline
    \end{tabular}
    \label{tab:subthreshold candidates}
\end{table*}

In addition to these events, we list in Table~\ref{tab:subthreshold candidates} the sub-threshold triggers of our search, defined as those with $0.1 < p_{\rm astro} < 0.5$. The sum of the $p_{\rm astro}$ of the events in this list exceeds unity; in fact, a candidate in bank \texttt{BBH (4,1)} has $p_{\rm astro} \approx 0.45$, which is close to the detection threshold (though it has a relatively high $\change{\rm IFAR}{{\rm FAR}^{-1}} \approx 0.8\,$O2). It is possible that an improved analysis, or rate-estimate, can push some of these candidates above the detection threshold. \change{Note that we have a threshold of $p_{\rm astro} > 0.1$ for candidates to appear in Table~\ref{tab:subthreshold candidates}. The set of loudest triggers in banks \texttt{BBH 2}, \texttt{BBH 1}, and \texttt{BBH 0} did not pass this cut.}

\section{Sensitivity of our pipeline}
\label{sec:sensitivity}

In the previous section, we described several additional events we detected that are not in the catalog of events published by the LVC. 
All of these events pass the thresholds for detection in \change{GWTC-1 \cite{LIGOScientific:2018mvr}}{Ref.~\cite{LIGOScientific:2018mvr}} (their FARs are above the threshold of 1 in 30 days, even accounting for the five banks in our BBH search, or eleven banks in a hypothetical binary neutron star and neutron-star--black-hole search~\cite{templatebankpaper}\change{, in the extreme scenario in which searches for different kinds of astrophysical systems are considered together}{}). 

\change{The values of the FAR and $p_{\rm astro}$ of the events come with error bars, and for the same event, their values can fluctuate even between two searches that use the same algorithms due to choices within. 
Given that searches operate with thresholds, the presence or absence of a near-threshold event in one search, {\em by itself}, is not proof of a difference in sensitivity. 
However, equally sensitive searches should agree on the overall population of events, and, on average, assign comparable values of FAR and $p_{\rm astro}$ to comparable events. 
Our results in Sec.~\ref{sec:previous_events} show that we assign systematically lower values of FAR to the events discovered by the LVC pipelines in general, and in particular, for GW170729, the least secure event in GWTC-1. 
Moreoever, we have a substantial number of events that individually clear the thresholds 
for detection; while it is certainly possible that some of the events with lower values of $p_{\rm astro}$ could have benefited from upward fluctuations in their test statistics, it is hard to explain the entire population away this way. 
These two effects suggest that our analysis has a substantially larger sensitive volume within the search space that we defined in Sec.~\ref{sec:changes}.}{This suggests that our search has a substantially larger sensitive volume. }

\change{In an idealized case, a simple way to compare sensitive volumes would be to inject a large number of signals in the data and measure the fraction recovered above the appropriate thresholds in the test statistics adopted. 
We will present the results of such a systematic injection campaign in a future paper; pending this, we would like to obtain a simple estimate of the additional sensitivity. 
In the rest of this section, we will consider the population of loud astrophysical events in this data akin to a common set of injected signals, and get a simple `back-of-the-envelope' level estimate of the order of magnitude of the sensitivity change. 
Since there are relatively few events, this is necessarily highly uncertain, and hence the results should not be over-interpreted.}{}

\begin{figure*}
    \centering
    \begin{subfigure}[t]{.47\linewidth}
        \includegraphics[width=\linewidth]{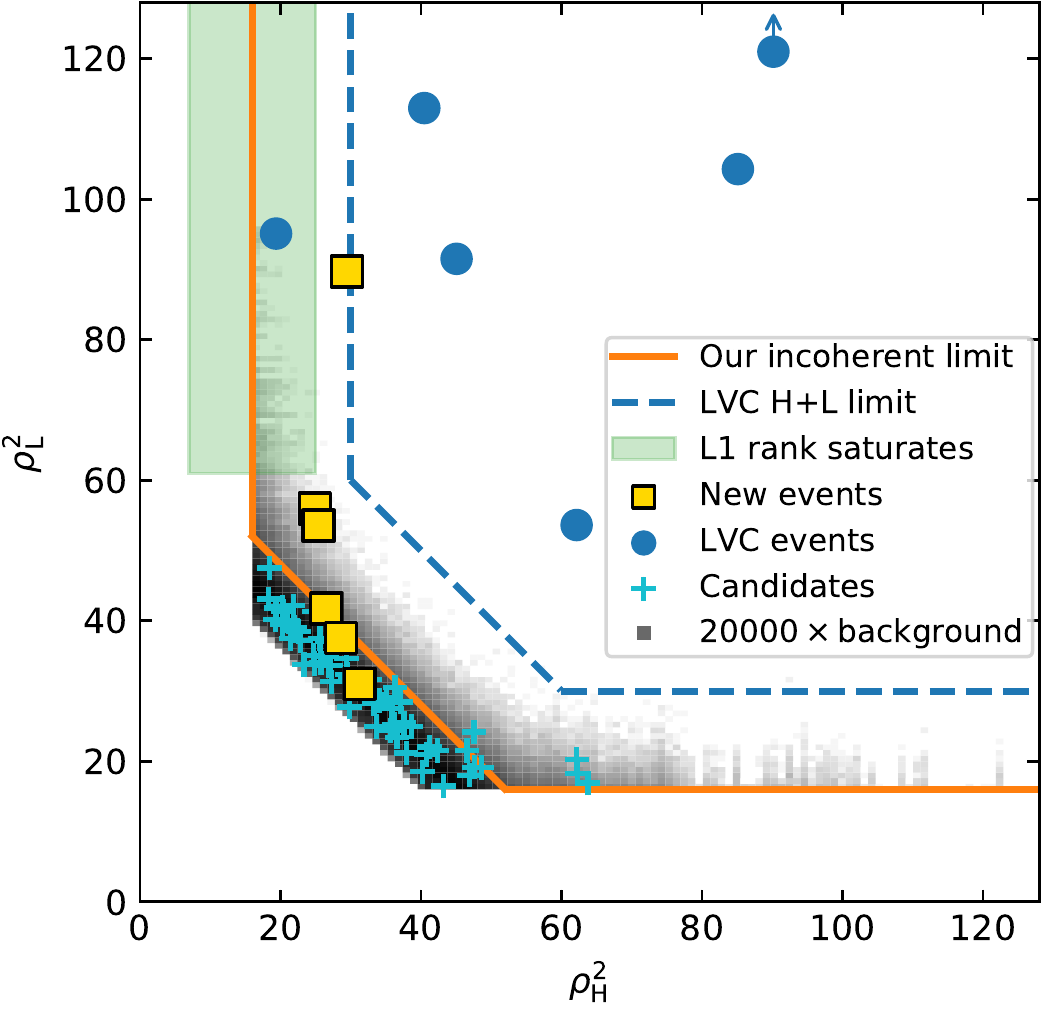}
        \caption{\label{fig:snr_limit} Incoherent $\snr^2$.}
    \end{subfigure}%
    \begin{subfigure}[t]{.53\linewidth}
        \includegraphics[width=\linewidth]{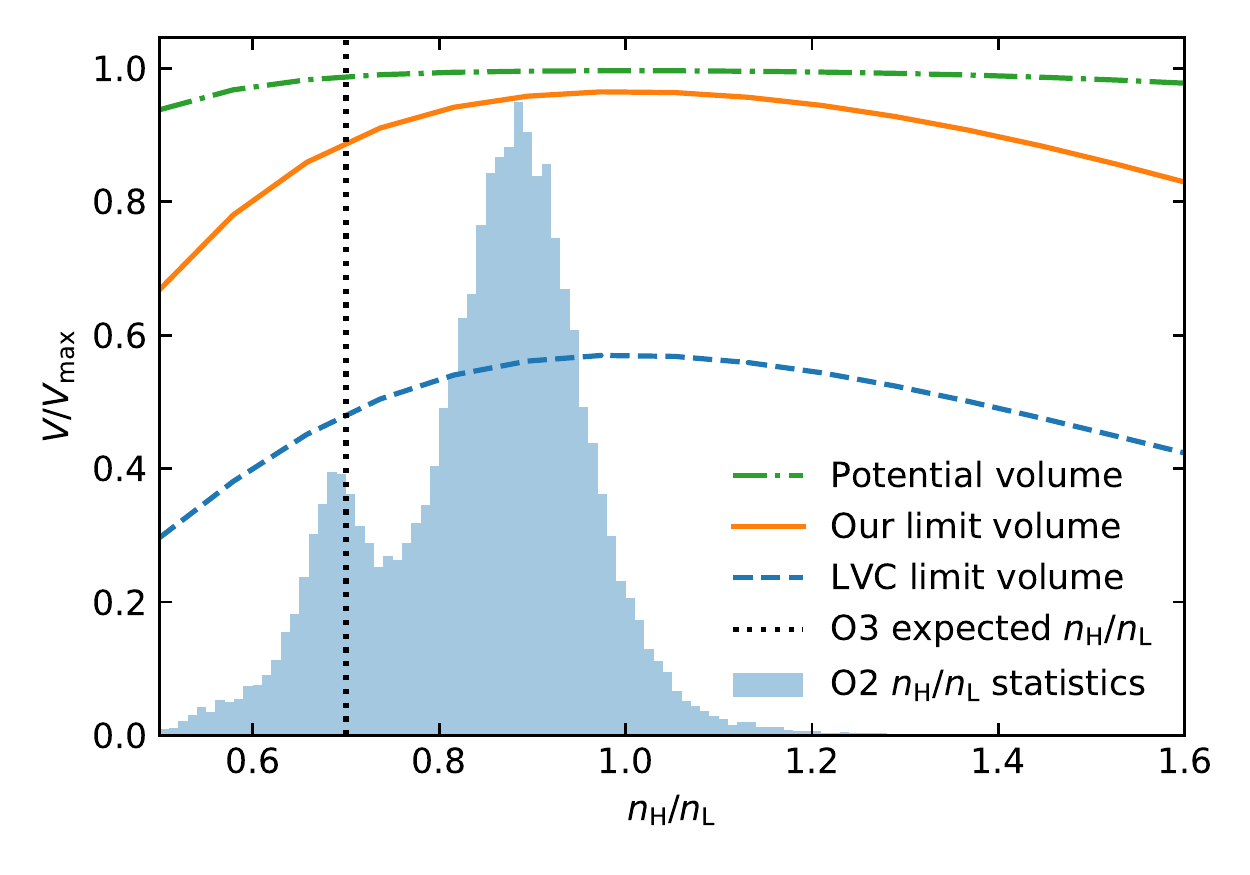}
        \caption{\label{fig:volume} Sensitive volume.}
    \end{subfigure}
    \caption{\textit{Left:} Incoherent Hanford and Livingston $\snr^2$ for coincident and background triggers (computed using \num{2e4} time slides), for all the sub-banks where there are events. The blue and orange lines are approximate incoherent detection limits for the LVC and the current analysis, respectively, restricted to using Hanford and Livingston data only (see text for caveats). \sk{The incoherent detection limits are inexact, since they do not account for the consistency of the time-delays, phases, and the relative sensitivities of the detectors. We show them only to qualitatively assess the relative volumes probed by different analyses.} GW170814 has $\rho_{\rm L}^2=170$, higher than shown here (indicated with an arrow). GW170608 is not shown, see Sec.~\ref{sec:previous_events}.
    \textit{Right:} The lines show the (incoherent) volume probed by different analyses as a function of the ratio of the Hanford and Livingston detector sensitivities, at fixed network total sensitivity. The orange (solid) and blue (dashed) curves show the estimated volume probed by the present and LVC analyses, respectively. The green, dashed-dotted curve shows the potential volume that can be opened up by analyzing interesting single-detector triggers. The shaded histogram shows the distribution of $V \times T$, i.e., the product of the sensitive volume and the time, as a function of the sensitivity ratios between the Hanford and Livingston detectors in the O2 run, as measured by our analysis. The O3 run, as reported in Ref.~\cite{gwosc_det_status}, has begun with a sensitivity ratio $n_{\rm H}/n_{\rm L} \approx 0.7$.}
    \label{fig:IncoherentPlots}
\end{figure*}

Figure~\ref{fig:snr_limit} shows the background triggers we collected using \num{20000} time slides in those BBH sub-banks in which all the events considered in this work, both from the LVC and our analysis, reside. This figure does not include the BBHs from the O1 run (GW150914, GW151012, GW151216, GW151226), nor GW170608, which was not included in the bulk data release we analyzed. This figure is not intended as a demonstration of how we compute the FAR or $p_{\rm astro}$ for particular events: firstly, it shows  $\rho_{\rm H}^2$ and $\rho_{\rm L}^2$, i.e., the incoherent H1 and L1 $\snr^2$, while we compute the FAR using a coherent score that takes into account the time-delays and the relative phases of the triggers, and the differing detector sensitivities; secondly, we estimate the FAR and the $p_{\rm astro}$ for a particular event using the background in its chirp-mass bank, and sub-bank, respectively. We include this figure only to easily visualize the sensitive volume. 

It is clear from Fig.~\ref{fig:IncoherentPlots} that our pipeline has substantially lower background in the relevant region: for example, we see no background triggers within the sensitive region of the \change{LVC pipelines}{standard pipelines}. 
All but one of the LVC reported events have values of $\rho_{\rm L}^2$ that are so large that we do not have even single-detector background triggers at their level (this is a consequence of our data-cleaning procedure, as well as our signal-quality vetoes). 
The only exception to this is GW170729, with $(\rho_{\rm H}^2,\rho_{\rm L}^2)=(62,53)$. This event had FARs of \SI{0.2}{yr^{-1}} and \SI{1.36}{yr^{-1}} in the \texttt{GstLAL} and \texttt{PyCBC} pipelines, respectively, but we have no background in its vicinity even incoherently (i.e., allowing for arbitrary phases, time-delays, and sensitivity ratios of the two detectors). 

\change{In order to calculate sensitive volumes, we need to estimate the detection limits of the analysis. 
With the above caveat on the validity of thresholds based on incoherent $\snr^2$, the solid and dashed lines show the approximate detection thresholds that we judged as appropriate for different analyses.}{
The solid and dashed lines show the approximate detection thresholds for different analyses (with the above caveat on the validity of incoherent thresholds).} The detection thresholds shown for the LVC catalog are approximate and conservative, they err on the side of reporting a better sensitivity\change{}{for the standard pipeline}. At the single-detector level, we set the threshold \change{to $\rho_{\rm H}^2,\rho_{\rm L}^2 > 30$ }{}by the non-detection of GW170121\change{.}{ (the \texttt{PyCBC} pipeline has an explicit cut on single-detector $\snr = 5.5$~\cite{PYCBCPipeline}).}
We set the minimum network $\snr^2 = \rho_{\rm H}^2 + \rho_{\rm L}^2 > 90$ by scaling the reported FAR of GW170729 to 1/O2, and rounding down. 
\change{We approximated our incoherent limit as $\rho_{\rm H}^2 + \rho_{\rm L}^2 > 68$, and $\rho_{\rm H}^2,\rho_{\rm L}^2 > 16$, based on our cuts, and the FAR we would assign to events at this level given our search background.}{}

The difference in the detection limits, at the same threshold on FAR, maps to a difference in the sensitive volume between the searches. The size of this difference depends on the ratio of the sensitivities of the two detectors we analyze, i.e., H1 and L1. The sensitivities $n_{\rm H}$ and $n_{\rm L}$ are proportional to the SNR with which a gravitational wave signal with a given strain amplitude is measured by H1 and L1, respectively.
Figure \ref{fig:volume} shows the volume as a function of the sensitivity ratio (scaled such that a volume of $V_{\rm max}$ corresponds to detecting all events with $\rho_{\rm H}^2 + \rho_{\rm L}^2 > 68$); the solid orange and the dashed blue lines are for the cuts shown in Fig.~\ref{fig:snr_limit}. The curves were obtained by randomizing the angular locations and inclinations of a large number of mergers on the sky and recording the incoherent scores (including the stochastic noise contribution) at the detectors with a given sensitivity-ratio. 
\change{We see that the gain between the blue and range lines depends significantly on the relative sensitivities of the H1 and L1 detectors.}{We see that the orange curve gains approximately a factor of two in sensitive volume. In particular, note that the gain is larger when the sensitivities of the detectors are very different.} The filled histogram in Figure \ref{fig:volume} shows the distribution of the ratio of the Hanford and Livingston sensitivities in our analysis of the O2 run, weighted by the momentary space-time volume. \change{We see that the orange curve gains approximately a factor of two in sensitive volume over the relevant range of sensitivities. 
As emphasized in the beginning of this section, this is only a rough, order-of-magnitude estimate of the relative sensitivity using a small number of events. This number comes with significant uncertainty, and as such, should not be over-interpreted.}{} \sk{We estimate this by approximating the detection regions of the different analyses. The LVC H+L limit was approximated by $\rho_{\rm H}^2 + \rho_{\rm L}^2 > 90$ and $\rho_{\rm H}^2,\rho_{\rm L}^2 > 30$. Our incoherent limit was approximated as $ \rho_{\rm H}^2 + \rho_{\rm L}^2 > 68$, and $\rho_{\rm H}^2,\rho_{\rm L}^2 > 16$.  }

\sk{we also show the distribution of sensitivity ratios in the O2 run for the relevant mass range.} 

Also of particular interest is the limit in which the $\snr$ is much larger in one detector than in the other. For part of the O2 run, the Livingston detector was substantially more sensitive than the Hanford one, and hence there is a substantial phase-space volume for astrophysical signals to have disparate $\snr$ in the detectors. Figure \ref{fig:snr_limit} shows that there is substantially less background in L1 in the high $\snr$ regime (see the teal shaded region). The few background events in this region come from the same small number of loud events in L1 matching with Gaussian fluctuations in H1 at different time shifts, and hence the ranking function $\tilde{\rho}_{\rm L}^2$ saturates and is severely affected by Poisson noise (see Fig.~\ref{fig:RankFunctions}). We need a different analysis to estimate a meaningful FAR in this regime; the green, dashed-dotted curve in Figure \ref{fig:volume} shows that we can gain a non-trivial amount of sensitive volume if we open up this region. We will study this regime in more detail in a subsequent paper, in which we will introduce a formalism for searching for events and assessing the FAR in this region. Notably, the LVC event GW170818, which was detected using Virgo data, belongs to this category.

\section{Conclusions}
\label{sec:conclusions}

In this paper we presented the results of our search for BBHs in the data from the O2 observing run of advanced LIGO, using the methods introduced in our work in Ref.~\cite{2019arXiv190210341V}. We report six new events above the detection thresholds defined by the LVC (in terms of FAR and $p_{\rm astro}$), three of whom have probability $p_{\rm astro} > 0.98$ of being of astrophysical origin. Interestingly, all the new events are in banks \texttt{BBH 3} and \texttt{BBH 4} (our heavy chirp-mass banks), as are most of the ones reported by the LVC.

The most significant new event (GW170121) prefers negative $\chi_{\rm eff}$ and is inconsistent with positive values at the $95\%$ level. The most marginal candidate event (GW170403), with ${p_{\rm astro}} \sim 0.5$, is inconsistent with zero or positive $\chi_{\rm eff}$. The spin of the merging BBHs is an important discriminator between formation channels~\cite{2018ApJ...854L...9F}. Hence, the new events presented in this work can throw light on the mechanisms by which BBHs are assembled\sk{ assembly one can use these measurements to infer the dependence of the merger rate on $\chi_{\rm eff}$}. 

More generally, with the increased number of events, the clear next step is to perform a population analysis that accounts for selection biases, which will map out the distribution of the intrinsic parameters of the mergers. In particular, including new events in population analyses can significantly inform us about the dependence of the merger-rate on mass and redshift.

The LVC recently started their third observing run (O3), and several new detections are expected in the near future\sk{, which can potentially shed light on some of the questions our new detections have raised}. The new events we report in this paper show that there will be additional information in the LIGO and Virgo data in addition to what the pipelines used by the LVC currently extract.
The development of our pipeline has been facilitated by access to the public O1 and O2 data, as well as the LIGO Algorithm Library~\cite{lalsuite}. We thank the LVC for releasing the data and tools to the community. We hope that data from current and future runs can be made available quickly to incentivize external groups to develop new analysis methods and maximize the scientific yield of the LVC data.
\sk{\jr{For the development of these new tools it has been crucial to get access to the public O1 and O2 data, as well as the LIGO Algorithm Library software \cite{lalsuite}. We take this opportunity to thank the LVC for releasing these to the community and encourage them to pursue these releases in the [near?] future, as they give external groups an incentive to develop new analysis methods and help maximize the yield of the LVC data.}}

\section*{Acknowledgements}

This research made use of data, software and/or web tools obtained from the Gravitational Wave Open Science Center (\url{https://www.gw-openscience.org}), a service of LIGO Laboratory, the LIGO Scientific Collaboration and the Virgo Collaboration. LIGO is funded by the U.S. National Science Foundation. Virgo is funded by the French Centre National de Recherche Scientifique (CNRS), the Italian Istituto Nazionale della Fisica Nucleare (INFN) and the Dutch Nikhef, with contributions by Polish and Hungarian institutes.

TV acknowledges support by the Friends of the Institute for Advanced Study.
BZ acknowledges the support of The Peter Svennilson Membership fund.
LD acknowledges the support from the Raymond and Beverly Sackler Foundation Fund.
MZ is supported by NSF grants AST-1409709,  PHY-1521097 and  PHY-1820775 the Canadian Institute for Advanced Research (CIFAR) program on Gravity and the Extreme Universe and the Simons Foundation Modern Inflationary Cosmology initiative.

\appendix

\section{\change{Properties of the new events}{Posteriors for the intrinsic parameters of the new events}}
\label{Appendix:posteriors}

\change{We performed full parameter estimation for all the new events presented in this paper. 
We coherently analyzed the data from all available detectors (including Virgo), in contrast to the search, which only used Hanford and Livingston data. 
We use a likelihood model that assumes the data is the sum of a gravitational wave signal and additive Gaussian random noise, and a prior that is uniform in the intrinsic source-frame parameters $m_1$ and $m_2$, the effective spin parameter $\chieff$, and luminosity volume (as detailed in Ref.~\cite{GW151216}). 
We use the \texttt{IMRPhenomD} approximant \cite{Khan2016} to generate waveforms, the relative binning method \cite{Zackay2018} to evaluate the likelihood, and \texttt{PyMultiNest} \cite{PyMultiNest} to generate samples from the posteriors.}{}

\begin{figure*}
     \centering
     \includegraphics[scale=0.75]{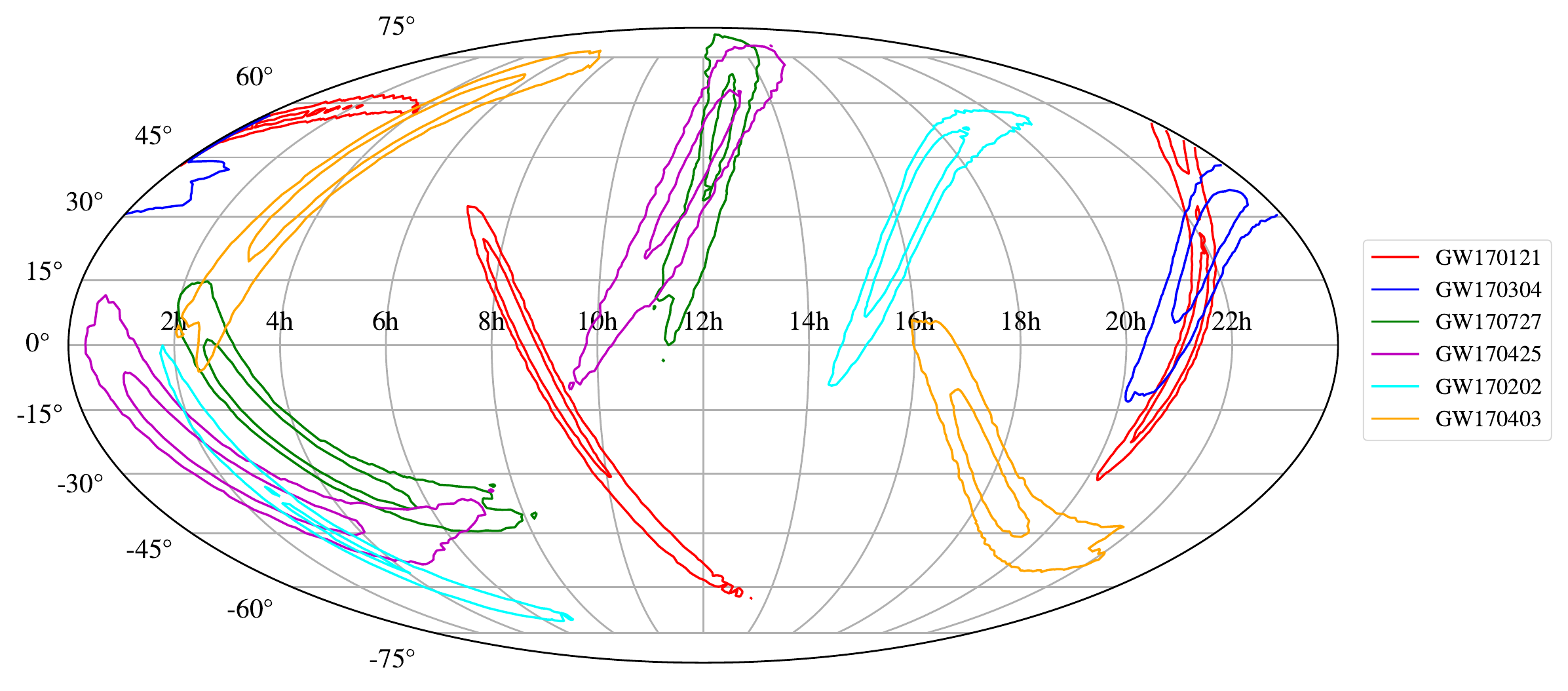}
     \caption{Contours of 90\% and 50\% credible regions for the sky locations of the six new BBH events shown in the Mollweide projection.}
     \label{fig:newbbhskyloc}
 \end{figure*}

Figures~\ref{fig:Posteriors1234} and \ref{fig:Posteriors56} show the \change{}{marginalized }posterior distributions of detector-frame chirp mass, mass-ratio, effective spin, and redshift for the new events reported with $p_{\rm astro}>0.5$\change{, marginalized over the extrinsic parameters of the mergers such as their sky location and inclination. Figure~\ref{fig:newbbhskyloc} shows the posteriors for the location of all the events on the sky, marginalized in turn over intrinsic parameters and inclination: we only show the 50\% and 90\% contours in order to present all the events on the same plot. 
Samples from the full posterior distribution (i.e., the joint distribution of the intrinsic and extrinsic parameters) are available at \url{https://github.com/jroulet/O2_samples}.}{}

\change{}{We obtained the distributions using a prior that is uniform in $m_1$, $m_2$, $\chi_{\rm eff}$ and luminosity volume, as detailed in Ref.~\cite{GW151216}. The search used only data from Hanford and Livingston, but we computed posteriors for each event by coherently analyzing the data from all detectors available (Hanford, Livingston and/or Virgo). We evaluated the likelihood using the relative binning method \cite{Zackay2018}, and used the \texttt{IMRPhenomD} waveform model \cite{Khan2016}. We used the \texttt{PyMultiNest} sampler to generate the posteriors \cite{PyMultiNest}.}

 \begin{figure*}[p]
     \centering
      \includegraphics[width=.5\linewidth]{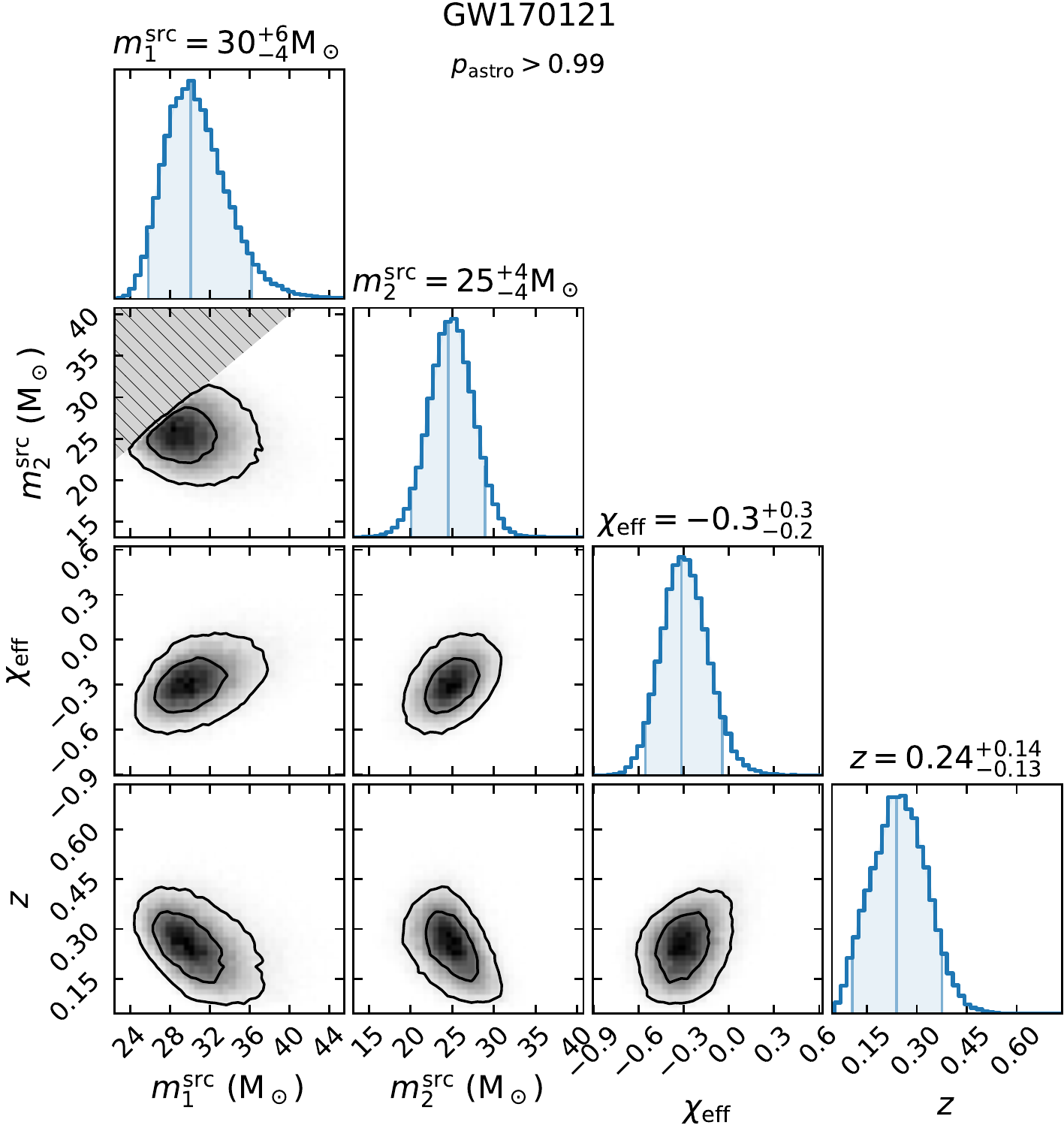}%
     \includegraphics[width=.5\linewidth]{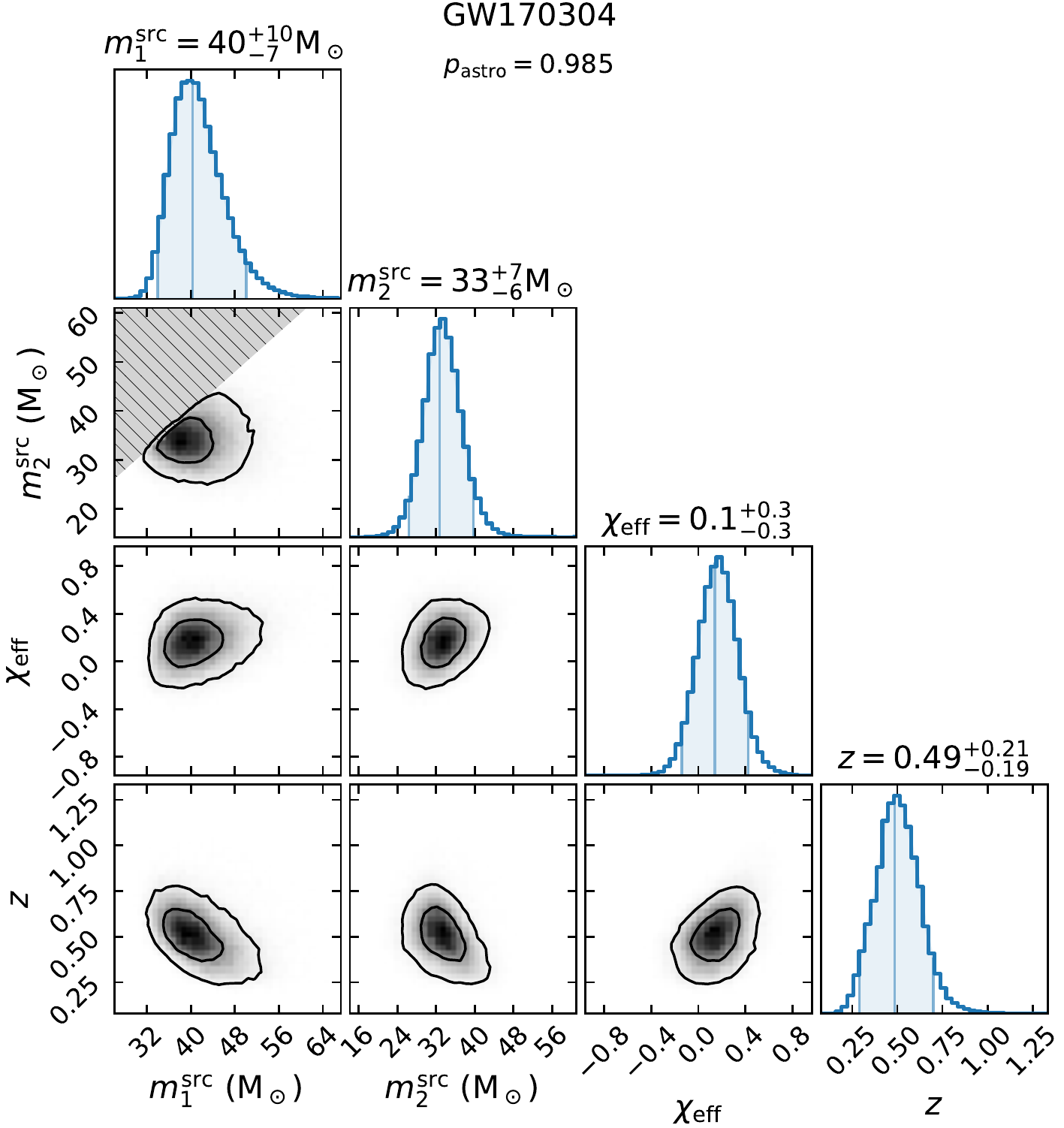}
     
     \includegraphics[width=.5\linewidth]{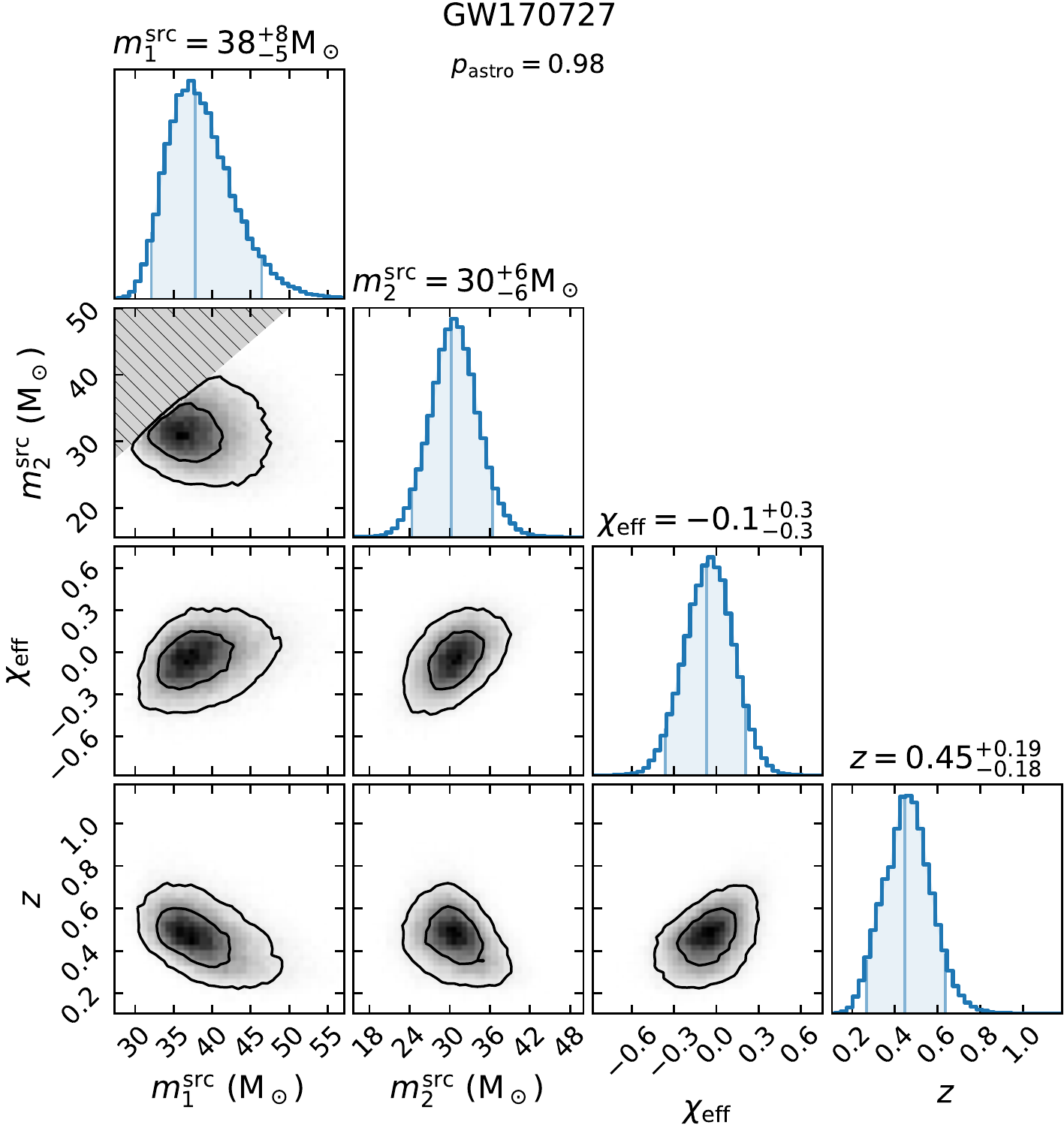}%
     \includegraphics[width=.5\linewidth]{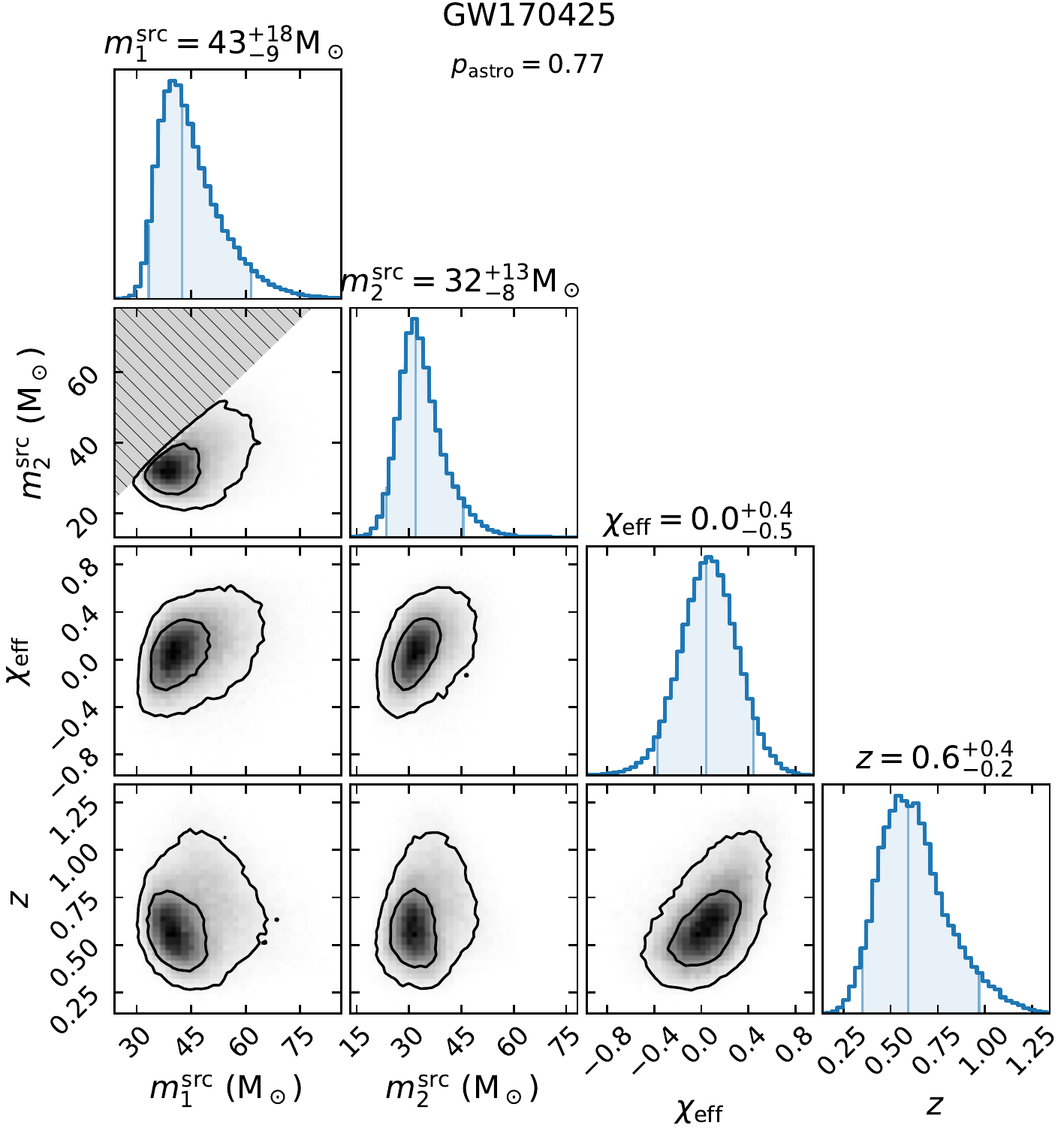}
     \caption{Marginalized posteriors for the new events reported in this work (continued in Fig.~\ref{fig:Posteriors56}). Two-dimensional contours enclose 50\% and 90\% of the distribution. In the one-dimensional posteriors, vertical lines show the 0.05, 0.5 and 0.95 quantiles. \change{The samples were generated using the \texttt{IMRPhenomD} approximant \cite{Khan2016}, and a prior that is uniform in detector-frame $m_1$, $m_2$, $\chi_{\rm eff}$ and luminosity volume~\cite{GW151216}.}{The prior used is uniform in detector-frame $m_1$, $m_2$, $\chi_{\rm eff}$ and luminosity volume. The waveform model used is \texttt{IMRPhenomD} \cite{Khan2016}.}}
     \label{fig:Posteriors1234}
 \end{figure*}

 \begin{figure*}
     \centering
     \includegraphics[width=.5\linewidth]{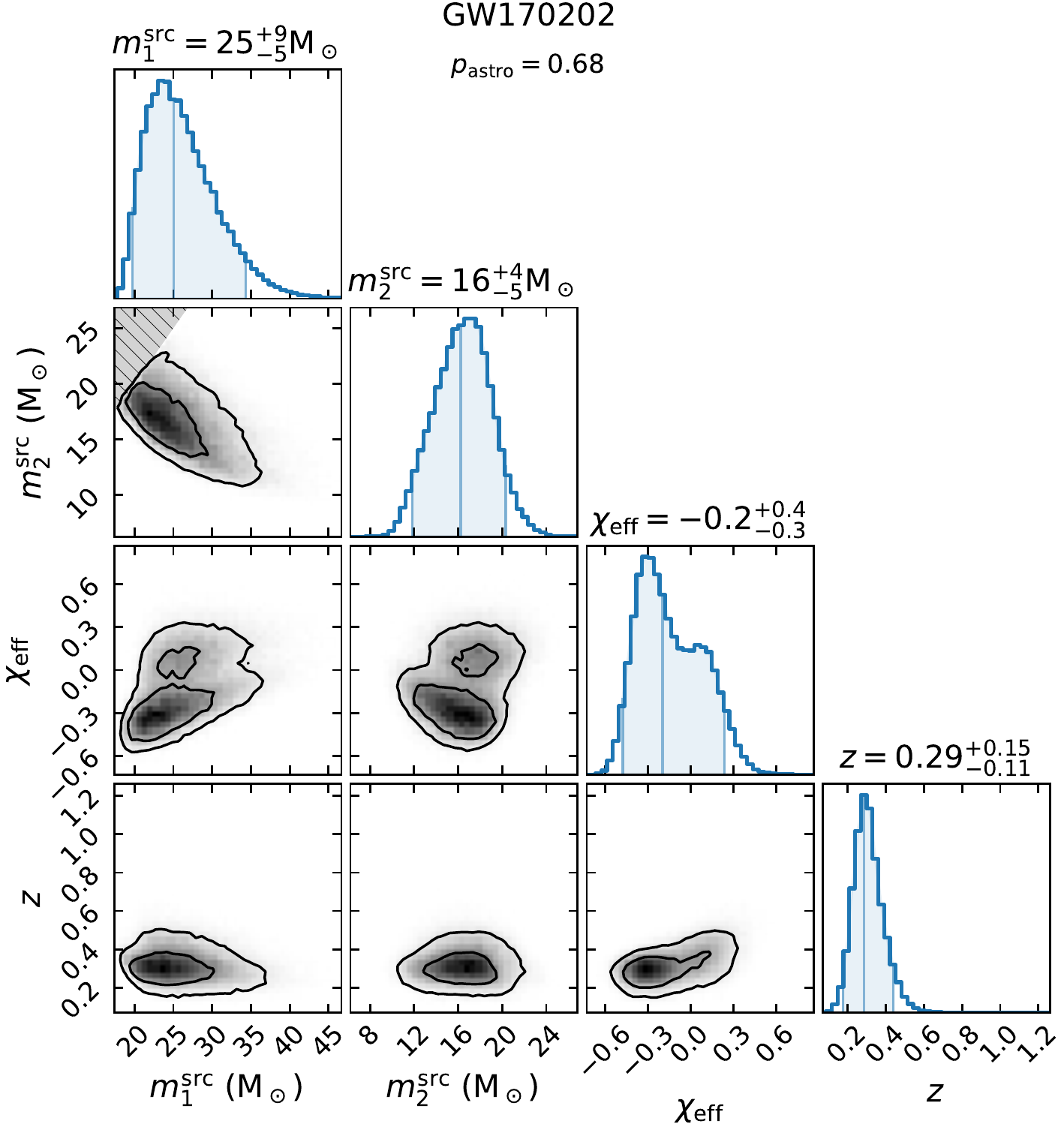}%
     \includegraphics[width=.5\linewidth]{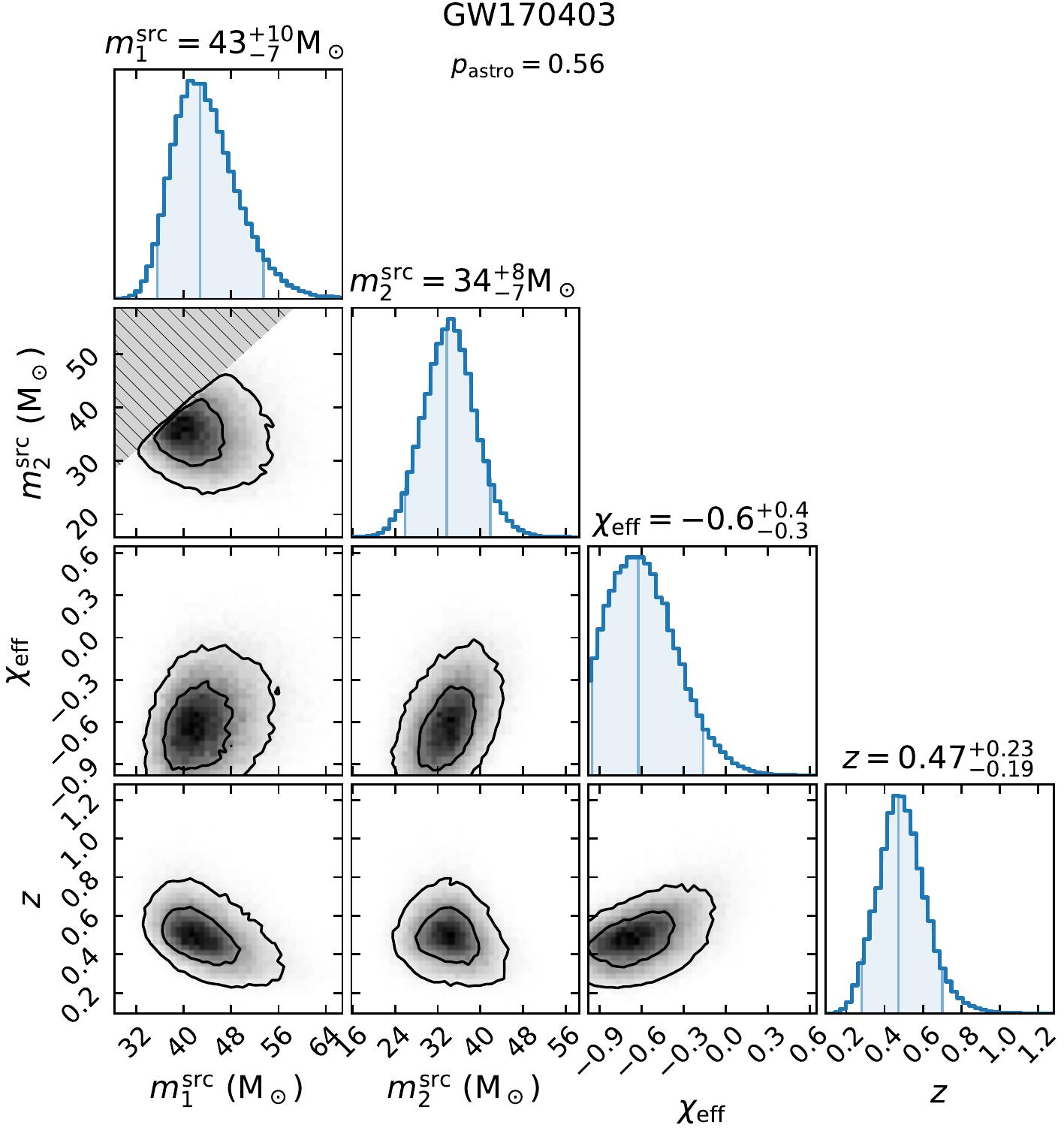}
     \caption{Continuation of Fig.~\ref{fig:Posteriors1234} displaying marginalized posteriors for GW170202 and GW170403.}
     \label{fig:Posteriors56}
 \end{figure*}
 
\section{Populating events in sub-banks and banks}
\label{Appendix:collection}

\change{In our analysis, we ultimately assign events to a single chirp-mass bank. 
We compute false alarm rates for an event within a particular bank by comparing its corresponding trigger to the triggers from background events (i.e., those generated by noise) in that bank. 
However, a single signal or noise transient produces triggers in several sub-banks, and even across chirp-mass banks. 
If we do not properly assign events (both real and background) to banks, we run the risk of losing sensitivity in our search, and hence we should carefully choose our criteria for bank assignment\footnote{\change{This is not unique to our search. For an extreme example, GW150914 produced triggers across the entire binary black hole parameter space in the LVC search~\cite{2016PhRvD..93l2003A}. If one of the fainter triggers had been picked as representing the event, it would have been assigned a different false alarm rate.}{}}.

We would like to perform the analysis as blindly as possible, and hence would like to avoid tweaking criteria after touching the data. 
Since we are analyzing a new dataset in this work, this gives us an opportunity to update the bank assignment procedure from the one we used for the O1 data (step 6 of Section II of TV19). 
We begin by considering the distribution of the triggers that a particular signal in the data produces across the banks. 
This distribution depends on the number of templates placed in each bank (effectively their density in parameter space), as well as the effective dimensionality of the parameter space in each bank/sub-bank (the dimensionality has a concrete meaning in the $c_\alpha$ parameter space, see Ref.~\cite{templatebankpaper}). 
Given this understanding, we use the set of triggers that an event throws up in a given bank to compute a net likelihood, and use this as a discriminator for assigning events to banks. 
In this rest of this appendix, we describe the details of this procedure.
}{A single signal or noise transient produces triggers in several sub-banks, and even across chirp-mass banks. We ultimately assign events to a single sub-bank within a single chirp-mass bank. In order to come up with a reasonable criterion for this assignment, we should consider the distribution of the triggers given a signal in the data.}

\change{}{In our search, we first collect coincident triggers (triggers with templates indexed by the same coefficients $c_\alpha$, and within \SI{10}{\milli\second} of each other) above a threshold. We then veto the H1 and L1 triggers, refine them on a finer $c_\alpha$ grid, and pick the best coincident trigger from the subsets of refined H1 and L1 triggers: every step involved shapes the distributions of the final triggers. }

For simplicity, let us start with triggers in a single detector. Let $d$ denote the strain data, $A$ be the signal amplitude, and $\mathbf{\Theta}$ be the other parameters of the signal: these include the coefficients $c_{\alpha, i}$ in sub-bank $i$ of chirp-mass bank $B$, and the phase and the merger-time. The likelihood of the sub-bank $i$ under the signal hypothesis is 
\begin{align}
    \mathcal{L}(d \mid i, B, \mathcal{S})
    & = \sum \iint {\rm d}\mathbf{\Theta} \, {\rm d}A \, p(\mathbf{\Theta}) p(A) \mathcal{L}(d \mid A, \mathbf{\Theta}), \label{eq:likelihoodbank}
\end{align}
where the sum runs over the grid points in $c_{\alpha, i}$, and the integral is over continuous parameters (time and phase). 
We can view the sum over the coefficients $c_{\alpha, i}$ as a Riemann sum for an integral over the underlying continuous space, $\mathcal{V}$, and thus approximate Eq.~\eqref{eq:likelihoodbank} as
\begin{align}
~~~ & \!\!\!\!
    \mathcal{L}(d \mid i, B, \mathcal{S}) \nonumber \\
    & \approx \frac{1}{(\Delta c_{\alpha, i})^{n_{c_{\alpha, i}}}} \iint {\rm d}\mathbf{\Theta} {{\rm d}A}\, p(A) \, p(\mathbf{\Theta})\change{}{ p(A)} \mathcal{L}(d \mid A, \mathbf{\Theta}), \label{eq:likelihoodbank_cont}
\end{align}
where $\Delta c_{\alpha, i}$ and $n_{c_{\alpha, i}}$ are the spacing and dimensionality of the template grid in sub-bank $i$. The factor in front of the integral is the volume per template in the discrete grid.

We adopt the following assumptions:
\begin{enumerate}
    \item In the prior $p(\mathbf{\Theta})$, astrophysical signals are equally likely to occur in two different chirp-mass banks.
    \item The astrophysical rate per template is uniform within a chirp-mass bank, i.e., given that a signal occurs within bank $B$, the probability that it occurs with a given template equals $1 / N_{{\rm temp}, B} = 1 / \sum_i N_{{\rm temp}, i}$, where $N_{{\rm temp}, B}$ and $N_{{\rm temp}, i}$ are the number of templates in the bank $B$ and sub-bank $i$, respectively (we make the same assumption when estimating the detector rate of events in Section~\ref{sec:changes}).
    \item The integral in Eq.~\eqref{eq:likelihoodbank_cont} receives most its contribution from around the best fit parameters, $\mathbf{\Theta}_*$, and amplitude $A_*$, which is valid in the limit of high $\snr$.
\end{enumerate}
The merger-time and phase are uniformly distributed within their ranges, while the amplitude has a prior distribution $p(A) \sim 1/A^4$ in a Euclidean universe. The integrand $\mathcal{L}(d \mid A, \mathbf{\Theta})$ has the form
\begin{align}
    \mathcal{L}(d \mid A, \mathbf{\Theta}) & = \exp \left[- \frac{\langle d - A \, t(\mathbf{\Theta}) \mid d - A \, t(\mathbf{\Theta}) \rangle}{2 }\right],
\end{align}
where $t(\mathbf{\Theta})$ is the template, and the inner product is weighted by the inverse PSD. Next, we integrate over the amplitude in Eq.~\eqref{eq:likelihoodbank_cont}. Under assumption 3 above, this gives us a prefactor $A_*^{-4}$, and simplifies the integrand to
\begin{align}
    \mathcal{L}(d \mid \mathbf{\Theta}) \propto \exp{\left[\frac{\langle d \mid t(\mathbf{\Theta}) \rangle^2}{2 \langle t(\mathbf{\Theta}) \mid t(\mathbf{\Theta}) \rangle}\right]}, \label{eq:likeintegrand}
\end{align}
where we have removed a term that does not depend on the templates (this step is identical to the standard derivation of the $\mathcal{F}$-statistic \cite{Jaranowski2012}). We express the best fit amplitude as $A_* = (\rho/\rho_0) A_0$, where $\rho_0$ is the $\snr$ of a merger at a fiducial distance and orientation, which captures the instantaneous sensitivity of the detector. 

By assumption 3 above, the integrand  in Eq.~\eqref{eq:likeintegrand} is sharply peaked around the bestfit parameters $\mathbf{\Theta}_0$, there it equals $\exp(\rho^2/2)$, where $\rho^2 = \langle A_* t(\mathbf{\Theta}) \mid A_* t(\mathbf{\Theta}) \rangle$ is the $\snr^2$. A nice feature of our template banks is that the deviation in coefficients, $\delta c_{\alpha}$, directly measures the degradation in the overlap between templates \cite{templatebankpaper}:
\begin{equation}
    \langle A_* t(c_{\alpha}) \mid A_* t(c_{\alpha, 0}) \rangle \approx \rho^2 \left( 1 - \frac{\delta c_{\alpha}^2}{2} \right).
\end{equation}
A similar relation holds for the other continuous parameters (time and phase), i.e the degradation in overlaps is quadratic with displacement (the principal directions are some linear combinations of time and phase), and the width is inversely related to $\rho$.
We simplify the integrand of Eq.~\eqref{eq:likeintegrand}, substitute it into Eq.~\eqref{eq:likelihoodbank_cont}, and use the prior on the templates from assumptions 1 and 2 to obtain
\begin{align}
~~~ & \!\!\!\!
    \mathcal{L}(d \mid i, B, \mathcal{S}) \notag \\
    & \propto \frac{(\rho/\rho_0)^{-4}}{N_{{\rm temp}, B} (\Delta c_{\alpha, i})^{n_{c_{\alpha, i}}}} \iint {\rm d}\delta\mathbf{\Theta} \, \exp{\left[ \frac{\rho^2}{2}(1 - \delta \Theta^2) \right]} \notag\\
    & = \frac{(\rho/\rho_0)^{-4}\exp{\left( \rho^2/2 \right)}}{N_{{\rm temp}, B} (\Delta c_{\alpha, i})^{n_{c_{\alpha, i}}}} \left( \frac{2\pi}{\rho^2} \right)^{(n_{\mathbf{\Theta}} \equiv (n_{c_{\alpha, i}} + 2))/2}.
\end{align}

When we refine coincident triggers between two detectors, the $\mathbf{\Theta}$ contains four extra parameters apart from the template coefficients: two times, and two phases. A complication is that astrophysical signals are not uniformly distributed in the space of the time delay and relative phase, and the distribution depends on the relative sensitivities of the two detectors, $\rho_{0, {\rm H}} / \rho_{0, {\rm L}}$. Thus there is a nontrivial prior in the space of parameters $\mathbf{\Theta}$.

We make progress by noting that the likelihood in the integrand in Eq.~\eqref{eq:likelihoodbank_cont} is independent of the relative times and phases. In this case, we are greatly helped by our assumption 3 above, which tells us that all we need is to evaluate the prior at the bestfit parameters $\mathbf{\Theta_*}$. We combine the prior and amplitude prefactor together into a function $p(\Delta t, \Delta \phi, \rho_{\rm H}^2, \rho_{\rm L}^2 \mid \rho_{0, {\rm H}}, \rho_{0, {\rm L}})$, which we evaluate by Montecarlo sampling methods.

The integral over the likelihood in Eq.~\eqref{eq:likelihoodbank_cont} can be evaluated in a similar manner as above, with the difference that now $\rho^2 = \rho_{\rm H}^2 + \rho_{\rm L}^2$. 
Thus we finally obtain
\begin{multline}
    \mathcal{L}(d_{\rm H}, d_{\rm L} \mid i, B, \mathcal{S}) \\
     \propto \frac{p(\Delta t, \Delta \phi, \rho_{\rm H}^2, \rho_{\rm L}^2 \mid \rho_{0, {\rm H}}, \rho_{0, {\rm L}}) e^{\rho^2/2}}{N_{{\rm temp}, B} (\Delta c_{\alpha, i})^{n_{c_{\alpha, i}}}} \left( \frac{2\pi}{\rho^2} \right)^{(n_{c_{\alpha, i}} + 4)/2}. \label{eq:likebank}
\end{multline}
Note that the exponent of the last term is different due to the extra degrees of freedom in the two-detector case. 

We use the ratio of the likelihoods given by Eq.~\eqref{eq:likebank} as a discriminator to compare triggers in the same location in different sub-banks (and possibly chirp-mass-banks).

\section{Definition of rank functions in different sub-banks}
\label{Appendix:normrank}

We compute FARs of events by comparing them to background triggers in their chirp-mass bank. An essential ingredient in this computation is the likelihood $\mathcal{L}(t\mid \mathcal{N})$ for a trigger $t$ under the noise hypothesis. Coincident triggers are produced with random time-delays and relative phases between the two detectors, and the likelihood $\mathcal{L}(t\mid \mathcal{N})$ depends only on the incoherent $\snr^2$ in the two detectors, and the template in sub-bank $i$ (which we denote by the set of coefficients $c_\alpha$ of the basis phase functions), i.e., $\mathcal{L}(t\mid \mathcal{N}) = P(\rho_{\rm H}^2,\rho_{\rm L}^2,c_\alpha,i)$. \sk{Our incoherent score should ideally be given by:  
\begin{equation}
    {\tilde{\rho}^2}=-2 \log P(\rho_{\rm H}^2,\rho_{\rm L}^2,c_\alpha,i). \label{eq:rhotildesq}
\end{equation}}
We can write:
\begin{align}
    P(\rho_{\rm H}^2,\rho_{\rm L}^2,c_\alpha,i) &= P(\rho_{\rm H}^2,\rho_{\rm L}^2, c_\alpha \mid i) P(i) \label{eq:likehoodnoise}\\
    &= \frac{P(\rho_{\rm H}^2 \mid i) P(\rho_{\rm L}^2 \mid i)}{N_{{\rm temp},i}} P(i), \label{eq:likehoodnoise_indep}
\end{align}
where $P(i)$ is the probability that noise produces a coincident trigger in sub-bank $i$, regardless of the $\snr^2$, and $N_{{\rm temp},i}$ is the number of templates in sub-bank $i$. 

Equation \eqref{eq:likehoodnoise_indep} assumes that a) the background is flat over templates within each sub-bank, and b) the triggers in different detectors are independent of each other. 
Both these assumptions fail to some degree (the latter happens because we refine coincident triggers, and pick the best common template). One solution would be to directly estimate the probability in Eq.~\eqref{eq:likehoodnoise} from the background, but in practice, the many-dimensional distribution is hard to sample finely enough, and thus real coincident triggers can receive spurious penalties to their scores.
We use the same assumptions to rank the time slides as well as the coincident triggers, and hence our FARs are not biased for the strategy we adopt; the price of the above assumptions is that our ranking is no longer strictly optimal. 

The probability $P(i)$ is
\begin{equation}
     P(i) = \frac{N_{{\rm trig},i}}{N_{{\rm trig}}},
\end{equation}
where $N_{{\rm trig},i}$ is the number of triggers in sub-bank $i$, and $N_{{\rm trig}}$ is the number of triggers summed over all sub-banks. 

We approximate the probability $P(\rho^2 \mid i)$ for each detector using the same ranking function that we adopted in our previous work~\cite{2019arXiv190210341V}, but compute it separately for each sub-bank: 
\begin{equation}
    \frac{P(\rho^2 \mid i)}{P(\rho^2_0 \mid i)}\approx\frac{{\rm Rank}(\rho^2 \mid i)}{{\rm Rank}(\rho^2_0 \mid i)},
\end{equation}
where ${\rm Rank}(\rho^2 \mid i)$ is the ranking of a given trigger in its sub-bank (with the lowest rank given to the loudest event) and $\rho^2_0$ is a normalization point that we set to $\rho_0^2\approx 30$. 
We estimate $P(\rho^2_0 \mid i)$ by taking the ratio between the number of triggers in a bin around $\rho^2_0$ in sub-bank $i$ over the total number of triggers in the sub-bank, i.e., 
\begin{align}
    P(\rho^2_0 \mid i) & \propto - \frac{1}{N_{{\rm trig},i}} \frac{\rm d}{\rm d\rho^2}{\rm Rank}(\rho^2 \mid i) \big|_{\rho = \rho_0}.
\end{align}

\bibliographystyle{apsrev4-1-etal}
\bibliography{gw}

\begin{thebibliography}{31}%
\makeatletter
\providecommand \@ifxundefined [1]{%
 \@ifx{#1\undefined}
}%
\providecommand \@ifnum [1]{%
 \ifnum #1\expandafter \@firstoftwo
 \else \expandafter \@secondoftwo
 \fi
}%
\providecommand \@ifx [1]{%
 \ifx #1\expandafter \@firstoftwo
 \else \expandafter \@secondoftwo
 \fi
}%
\providecommand \natexlab [1]{#1}%
\providecommand \enquote  [1]{``#1''}%
\providecommand \bibnamefont  [1]{#1}%
\providecommand \bibfnamefont [1]{#1}%
\providecommand \citenamefont [1]{#1}%
\providecommand \href@noop [0]{\@secondoftwo}%
\providecommand \href [0]{\begingroup \@sanitize@url \@href}%
\providecommand \@href[1]{\@@startlink{#1}\@@href}%
\providecommand \@@href[1]{\endgroup#1\@@endlink}%
\providecommand \@sanitize@url [0]{\catcode `\\12\catcode `\$12\catcode
  `\&12\catcode `\#12\catcode `\^12\catcode `\_12\catcode `\%12\relax}%
\providecommand \@@startlink[1]{}%
\providecommand \@@endlink[0]{}%
\providecommand \url  [0]{\begingroup\@sanitize@url \@url }%
\providecommand \@url [1]{\endgroup\@href {#1}{\urlprefix }}%
\providecommand \urlprefix  [0]{URL }%
\providecommand \Eprint [0]{\href }%
\providecommand \doibase [0]{http://dx.doi.org/}%
\providecommand \selectlanguage [0]{\@gobble}%
\providecommand \bibinfo  [0]{\@secondoftwo}%
\providecommand \bibfield  [0]{\@secondoftwo}%
\providecommand \translation [1]{[#1]}%
\providecommand \BibitemOpen [0]{}%
\providecommand \bibitemStop [0]{}%
\providecommand \bibitemNoStop [0]{.\EOS\space}%
\providecommand \EOS [0]{\spacefactor3000\relax}%
\providecommand \BibitemShut  [1]{\csname bibitem#1\endcsname}%
\let\auto@bib@innerbib\@empty
\bibitem [{\citenamefont {{Venumadhav}}\ \emph
  {et~al.}(2019{\natexlab{a}})\citenamefont {{Venumadhav}}, \citenamefont
  {{Zackay}}, \citenamefont {{Roulet}}, \citenamefont {{Dai}},\ and\
  \citenamefont {{Zaldarriaga}}}]{2019PhRvD.100b3011V}%
  \BibitemOpen
  \bibfield  {author} {\bibinfo {author} {\bibfnamefont {T.}~\bibnamefont
  {{Venumadhav}}}, \bibinfo {author} {\bibfnamefont {B.}~\bibnamefont
  {{Zackay}}}, \bibinfo {author} {\bibfnamefont {J.}~\bibnamefont {{Roulet}}},
  \bibinfo {author} {\bibfnamefont {L.}~\bibnamefont {{Dai}}}, \ and\ \bibinfo
  {author} {\bibfnamefont {M.}~\bibnamefont {{Zaldarriaga}}},\ }\href {\doibase
  10.1103/PhysRevD.100.023011} {\bibfield  {journal} {\bibinfo  {journal}
  {\prd}\ }\textbf {\bibinfo {volume} {100}},\ \bibinfo {eid} {023011}
  (\bibinfo {year} {2019}{\natexlab{a}})},\ \Eprint
  {http://arxiv.org/abs/1902.10341} {arXiv:1902.10341 [astro-ph.IM]}
  \BibitemShut {NoStop}%
\bibitem [{\citenamefont {Abbott}\ \emph {et~al.}(2018)\citenamefont {Abbott}
  \emph {et~al.}}]{LIGOScientific:2018mvr}%
  \BibitemOpen
  \bibfield  {author} {\bibinfo {author} {\bibfnamefont {B.~P.}\ \bibnamefont
  {Abbott}} \emph {et~al.} (\bibinfo {collaboration} {LIGO Scientific,
  Virgo}),\ }\href@noop {} {\  (\bibinfo {year} {2018})},\ \Eprint
  {http://arxiv.org/abs/1811.12907} {arXiv:1811.12907 [astro-ph.HE]}
  \BibitemShut {NoStop}%
\bibitem [{\citenamefont {Abbott}\ \emph
  {et~al.}(2016{\natexlab{a}})\citenamefont {Abbott} \emph
  {et~al.}}]{GW150914}%
  \BibitemOpen
  \bibfield  {author} {\bibinfo {author} {\bibfnamefont {B.~P.}\ \bibnamefont
  {Abbott}} \emph {et~al.} (\bibinfo {collaboration} {LIGO Scientific
  Collaboration and Virgo Collaboration}),\ }\href {\doibase
  10.1103/PhysRevLett.116.061102} {\bibfield  {journal} {\bibinfo  {journal}
  {Physical Review Letters}\ }\textbf {\bibinfo {volume} {116}},\ \bibinfo
  {pages} {061102} (\bibinfo {year} {2016}{\natexlab{a}})}\BibitemShut
  {NoStop}%
\bibitem [{\citenamefont {Abbott}\ \emph
  {et~al.}(2016{\natexlab{b}})\citenamefont {Abbott} \emph
  {et~al.}}]{GW151226}%
  \BibitemOpen
  \bibfield  {author} {\bibinfo {author} {\bibfnamefont {B.~P.}\ \bibnamefont
  {Abbott}} \emph {et~al.} (\bibinfo {collaboration} {LIGO Scientific
  Collaboration and Virgo Collaboration}),\ }\href {\doibase
  10.1103/PhysRevLett.116.241103} {\bibfield  {journal} {\bibinfo  {journal}
  {Physical Review Letters}\ }\textbf {\bibinfo {volume} {116}},\ \bibinfo
  {pages} {241103} (\bibinfo {year} {2016}{\natexlab{b}})}\BibitemShut
  {NoStop}%
\bibitem [{\citenamefont {Abbott}\ \emph
  {et~al.}(2016{\natexlab{c}})\citenamefont {Abbott} \emph
  {et~al.}}]{O1catalog}%
  \BibitemOpen
  \bibfield  {author} {\bibinfo {author} {\bibfnamefont {B.~P.}\ \bibnamefont
  {Abbott}} \emph {et~al.} (\bibinfo {collaboration} {LIGO Scientific
  Collaboration and Virgo Collaboration}),\ }\href {\doibase
  10.1103/PhysRevX.6.041015} {\bibfield  {journal} {\bibinfo  {journal} {Phys.
  Rev. X}\ }\textbf {\bibinfo {volume} {6}},\ \bibinfo {pages} {041015}
  (\bibinfo {year} {2016}{\natexlab{c}})}\BibitemShut {NoStop}%
\bibitem [{\citenamefont {Abbott}\ \emph
  {et~al.}(2017{\natexlab{a}})\citenamefont {Abbott} \emph
  {et~al.}}]{GW170104}%
  \BibitemOpen
  \bibfield  {author} {\bibinfo {author} {\bibfnamefont {B.~P.}\ \bibnamefont
  {Abbott}} \emph {et~al.} (\bibinfo {collaboration} {LIGO Scientific and Virgo
  Collaboration}),\ }\href {\doibase 10.1103/PhysRevLett.118.221101} {\bibfield
   {journal} {\bibinfo  {journal} {Phys. Rev. Lett.}\ }\textbf {\bibinfo
  {volume} {118}},\ \bibinfo {pages} {221101} (\bibinfo {year}
  {2017}{\natexlab{a}})}\BibitemShut {NoStop}%
\bibitem [{\citenamefont {Abbott}\ \emph
  {et~al.}(2017{\natexlab{b}})\citenamefont {Abbott} \emph
  {et~al.}}]{GW170608}%
  \BibitemOpen
  \bibfield  {author} {\bibinfo {author} {\bibfnamefont {B.~P.}\ \bibnamefont
  {Abbott}} \emph {et~al.},\ }\href {\doibase 10.3847/2041-8213/aa9f0c}
  {\bibfield  {journal} {\bibinfo  {journal} {The Astrophysical Journal}\
  }\textbf {\bibinfo {volume} {851}},\ \bibinfo {pages} {L35} (\bibinfo {year}
  {2017}{\natexlab{b}})}\BibitemShut {NoStop}%
\bibitem [{\citenamefont {Abbott}\ \emph
  {et~al.}(2017{\natexlab{c}})\citenamefont {Abbott} \emph
  {et~al.}}]{GW170814}%
  \BibitemOpen
  \bibfield  {author} {\bibinfo {author} {\bibfnamefont {B.~P.}\ \bibnamefont
  {Abbott}} \emph {et~al.} (\bibinfo {collaboration} {LIGO Scientific
  Collaboration and Virgo Collaboration}),\ }\href {\doibase
  10.1103/PhysRevLett.119.141101} {\bibfield  {journal} {\bibinfo  {journal}
  {Physical Review Letters}\ }\textbf {\bibinfo {volume} {119}},\ \bibinfo
  {pages} {141101} (\bibinfo {year} {2017}{\natexlab{c}})}\BibitemShut
  {NoStop}%
\bibitem [{\citenamefont {Abbott}\ \emph
  {et~al.}(2017{\natexlab{d}})\citenamefont {Abbott} \emph
  {et~al.}}]{GW170817}%
  \BibitemOpen
  \bibfield  {author} {\bibinfo {author} {\bibfnamefont {B.~P.}\ \bibnamefont
  {Abbott}} \emph {et~al.} (\bibinfo {collaboration} {LIGO Scientific
  Collaboration and Virgo Collaboration}),\ }\href {\doibase
  10.1103/PhysRevLett.119.161101} {\bibfield  {journal} {\bibinfo  {journal}
  {Phys. Rev. Lett.}\ }\textbf {\bibinfo {volume} {119}},\ \bibinfo {pages}
  {161101} (\bibinfo {year} {2017}{\natexlab{d}})}\BibitemShut {NoStop}%
\bibitem [{\citenamefont {{Chatziioannou}}\ \emph {et~al.}(2019)\citenamefont
  {{Chatziioannou}}, \citenamefont {{Cotesta}}, \citenamefont {{Ghonge}},
  \citenamefont {{Lange}}, \citenamefont {{Ng}}, \citenamefont {{Bustillo}},
  \citenamefont {{Clark}}, \citenamefont {{Haster}}, \citenamefont {{Khan}},
  \citenamefont {{Puerrer}}, \citenamefont {{Raymond}}, \citenamefont
  {{Vitale}}, \citenamefont {{Afshari}}, \citenamefont {{Babak}}, \citenamefont
  {{Barkett}}, \citenamefont {{Blackman}}, \citenamefont {{Bohe}},
  \citenamefont {{Boyle}}, \citenamefont {{Buonanno}}, \citenamefont
  {{Campanelli}}, \citenamefont {{Carullo}}, \citenamefont {{Chu}},
  \citenamefont {{Flynn}}, \citenamefont {{Fong}}, \citenamefont {{Garcia}},
  \citenamefont {{Giesler}}, \citenamefont {{Haney}}, \citenamefont {{Hannam}},
  \citenamefont {{Harry}}, \citenamefont {{Healy}}, \citenamefont
  {{Hemberger}}, \citenamefont {{Hinder}}, \citenamefont {{Jani}},
  \citenamefont {{Khamersa}}, \citenamefont {{Kidder}}, \citenamefont
  {{Kumar}}, \citenamefont {{Laguna}}, \citenamefont {{Lousto}}, \citenamefont
  {{Lovelace}}, \citenamefont {{Littenberg}}, \citenamefont {{London}},
  \citenamefont {{Millhouse}}, \citenamefont {{Nuttall}}, \citenamefont
  {{Ohme}}, \citenamefont {{O'Shaughnessy}}, \citenamefont {{Ossokine}},
  \citenamefont {{Pannarale}}, \citenamefont {{Schmidt}}, \citenamefont
  {{Pfeiffer}}, \citenamefont {{Scheel}}, \citenamefont {{Shao}}, \citenamefont
  {{Shoemaker}}, \citenamefont {{Szilagyi}}, \citenamefont {{Taracchini}},
  \citenamefont {{Teukolsky}},\ and\ \citenamefont
  {{Zlochower}}}]{GW170729HigherModes}%
  \BibitemOpen
  \bibfield  {author} {\bibinfo {author} {\bibfnamefont {K.}~\bibnamefont
  {{Chatziioannou}}}, \bibinfo {author} {\bibfnamefont {R.}~\bibnamefont
  {{Cotesta}}}, \bibinfo {author} {\bibfnamefont {S.}~\bibnamefont {{Ghonge}}},
  \bibinfo {author} {\bibfnamefont {J.}~\bibnamefont {{Lange}}},  \emph
  {et~al.},\ }\href@noop {} {\bibfield  {journal} {\bibinfo  {journal} {arXiv
  e-prints}\ ,\ \bibinfo {eid} {arXiv:1903.06742}} (\bibinfo {year} {2019})},\
  \Eprint {http://arxiv.org/abs/1903.06742} {arXiv:1903.06742 [gr-qc]}
  \BibitemShut {NoStop}%
\bibitem [{\citenamefont {{Usman}}\ \emph {et~al.}(2016)\citenamefont
  {{Usman}}, \citenamefont {{Nitz}}, \citenamefont {{Harry}}, \citenamefont
  {{Biwer}}, \citenamefont {{Brown}}, \citenamefont {{Cabero}}, \citenamefont
  {{Capano}}, \citenamefont {{Dal Canton}}, \citenamefont {{Dent}},
  \citenamefont {{Fairhurst}}, \citenamefont {{Kehl}}, \citenamefont
  {{Keppel}}, \citenamefont {{Krishnan}}, \citenamefont {{Lenon}},
  \citenamefont {{Lundgren}}, \citenamefont {{Nielsen}}, \citenamefont
  {{Pekowsky}}, \citenamefont {{Pfeiffer}}, \citenamefont {{Saulson}},
  \citenamefont {{West}},\ and\ \citenamefont {{Willis}}}]{PYCBCPipeline}%
  \BibitemOpen
  \bibfield  {author} {\bibinfo {author} {\bibfnamefont {S.~A.}\ \bibnamefont
  {{Usman}}}, \bibinfo {author} {\bibfnamefont {A.~H.}\ \bibnamefont {{Nitz}}},
  \bibinfo {author} {\bibfnamefont {I.~W.}\ \bibnamefont {{Harry}}}, \bibinfo
  {author} {\bibfnamefont {C.~M.}\ \bibnamefont {{Biwer}}},  \emph {et~al.},\
  }\href {\doibase 10.1088/0264-9381/33/21/215004} {\bibfield  {journal}
  {\bibinfo  {journal} {Classical and Quantum Gravity}\ }\textbf {\bibinfo
  {volume} {33}},\ \bibinfo {eid} {215004} (\bibinfo {year} {2016})},\ \Eprint
  {http://arxiv.org/abs/1508.02357} {arXiv:1508.02357 [gr-qc]} \BibitemShut
  {NoStop}%
\bibitem [{\citenamefont {Sachdev}\ \emph {et~al.}(2019)\citenamefont
  {Sachdev}, \citenamefont {Caudill}, \citenamefont {Fong}, \citenamefont {Lo},
  \citenamefont {Messick}, \citenamefont {Mukherjee}, \citenamefont {Magee},
  \citenamefont {Tsukada}, \citenamefont {Blackburn}, \citenamefont {Brady},
  \citenamefont {Brockill}, \citenamefont {Cannon}, \citenamefont {Chamberlin},
  \citenamefont {Chatterjee}, \citenamefont {Creighton}, \citenamefont
  {Godwin}, \citenamefont {Gupta}, \citenamefont {Hanna}, \citenamefont
  {Kapadia}, \citenamefont {Lang}, \citenamefont {Li}, \citenamefont {Meacher},
  \citenamefont {Pace}, \citenamefont {Privitera}, \citenamefont {Sadeghian},
  \citenamefont {Wade}, \citenamefont {Wade}, \citenamefont {Weinstein},\ and\
  \citenamefont {Xiao}}]{gstlal}%
  \BibitemOpen
  \bibfield  {author} {\bibinfo {author} {\bibfnamefont {S.}~\bibnamefont
  {Sachdev}}, \bibinfo {author} {\bibfnamefont {S.}~\bibnamefont {Caudill}},
  \bibinfo {author} {\bibfnamefont {H.}~\bibnamefont {Fong}}, \bibinfo {author}
  {\bibfnamefont {R.~K.~L.}\ \bibnamefont {Lo}},  \emph {et~al.},\ }\href
  {http://arxiv.org/abs/1901.08580v1} {\  (\bibinfo {year} {2019})},\ \Eprint
  {http://arxiv.org/abs/1901.08580v1} {arXiv:1901.08580v1 [gr-qc]} \BibitemShut
  {NoStop}%
\bibitem [{\citenamefont {Klimenko}\ \emph {et~al.}(2016)\citenamefont
  {Klimenko}, \citenamefont {Vedovato}, \citenamefont {Drago}, \citenamefont
  {Salemi}, \citenamefont {Tiwari}, \citenamefont {Prodi}, \citenamefont
  {Lazzaro}, \citenamefont {Ackley}, \citenamefont {Tiwari}, \citenamefont
  {Da~Silva},\ and\ \citenamefont {Mitselmakher}}]{cWB}%
  \BibitemOpen
  \bibfield  {author} {\bibinfo {author} {\bibfnamefont {S.}~\bibnamefont
  {Klimenko}}, \bibinfo {author} {\bibfnamefont {G.}~\bibnamefont {Vedovato}},
  \bibinfo {author} {\bibfnamefont {M.}~\bibnamefont {Drago}}, \bibinfo
  {author} {\bibfnamefont {F.}~\bibnamefont {Salemi}},  \emph {et~al.},\ }\href
  {\doibase 10.1103/PhysRevD.93.042004} {\bibfield  {journal} {\bibinfo
  {journal} {Phys. Rev. D}\ }\textbf {\bibinfo {volume} {93}},\ \bibinfo
  {pages} {042004} (\bibinfo {year} {2016})}\BibitemShut {NoStop}%
\bibitem [{\citenamefont {{Nitz}}\ \emph {et~al.}(2018)\citenamefont {{Nitz}},
  \citenamefont {{Capano}}, \citenamefont {{Nielsen}}, \citenamefont {{Reyes}},
  \citenamefont {{White}}, \citenamefont {{Brown}},\ and\ \citenamefont
  {{Krishnan}}}]{NitzCatalog}%
  \BibitemOpen
  \bibfield  {author} {\bibinfo {author} {\bibfnamefont {A.~H.}\ \bibnamefont
  {{Nitz}}}, \bibinfo {author} {\bibfnamefont {C.}~\bibnamefont {{Capano}}},
  \bibinfo {author} {\bibfnamefont {A.~B.}\ \bibnamefont {{Nielsen}}}, \bibinfo
  {author} {\bibfnamefont {S.}~\bibnamefont {{Reyes}}}, \bibinfo {author}
  {\bibfnamefont {R.}~\bibnamefont {{White}}}, \bibinfo {author} {\bibfnamefont
  {D.~A.}\ \bibnamefont {{Brown}}}, \ and\ \bibinfo {author} {\bibfnamefont
  {B.}~\bibnamefont {{Krishnan}}},\ }\href@noop {} {\bibfield  {journal}
  {\bibinfo  {journal} {arXiv e-prints}\ ,\ \bibinfo {eid} {arXiv:1811.01921}}
  (\bibinfo {year} {2018})},\ \Eprint {http://arxiv.org/abs/1811.01921}
  {arXiv:1811.01921 [gr-qc]} \BibitemShut {NoStop}%
\bibitem [{\citenamefont {{Venumadhav}}\ \emph
  {et~al.}(2019{\natexlab{b}})\citenamefont {{Venumadhav}}, \citenamefont
  {{Zackay}}, \citenamefont {{Roulet}}, \citenamefont {{Dai}},\ and\
  \citenamefont {{Zaldarriaga}}}]{2019arXiv190210341V}%
  \BibitemOpen
  \bibfield  {author} {\bibinfo {author} {\bibfnamefont {T.}~\bibnamefont
  {{Venumadhav}}}, \bibinfo {author} {\bibfnamefont {B.}~\bibnamefont
  {{Zackay}}}, \bibinfo {author} {\bibfnamefont {J.}~\bibnamefont {{Roulet}}},
  \bibinfo {author} {\bibfnamefont {L.}~\bibnamefont {{Dai}}}, \ and\ \bibinfo
  {author} {\bibfnamefont {M.}~\bibnamefont {{Zaldarriaga}}},\ }\href@noop {}
  {\bibfield  {journal} {\bibinfo  {journal} {arXiv e-prints}\ ,\ \bibinfo
  {eid} {arXiv:1902.10341}} (\bibinfo {year} {2019}{\natexlab{b}})},\ \Eprint
  {http://arxiv.org/abs/1902.10341} {arXiv:1902.10341 [astro-ph.IM]}
  \BibitemShut {NoStop}%
\bibitem [{gwo(2 27)}]{gwosc_url}%
  \BibitemOpen
  \href@noop {} {\enquote {\bibinfo {title} {{Gravitational Wave Open Science
  Center (GWOSC)}},}\ }\bibinfo {howpublished}
  {\url{www.gw-openscience.org/O2/}} (\bibinfo {year} {accessed
  2019-02-27})\BibitemShut {NoStop}%
\bibitem [{\citenamefont {{Vallisneri}}\ \emph {et~al.}(2015)\citenamefont
  {{Vallisneri}}, \citenamefont {{Kanner}}, \citenamefont {{Williams}},
  \citenamefont {{Weinstein}},\ and\ \citenamefont {{Stephens}}}]{GWOSC}%
  \BibitemOpen
  \bibfield  {author} {\bibinfo {author} {\bibfnamefont {M.}~\bibnamefont
  {{Vallisneri}}}, \bibinfo {author} {\bibfnamefont {J.}~\bibnamefont
  {{Kanner}}}, \bibinfo {author} {\bibfnamefont {R.}~\bibnamefont
  {{Williams}}}, \bibinfo {author} {\bibfnamefont {A.}~\bibnamefont
  {{Weinstein}}}, \ and\ \bibinfo {author} {\bibfnamefont {B.}~\bibnamefont
  {{Stephens}}},\ }in\ \href {\doibase 10.1088/1742-6596/610/1/012021} {\emph
  {\bibinfo {booktitle} {Journal of Physics Conference Series}}},\ \bibinfo
  {series} {Journal of Physics Conference Series}, Vol.\ \bibinfo {volume}
  {610}\ (\bibinfo {year} {2015})\ p.\ \bibinfo {pages} {012021},\ \Eprint
  {http://arxiv.org/abs/1410.4839} {arXiv:1410.4839 [gr-qc]} \BibitemShut
  {NoStop}%
\bibitem [{\citenamefont {{Zackay}}\ \emph {et~al.}(2019)\citenamefont
  {{Zackay}}, \citenamefont {{Venumadhav}}, \citenamefont {{Dai}},
  \citenamefont {{Roulet}},\ and\ \citenamefont {{Zaldarriaga}}}]{GW151216}%
  \BibitemOpen
  \bibfield  {author} {\bibinfo {author} {\bibfnamefont {B.}~\bibnamefont
  {{Zackay}}}, \bibinfo {author} {\bibfnamefont {T.}~\bibnamefont
  {{Venumadhav}}}, \bibinfo {author} {\bibfnamefont {L.}~\bibnamefont {{Dai}}},
  \bibinfo {author} {\bibfnamefont {J.}~\bibnamefont {{Roulet}}}, \ and\
  \bibinfo {author} {\bibfnamefont {M.}~\bibnamefont {{Zaldarriaga}}},\
  }\href@noop {} {\bibfield  {journal} {\bibinfo  {journal} {arXiv e-prints}\
  ,\ \bibinfo {eid} {arXiv:1902.10331}} (\bibinfo {year} {2019})},\ \Eprint
  {http://arxiv.org/abs/1902.10331} {arXiv:1902.10331 [astro-ph.HE]}
  \BibitemShut {NoStop}%
\bibitem [{\citenamefont {{Khan}}\ \emph {et~al.}(2016)\citenamefont {{Khan}},
  \citenamefont {{Husa}}, \citenamefont {{Hannam}}, \citenamefont {{Ohme}},
  \citenamefont {{P{\"u}rrer}}, \citenamefont {{Forteza}},\ and\ \citenamefont
  {{Boh{\'e}}}}]{2016PhRvD..93d4007K}%
  \BibitemOpen
  \bibfield  {author} {\bibinfo {author} {\bibfnamefont {S.}~\bibnamefont
  {{Khan}}}, \bibinfo {author} {\bibfnamefont {S.}~\bibnamefont {{Husa}}},
  \bibinfo {author} {\bibfnamefont {M.}~\bibnamefont {{Hannam}}}, \bibinfo
  {author} {\bibfnamefont {F.}~\bibnamefont {{Ohme}}}, \bibinfo {author}
  {\bibfnamefont {M.}~\bibnamefont {{P{\"u}rrer}}}, \bibinfo {author}
  {\bibfnamefont {X.~J.}\ \bibnamefont {{Forteza}}}, \ and\ \bibinfo {author}
  {\bibfnamefont {A.}~\bibnamefont {{Boh{\'e}}}},\ }\href {\doibase
  10.1103/PhysRevD.93.044007} {\bibfield  {journal} {\bibinfo  {journal}
  {\prd}\ }\textbf {\bibinfo {volume} {93}},\ \bibinfo {eid} {044007} (\bibinfo
  {year} {2016})},\ \Eprint {http://arxiv.org/abs/1508.07253} {arXiv:1508.07253
  [gr-qc]} \BibitemShut {NoStop}%
\bibitem [{\citenamefont {Roulet}\ \emph {et~al.}(2019)\citenamefont {Roulet},
  \citenamefont {Dai}, \citenamefont {Venumadhav}, \citenamefont {Zackay},\
  and\ \citenamefont {Zaldarriaga}}]{templatebankpaper}%
  \BibitemOpen
  \bibfield  {author} {\bibinfo {author} {\bibfnamefont {J.}~\bibnamefont
  {Roulet}}, \bibinfo {author} {\bibfnamefont {L.}~\bibnamefont {Dai}},
  \bibinfo {author} {\bibfnamefont {T.}~\bibnamefont {Venumadhav}}, \bibinfo
  {author} {\bibfnamefont {B.}~\bibnamefont {Zackay}}, \ and\ \bibinfo {author}
  {\bibfnamefont {M.}~\bibnamefont {Zaldarriaga}},\ }\href {\doibase
  10.1103/physrevd.99.123022} {\bibfield  {journal} {\bibinfo  {journal}
  {Physical Review D}\ }\textbf {\bibinfo {volume} {99}} (\bibinfo {year}
  {2019}),\ 10.1103/physrevd.99.123022}\BibitemShut {NoStop}%
\bibitem [{\citenamefont {{Nitz}}\ \emph {et~al.}(2017)\citenamefont {{Nitz}},
  \citenamefont {{Dent}}, \citenamefont {{Dal Canton}}, \citenamefont
  {{Fairhurst}},\ and\ \citenamefont {{Brown}}}]{CoherentScore}%
  \BibitemOpen
  \bibfield  {author} {\bibinfo {author} {\bibfnamefont {A.~H.}\ \bibnamefont
  {{Nitz}}}, \bibinfo {author} {\bibfnamefont {T.}~\bibnamefont {{Dent}}},
  \bibinfo {author} {\bibfnamefont {T.}~\bibnamefont {{Dal Canton}}}, \bibinfo
  {author} {\bibfnamefont {S.}~\bibnamefont {{Fairhurst}}}, \ and\ \bibinfo
  {author} {\bibfnamefont {D.~A.}\ \bibnamefont {{Brown}}},\ }\href {\doibase
  10.3847/1538-4357/aa8f50} {\bibfield  {journal} {\bibinfo  {journal} {\apj}\
  }\textbf {\bibinfo {volume} {849}},\ \bibinfo {eid} {118} (\bibinfo {year}
  {2017})},\ \Eprint {http://arxiv.org/abs/1705.01513} {arXiv:1705.01513
  [gr-qc]} \BibitemShut {NoStop}%
\bibitem [{\citenamefont {{Dal Canton}}\ \emph {et~al.}(2014)\citenamefont
  {{Dal Canton}}, \citenamefont {{Bhagwat}}, \citenamefont {{Dhurandhar}},\
  and\ \citenamefont {{Lundgren}}}]{2014CQGra..31a5016D}%
  \BibitemOpen
  \bibfield  {author} {\bibinfo {author} {\bibfnamefont {T.}~\bibnamefont {{Dal
  Canton}}}, \bibinfo {author} {\bibfnamefont {S.}~\bibnamefont {{Bhagwat}}},
  \bibinfo {author} {\bibfnamefont {S.~V.}\ \bibnamefont {{Dhurandhar}}}, \
  and\ \bibinfo {author} {\bibfnamefont {A.}~\bibnamefont {{Lundgren}}},\
  }\href {\doibase 10.1088/0264-9381/31/1/015016} {\bibfield  {journal}
  {\bibinfo  {journal} {Classical and Quantum Gravity}\ }\textbf {\bibinfo
  {volume} {31}},\ \bibinfo {eid} {015016} (\bibinfo {year} {2014})},\ \Eprint
  {http://arxiv.org/abs/1304.0008} {arXiv:1304.0008 [gr-qc]} \BibitemShut
  {NoStop}%
\bibitem [{\citenamefont {{Bose}}\ \emph {et~al.}(2016)\citenamefont {{Bose}},
  \citenamefont {{Dhurandhar}}, \citenamefont {{Gupta}},\ and\ \citenamefont
  {{Lundgren}}}]{2016PhRvD..94l2004B}%
  \BibitemOpen
  \bibfield  {author} {\bibinfo {author} {\bibfnamefont {S.}~\bibnamefont
  {{Bose}}}, \bibinfo {author} {\bibfnamefont {S.}~\bibnamefont
  {{Dhurandhar}}}, \bibinfo {author} {\bibfnamefont {A.}~\bibnamefont
  {{Gupta}}}, \ and\ \bibinfo {author} {\bibfnamefont {A.}~\bibnamefont
  {{Lundgren}}},\ }\href {\doibase 10.1103/PhysRevD.94.122004} {\bibfield
  {journal} {\bibinfo  {journal} {\prd}\ }\textbf {\bibinfo {volume} {94}},\
  \bibinfo {eid} {122004} (\bibinfo {year} {2016})},\ \Eprint
  {http://arxiv.org/abs/1606.06096} {arXiv:1606.06096 [gr-qc]} \BibitemShut
  {NoStop}%
\bibitem [{gwo(4 12)}]{gwosc_det_status}%
  \BibitemOpen
  \href@noop {} {\enquote {\bibinfo {title} {{Gravitational Wave Open Science
  Center} detector status},}\ }\bibinfo {howpublished}
  {\url{www.gw-openscience.org/detector_status/day/20190412/}} (\bibinfo {year}
  {accessed 2019-04-12})\BibitemShut {NoStop}%
\bibitem [{\citenamefont {{Farr}}\ \emph {et~al.}(2018)\citenamefont {{Farr}},
  \citenamefont {{Holz}},\ and\ \citenamefont {{Farr}}}]{2018ApJ...854L...9F}%
  \BibitemOpen
  \bibfield  {author} {\bibinfo {author} {\bibfnamefont {B.}~\bibnamefont
  {{Farr}}}, \bibinfo {author} {\bibfnamefont {D.~E.}\ \bibnamefont {{Holz}}},
  \ and\ \bibinfo {author} {\bibfnamefont {W.~M.}\ \bibnamefont {{Farr}}},\
  }\href {\doibase 10.3847/2041-8213/aaaa64} {\bibfield  {journal} {\bibinfo
  {journal} {\apj}\ }\textbf {\bibinfo {volume} {854}},\ \bibinfo {eid} {L9}
  (\bibinfo {year} {2018})},\ \Eprint {http://arxiv.org/abs/1709.07896}
  {arXiv:1709.07896 [astro-ph.HE]} \BibitemShut {NoStop}%
\bibitem [{\citenamefont {{LIGO Scientific Collaboration}}(2018)}]{lalsuite}%
  \BibitemOpen
  \bibfield  {author} {\bibinfo {author} {\bibnamefont {{LIGO Scientific
  Collaboration}}},\ }\href {\doibase 10.7935/GT1W-FZ16} {\enquote {\bibinfo
  {title} {{LIGO} {A}lgorithm {L}ibrary - {LALS}uite},}\ }\bibinfo
  {howpublished} {free software (GPL)} (\bibinfo {year} {2018})\BibitemShut
  {NoStop}%
\bibitem [{\citenamefont {Khan}\ \emph {et~al.}(2016)\citenamefont {Khan},
  \citenamefont {Husa}, \citenamefont {Hannam}, \citenamefont {Ohme},
  \citenamefont {P\"{u}rrer}, \citenamefont {Forteza},\ and\ \citenamefont
  {Boh{\'{e}}}}]{Khan2016}%
  \BibitemOpen
  \bibfield  {author} {\bibinfo {author} {\bibfnamefont {S.}~\bibnamefont
  {Khan}}, \bibinfo {author} {\bibfnamefont {S.}~\bibnamefont {Husa}}, \bibinfo
  {author} {\bibfnamefont {M.}~\bibnamefont {Hannam}}, \bibinfo {author}
  {\bibfnamefont {F.}~\bibnamefont {Ohme}}, \bibinfo {author} {\bibfnamefont
  {M.}~\bibnamefont {P\"{u}rrer}}, \bibinfo {author} {\bibfnamefont {X.~J.}\
  \bibnamefont {Forteza}}, \ and\ \bibinfo {author} {\bibfnamefont
  {A.}~\bibnamefont {Boh{\'{e}}}},\ }\href {\doibase
  10.1103/physrevd.93.044007} {\bibfield  {journal} {\bibinfo  {journal}
  {Physical Review D}\ }\textbf {\bibinfo {volume} {93}} (\bibinfo {year}
  {2016}),\ 10.1103/physrevd.93.044007}\BibitemShut {NoStop}%
\bibitem [{\citenamefont {{Zackay}}\ \emph {et~al.}(2018)\citenamefont
  {{Zackay}}, \citenamefont {{Dai}},\ and\ \citenamefont
  {{Venumadhav}}}]{Zackay2018}%
  \BibitemOpen
  \bibfield  {author} {\bibinfo {author} {\bibfnamefont {B.}~\bibnamefont
  {{Zackay}}}, \bibinfo {author} {\bibfnamefont {L.}~\bibnamefont {{Dai}}}, \
  and\ \bibinfo {author} {\bibfnamefont {T.}~\bibnamefont {{Venumadhav}}},\
  }\href@noop {} {\bibfield  {journal} {\bibinfo  {journal} {arXiv e-prints}\ }
  (\bibinfo {year} {2018})},\ \Eprint {http://arxiv.org/abs/1806.08792}
  {arXiv:1806.08792 [astro-ph.IM]} \BibitemShut {NoStop}%
\bibitem [{\citenamefont {Buchner}\ \emph {et~al.}(2014)\citenamefont
  {Buchner}, \citenamefont {Georgakakis}, \citenamefont {Nandra}, \citenamefont
  {Hsu}, \citenamefont {Rangel}, \citenamefont {Brightman}, \citenamefont
  {Merloni}, \citenamefont {Salvato}, \citenamefont {Donley},\ and\
  \citenamefont {Kocevski}}]{PyMultiNest}%
  \BibitemOpen
  \bibfield  {author} {\bibinfo {author} {\bibfnamefont {J.}~\bibnamefont
  {Buchner}}, \bibinfo {author} {\bibfnamefont {A.}~\bibnamefont
  {Georgakakis}}, \bibinfo {author} {\bibfnamefont {K.}~\bibnamefont {Nandra}},
  \bibinfo {author} {\bibfnamefont {L.}~\bibnamefont {Hsu}},  \emph {et~al.},\
  }\href {\doibase 10.1051/0004-6361/201322971} {\bibfield  {journal} {\bibinfo
   {journal} {Astronomy {\&} Astrophysics}\ }\textbf {\bibinfo {volume}
  {564}},\ \bibinfo {pages} {A125} (\bibinfo {year} {2014})}\BibitemShut
  {NoStop}%
\bibitem [{\citenamefont {{Abbott}}\ \emph {et~al.}(2016)\citenamefont
  {{Abbott}}, \citenamefont {{Abbott}}, \citenamefont {{Abbott}}, \citenamefont
  {{Abernathy}}, \citenamefont {{Acernese}}, \citenamefont {{Ackley}},
  \citenamefont {{Adams}}, \citenamefont {{Adams}}, \citenamefont {{Addesso}},
  \citenamefont {{Adhikari}}, \citenamefont {{Adya}}, \citenamefont
  {{Affeldt}}, \citenamefont {{Agathos}}, \citenamefont {{Agatsuma}},
  \citenamefont {{Aggarwal}}, \citenamefont {{Aguiar}}, \citenamefont
  {{Aiello}}, \citenamefont {{Ain}}, \citenamefont {{Ajith}}, \citenamefont
  {{Allen}}, \citenamefont {{Allocca}}, \citenamefont {{Altin}}, \citenamefont
  {{Anderson}}, \citenamefont {{Anderson}}, \citenamefont {{Arai}},
  \citenamefont {{Araya}}, \citenamefont {{Arceneaux}}, \citenamefont
  {{Areeda}}, \citenamefont {{Arnaud}}, \citenamefont {{Arun}}, \citenamefont
  {{Ascenzi}}, \citenamefont {{Ashton}}, \citenamefont {{Ast}}, \citenamefont
  {{Aston}}, \citenamefont {{Astone}}, \citenamefont {{Aufmuth}}, \citenamefont
  {{Aulbert}}, \citenamefont {{Babak}}, \citenamefont {{Bacon}}, \citenamefont
  {{Bader}}, \citenamefont {{Baker}}, \citenamefont {{Baldaccini}},
  \citenamefont {{Ballardin}}, \citenamefont {{Ballmer}}, \citenamefont
  {{Barayoga}}, \citenamefont {{Barclay}}, \citenamefont {{Barish}},
  \citenamefont {{Barker}}, \citenamefont {{Barone}}, \citenamefont {{Barr}},
  \citenamefont {{Barsotti}}, \citenamefont {{Barsuglia}}, \citenamefont
  {{Barta}}, \citenamefont {{Bartlett}}, \citenamefont {{Bartos}},
  \citenamefont {{Bassiri}}, \citenamefont {{Basti}}, \citenamefont {{Batch}},
  \citenamefont {{Baune}}, \citenamefont {{Bavigadda}}, \citenamefont
  {{Bazzan}}, \citenamefont {{Behnke}}, \citenamefont {{Bejger}}, \citenamefont
  {{Bell}}, \citenamefont {{Bell}}, \citenamefont {{Berger}}, \citenamefont
  {{Bergman}}, \citenamefont {{Bergmann}}, \citenamefont {{Berry}},
  \citenamefont {{Bersanetti}}, \citenamefont {{Bertolini}}, \citenamefont
  {{Betzwieser}}, \citenamefont {{Bhagwat}}, \citenamefont {{Bhand are}},
  \citenamefont {{Bilenko}}, \citenamefont {{Billingsley}}, \citenamefont
  {{Birch}}, \citenamefont {{Birney}}, \citenamefont {{Biscans}}, \citenamefont
  {{Bisht}}, \citenamefont {{Bitossi}}, \citenamefont {{Biwer}}, \citenamefont
  {{Bizouard}}, \citenamefont {{Blackburn}}, \citenamefont {{Blair}},
  \citenamefont {{Blair}}, \citenamefont {{Blair}}, \citenamefont {{Bloemen}},
  \citenamefont {{Bock}}, \citenamefont {{Bodiya}}, \citenamefont {{Boer}},
  \citenamefont {{Bogaert}}, \citenamefont {{Bogan}}, \citenamefont {{Bohe}},
  \citenamefont {{Boh{\'e}mier}}, \citenamefont {{Bojtos}}, \citenamefont
  {{Bond}}, \citenamefont {{Bondu}}, \citenamefont {{Bonnand}}, \citenamefont
  {{Boom}}, \citenamefont {{Bork}}, \citenamefont {{Boschi}}, \citenamefont
  {{Bose}}, \citenamefont {{Bouffanais}}, \citenamefont {{Bozzi}},
  \citenamefont {{Bradaschia}}, \citenamefont {{Brady}}, \citenamefont
  {{Braginsky}}, \citenamefont {{Branchesi}}, \citenamefont {{Brau}},
  \citenamefont {{Briant}}, \citenamefont {{Brillet}}, \citenamefont
  {{Brinkmann}}, \citenamefont {{Brisson}}, \citenamefont {{Brockill}},
  \citenamefont {{Brooks}}, \citenamefont {{Brown}}, \citenamefont {{Brown}},
  \citenamefont {{Brown}}, \citenamefont {{Buchanan}}, \citenamefont
  {{Buikema}}, \citenamefont {{Bulik}}, \citenamefont {{Bulten}}, \citenamefont
  {{Buonanno}}, \citenamefont {{Buskulic}}, \citenamefont {{Buy}},
  \citenamefont {{Byer}}, \citenamefont {{Cabero}}, \citenamefont {{Cadonati}},
  \citenamefont {{Cagnoli}}, \citenamefont {{Cahillane}}, \citenamefont
  {{Calder{\'o}n Bustillo}}, \citenamefont {{Callister}}, \citenamefont
  {{Calloni}}, \citenamefont {{Camp}}, \citenamefont {{Cannon}}, \citenamefont
  {{Cao}}, \citenamefont {{Capano}}, \citenamefont {{Capocasa}}, \citenamefont
  {{Carbognani}}, \citenamefont {{Caride}}, \citenamefont {{Casanueva Diaz}},
  \citenamefont {{Casentini}}, \citenamefont {{Caudill}}, \citenamefont
  {{Cavagli{\`a}}}, \citenamefont {{Cavalier}}, \citenamefont {{Cavalieri}},
  \citenamefont {{Cella}}, \citenamefont {{Cepeda}}, \citenamefont {{Cerboni
  Baiardi}}, \citenamefont {{Cerretani}}, \citenamefont {{Cesarini}},
  \citenamefont {{Chakraborty}}, \citenamefont {{Chalermsongsak}},
  \citenamefont {{Chamberlin}}, \citenamefont {{Chan}}, \citenamefont {{Chao}},
  \citenamefont {{Charlton}}, \citenamefont {{Chassande-Mottin}}, \citenamefont
  {{Chen}}, \citenamefont {{Chen}}, \citenamefont {{Cheng}}, \citenamefont
  {{Chincarini}}, \citenamefont {{Chiummo}}, \citenamefont {{Cho}},
  \citenamefont {{Cho}}, \citenamefont {{Chow}}, \citenamefont {{Christensen}},
  \citenamefont {{Chu}}, \citenamefont {{Chua}}, \citenamefont {{Chung}},
  \citenamefont {{Ciani}}, \citenamefont {{Clara}}, \citenamefont {{Clark}},
  \citenamefont {{Clayton}}, \citenamefont {{Cleva}}, \citenamefont {{Coccia}},
  \citenamefont {{Cohadon}}, \citenamefont {{Cokelaer}}, \citenamefont
  {{Colla}}, \citenamefont {{Collette}}, \citenamefont {{Cominsky}},
  \citenamefont {{Constancio}}, \citenamefont {{Conte}}, \citenamefont
  {{Conti}}, \citenamefont {{Cook}}, \citenamefont {{Corbitt}}, \citenamefont
  {{Cornish}}, \citenamefont {{Corsi}}, \citenamefont {{Cortese}},
  \citenamefont {{Costa}}, \citenamefont {{Coughlin}}, \citenamefont
  {{Coughlin}}, \citenamefont {{Coulon}}, \citenamefont {{Countryman}},
  \citenamefont {{Couvares}}, \citenamefont {{Cowan}}, \citenamefont
  {{Coward}}, \citenamefont {{Cowart}}, \citenamefont {{Coyne}}, \citenamefont
  {{Coyne}}, \citenamefont {{Craig}}, \citenamefont {{Creighton}},
  \citenamefont {{Creighton}}, \citenamefont {{Cripe}}, \citenamefont
  {{Crowder}}, \citenamefont {{Cumming}}, \citenamefont {{Cunningham}},
  \citenamefont {{Cuoco}}, \citenamefont {{Dal Canton}}, \citenamefont
  {{Danilishin}}, \citenamefont {{D'Antonio}}, \citenamefont {{Danzmann}},
  \citenamefont {{Darman}}, \citenamefont {{Dattilo}}, \citenamefont {{Dave}},
  \citenamefont {{Daveloza}}, \citenamefont {{Davier}}, \citenamefont
  {{Davies}}, \citenamefont {{Daw}}, \citenamefont {{Day}}, \citenamefont
  {{De}}, \citenamefont {{DeBra}}, \citenamefont {{Debreczeni}}, \citenamefont
  {{Degallaix}}, \citenamefont {{De Laurentis}}, \citenamefont
  {{Del{\'e}glise}}, \citenamefont {{Del Pozzo}}, \citenamefont {{Denker}},
  \citenamefont {{Dent}}, \citenamefont {{Dereli}}, \citenamefont
  {{Dergachev}}, \citenamefont {{DeRosa}}, \citenamefont {{De Rosa}},
  \citenamefont {{DeSalvo}}, \citenamefont {{Dhurand har}}, \citenamefont
  {{D{\'\i}az}}, \citenamefont {{Dietz}}, \citenamefont {{Di Fiore}},
  \citenamefont {{Di Giovanni}}, \citenamefont {{Di Lieto}}, \citenamefont {{Di
  Pace}}, \citenamefont {{Di Palma}}, \citenamefont {{Di Virgilio}},
  \citenamefont {{Dojcinoski}}, \citenamefont {{Dolique}}, \citenamefont
  {{Donovan}}, \citenamefont {{Dooley}}, \citenamefont {{Doravari}},
  \citenamefont {{Douglas}}, \citenamefont {{Downes}}, \citenamefont {{Drago}},
  \citenamefont {{Drever}}, \citenamefont {{Driggers}}, \citenamefont {{Du}},
  \citenamefont {{Ducrot}}, \citenamefont {{Dwyer}}, \citenamefont {{Edo}},
  \citenamefont {{Edwards}}, \citenamefont {{Effler}}, \citenamefont
  {{Eggenstein}}, \citenamefont {{Ehrens}}, \citenamefont {{Eichholz}},
  \citenamefont {{Eikenberry}}, \citenamefont {{Engels}}, \citenamefont
  {{Essick}}, \citenamefont {{Etzel}}, \citenamefont {{Evans}}, \citenamefont
  {{Evans}}, \citenamefont {{Everett}}, \citenamefont {{Factourovich}},
  \citenamefont {{Fafone}}, \citenamefont {{Fair}}, \citenamefont
  {{Fairhurst}}, \citenamefont {{Fan}}, \citenamefont {{Fang}}, \citenamefont
  {{Farinon}}, \citenamefont {{Farr}}, \citenamefont {{Farr}}, \citenamefont
  {{Favata}}, \citenamefont {{Fays}}, \citenamefont {{Fehrmann}}, \citenamefont
  {{Fejer}}, \citenamefont {{Ferrante}}, \citenamefont {{Ferreira}},
  \citenamefont {{Ferrini}}, \citenamefont {{Fidecaro}}, \citenamefont
  {{Fiori}}, \citenamefont {{Fiorucci}}, \citenamefont {{Fisher}},
  \citenamefont {{Flaminio}}, \citenamefont {{Fletcher}}, \citenamefont
  {{Fotopoulos}}, \citenamefont {{Fournier}}, \citenamefont {{Franco}},
  \citenamefont {{Frasca}}, \citenamefont {{Frasconi}}, \citenamefont {{Frei}},
  \citenamefont {{Frei}}, \citenamefont {{Freise}}, \citenamefont {{Frey}},
  \citenamefont {{Frey}}, \citenamefont {{Fricke}}, \citenamefont
  {{Fritschel}}, \citenamefont {{Frolov}}, \citenamefont {{Fulda}},
  \citenamefont {{Fyffe}}, \citenamefont {{Gabbard}}, \citenamefont {{Gair}},
  \citenamefont {{Gammaitoni}}, \citenamefont {{Gaonkar}}, \citenamefont
  {{Garufi}}, \citenamefont {{Gatto}}, \citenamefont {{Gaur}}, \citenamefont
  {{Gehrels}}, \citenamefont {{Gemme}}, \citenamefont {{Gendre}}, \citenamefont
  {{Genin}}, \citenamefont {{Gennai}}, \citenamefont {{George}}, \citenamefont
  {{Gergely}}, \citenamefont {{Germain}}, \citenamefont {{Ghosh}},
  \citenamefont {{Ghosh}}, \citenamefont {{Giaime}}, \citenamefont
  {{Giardina}}, \citenamefont {{Giazotto}}, \citenamefont {{Gill}},
  \citenamefont {{Glaefke}}, \citenamefont {{Goetz}}, \citenamefont {{Goetz}},
  \citenamefont {{Goggin}}, \citenamefont {{Gondan}}, \citenamefont
  {{Gonz{\'a}lez}}, \citenamefont {{Gonzalez Castro}}, \citenamefont
  {{Gopakumar}}, \citenamefont {{Gordon}}, \citenamefont {{Gorodetsky}},
  \citenamefont {{Gossan}}, \citenamefont {{Gosselin}}, \citenamefont
  {{Gouaty}}, \citenamefont {{Graef}}, \citenamefont {{Graff}}, \citenamefont
  {{Granata}}, \citenamefont {{Grant}}, \citenamefont {{Gras}}, \citenamefont
  {{Gray}}, \citenamefont {{Greco}}, \citenamefont {{Green}}, \citenamefont
  {{Groot}}, \citenamefont {{Grote}}, \citenamefont {{Grunewald}},
  \citenamefont {{Guidi}}, \citenamefont {{Guo}}, \citenamefont {{Gupta}},
  \citenamefont {{Gupta}}, \citenamefont {{Gushwa}}, \citenamefont
  {{Gustafson}}, \citenamefont {{Gustafson}}, \citenamefont {{Hacker}},
  \citenamefont {{Hall}}, \citenamefont {{Hall}}, \citenamefont {{Hammond}},
  \citenamefont {{Haney}}, \citenamefont {{Hanke}}, \citenamefont {{Hanks}},
  \citenamefont {{Hanna}}, \citenamefont {{Hannam}}, \citenamefont {{Hanson}},
  \citenamefont {{Hardwick}}, \citenamefont {{Harms}}, \citenamefont {{Harry}},
  \citenamefont {{Harry}}, \citenamefont {{Hart}}, \citenamefont {{Hartman}},
  \citenamefont {{Haster}}, \citenamefont {{Haughian}}, \citenamefont
  {{Heidmann}}, \citenamefont {{Heintze}}, \citenamefont {{Heitmann}},
  \citenamefont {{Hello}}, \citenamefont {{Hemming}}, \citenamefont {{Hendry}},
  \citenamefont {{Heng}}, \citenamefont {{Hennig}}, \citenamefont
  {{Heptonstall}}, \citenamefont {{Heurs}}, \citenamefont {{Hild}},
  \citenamefont {{Hoak}}, \citenamefont {{Hodge}}, \citenamefont {{Hofman}},
  \citenamefont {{Hollitt}}, \citenamefont {{Holt}}, \citenamefont {{Holz}},
  \citenamefont {{Hopkins}}, \citenamefont {{Hosken}}, \citenamefont {{Hough}},
  \citenamefont {{Houston}}, \citenamefont {{Howell}}, \citenamefont {{Hu}},
  \citenamefont {{Huang}}, \citenamefont {{Huerta}}, \citenamefont {{Huet}},
  \citenamefont {{Hughey}}, \citenamefont {{Husa}}, \citenamefont {{Huttner}},
  \citenamefont {{Huynh-Dinh}}, \citenamefont {{Idrisy}}, \citenamefont
  {{Indik}}, \citenamefont {{Ingram}}, \citenamefont {{Inta}}, \citenamefont
  {{Isa}}, \citenamefont {{Isac}}, \citenamefont {{Isi}}, \citenamefont
  {{Islas}}, \citenamefont {{Isogai}}, \citenamefont {{Iyer}}, \citenamefont
  {{Izumi}}, \citenamefont {{Jacqmin}}, \citenamefont {{Jang}}, \citenamefont
  {{Jani}}, \citenamefont {{Jaranowski}}, \citenamefont {{Jawahar}},
  \citenamefont {{Jim{\'e}nez-Forteza}}, \citenamefont {{Johnson}},
  \citenamefont {{Jones}}, \citenamefont {{Jones}}, \citenamefont {{Jones}},
  \citenamefont {{Jonker}}, \citenamefont {{Ju}}, \citenamefont {{Haris}},
  \citenamefont {{Kalaghatgi}}, \citenamefont {{Kalogera}}, \citenamefont
  {{Kandhasamy}}, \citenamefont {{Kang}}, \citenamefont {{Kanner}},
  \citenamefont {{Karki}}, \citenamefont {{Kasprzack}}, \citenamefont
  {{Katsavounidis}}, \citenamefont {{Katzman}}, \citenamefont {{Kaufer}},
  \citenamefont {{Kaur}}, \citenamefont {{Kawabe}}, \citenamefont {{Kawazoe}},
  \citenamefont {{K{\'e}f{\'e}lian}}, \citenamefont {{Kehl}}, \citenamefont
  {{Keitel}}, \citenamefont {{Kelley}}, \citenamefont {{Kells}}, \citenamefont
  {{Keppel}}, \citenamefont {{Kennedy}}, \citenamefont {{Key}}, \citenamefont
  {{Khalaidovski}}, \citenamefont {{Khalili}}, \citenamefont {{Khan}},
  \citenamefont {{Khan}}, \citenamefont {{Khan}}, \citenamefont {{Khazanov}},
  \citenamefont {{Kijbunchoo}}, \citenamefont {{Kim}}, \citenamefont {{Kim}},
  \citenamefont {{Kim}}, \citenamefont {{Kim}}, \citenamefont {{Kim}},
  \citenamefont {{Kim}}, \citenamefont {{King}}, \citenamefont {{King}},
  \citenamefont {{Kinzel}}, \citenamefont {{Kissel}}, \citenamefont
  {{Kleybolte}}, \citenamefont {{Klimenko}}, \citenamefont {{Koehlenbeck}},
  \citenamefont {{Kokeyama}}, \citenamefont {{Koley}}, \citenamefont
  {{Kondrashov}}, \citenamefont {{Kontos}}, \citenamefont {{Korobko}},
  \citenamefont {{Korth}}, \citenamefont {{Kowalska}}, \citenamefont {{Kozak}},
  \citenamefont {{Kringel}}, \citenamefont {{Krishnan}}, \citenamefont
  {{Kr{\'o}lak}}, \citenamefont {{Krueger}}, \citenamefont {{Kuehn}},
  \citenamefont {{Kumar}}, \citenamefont {{Kuo}}, \citenamefont {{Kutynia}},
  \citenamefont {{Lackey}}, \citenamefont {{Landry}}, \citenamefont {{Lange}},
  \citenamefont {{Lantz}}, \citenamefont {{Lasky}}, \citenamefont
  {{Lazzarini}}, \citenamefont {{Lazzaro}}, \citenamefont {{Leaci}},
  \citenamefont {{Leavey}}, \citenamefont {{Lebigot}}, \citenamefont {{Lee}},
  \citenamefont {{Lee}}, \citenamefont {{Lee}}, \citenamefont {{Lee}},
  \citenamefont {{Lenon}}, \citenamefont {{Leonardi}}, \citenamefont {{Leong}},
  \citenamefont {{Leroy}}, \citenamefont {{Letendre}}, \citenamefont {{Levin}},
  \citenamefont {{Levine}}, \citenamefont {{Li}}, \citenamefont {{Libson}},
  \citenamefont {{Littenberg}}, \citenamefont {{Lockerbie}}, \citenamefont
  {{Logue}}, \citenamefont {{Lombardi}}, \citenamefont {{Lord}}, \citenamefont
  {{Lorenzini}}, \citenamefont {{Loriette}}, \citenamefont {{Lormand}},
  \citenamefont {{Losurdo}}, \citenamefont {{Lough}}, \citenamefont
  {{L{\"u}ck}}, \citenamefont {{Lundgren}}, \citenamefont {{Luo}},
  \citenamefont {{Lynch}}, \citenamefont {{Ma}}, \citenamefont {{MacDonald}},
  \citenamefont {{Machenschalk}}, \citenamefont {{MacInnis}}, \citenamefont
  {{Macleod}}, \citenamefont {{Maga{\~n}a-Sandoval}}, \citenamefont {{Magee}},
  \citenamefont {{Mageswaran}}, \citenamefont {{Majorana}}, \citenamefont
  {{Maksimovic}}, \citenamefont {{Malvezzi}}, \citenamefont {{Man}},
  \citenamefont {{Mandel}}, \citenamefont {{Mandic}}, \citenamefont
  {{Mangano}}, \citenamefont {{Mansell}}, \citenamefont {{Manske}},
  \citenamefont {{Mantovani}}, \citenamefont {{Marchesoni}}, \citenamefont
  {{Marion}}, \citenamefont {{M{\'a}rka}}, \citenamefont {{M{\'a}rka}},
  \citenamefont {{Markosyan}}, \citenamefont {{Maros}}, \citenamefont
  {{Martelli}}, \citenamefont {{Martellini}}, \citenamefont {{Martin}},
  \citenamefont {{Martin}}, \citenamefont {{Martynov}}, \citenamefont {{Marx}},
  \citenamefont {{Mason}}, \citenamefont {{Masserot}}, \citenamefont
  {{Massinger}}, \citenamefont {{Masso-Reid}}, \citenamefont {{Matichard}},
  \citenamefont {{Matone}}, \citenamefont {{Mavalvala}}, \citenamefont
  {{Mazumder}}, \citenamefont {{Mazzolo}}, \citenamefont {{McCarthy}},
  \citenamefont {{McClelland}}, \citenamefont {{McCormick}}, \citenamefont
  {{McGuire}}, \citenamefont {{McIntyre}}, \citenamefont {{McIver}},
  \citenamefont {{McKechan}}, \citenamefont {{McManus}}, \citenamefont
  {{McWilliams}}, \citenamefont {{Meacher}}, \citenamefont {{Meadors}},
  \citenamefont {{Meidam}}, \citenamefont {{Melatos}}, \citenamefont
  {{Mendell}}, \citenamefont {{Mendoza-Gandara}}, \citenamefont {{Mercer}},
  \citenamefont {{Merilh}}, \citenamefont {{Merzougui}}, \citenamefont
  {{Meshkov}}, \citenamefont {{Messaritaki}}, \citenamefont {{Messenger}},
  \citenamefont {{Messick}}, \citenamefont {{Meyers}}, \citenamefont
  {{Mezzani}}, \citenamefont {{Miao}}, \citenamefont {{Michel}}, \citenamefont
  {{Middleton}}, \citenamefont {{Mikhailov}}, \citenamefont {{Milano}},
  \citenamefont {{Miller}}, \citenamefont {{Millhouse}}, \citenamefont
  {{Minenkov}}, \citenamefont {{Ming}}, \citenamefont {{Mirshekari}},
  \citenamefont {{Mishra}}, \citenamefont {{Mitra}}, \citenamefont
  {{Mitrofanov}}, \citenamefont {{Mitselmakher}}, \citenamefont {{Mittleman}},
  \citenamefont {{Moggi}}, \citenamefont {{Mohan}}, \citenamefont
  {{Mohapatra}}, \citenamefont {{Montani}}, \citenamefont {{Moore}},
  \citenamefont {{Moore}}, \citenamefont {{Moraru}}, \citenamefont {{Moreno}},
  \citenamefont {{Morriss}}, \citenamefont {{Mossavi}}, \citenamefont
  {{Mours}}, \citenamefont {{Mow-Lowry}}, \citenamefont {{Mueller}},
  \citenamefont {{Mueller}}, \citenamefont {{Muir}}, \citenamefont
  {{Mukherjee}}, \citenamefont {{Mukherjee}}, \citenamefont {{Mukherjee}},
  \citenamefont {{Mukund}}, \citenamefont {{Mullavey}}, \citenamefont
  {{Munch}}, \citenamefont {{Murphy}}, \citenamefont {{Murray}}, \citenamefont
  {{Mytidis}}, \citenamefont {{Nardecchia}}, \citenamefont {{Naticchioni}},
  \citenamefont {{Nayak}}, \citenamefont {{Necula}}, \citenamefont {{Nedkova}},
  \citenamefont {{Nelemans}}, \citenamefont {{Neri}}, \citenamefont
  {{Neunzert}}, \citenamefont {{Newton}}, \citenamefont {{Nguyen}},
  \citenamefont {{Nielsen}}, \citenamefont {{Nissanke}}, \citenamefont
  {{Nitz}}, \citenamefont {{Nocera}}, \citenamefont {{Nolting}}, \citenamefont
  {{Normandin}}, \citenamefont {{Nuttall}}, \citenamefont {{Oberling}},
  \citenamefont {{Ochsner}}, \citenamefont {{O'Dell}}, \citenamefont
  {{Oelker}}, \citenamefont {{Ogin}}, \citenamefont {{Oh}}, \citenamefont
  {{Oh}}, \citenamefont {{Ohme}}, \citenamefont {{Oliver}}, \citenamefont
  {{Oppermann}}, \citenamefont {{Oram}}, \citenamefont {{O'Reilly}},
  \citenamefont {{O'Shaughnessy}}, \citenamefont {{Ottaway}}, \citenamefont
  {{Ottens}}, \citenamefont {{Overmier}}, \citenamefont {{Owen}}, \citenamefont
  {{Pai}}, \citenamefont {{Pai}}, \citenamefont {{Palamos}}, \citenamefont
  {{Palashov}}, \citenamefont {{Palomba}}, \citenamefont {{Pal-Singh}},
  \citenamefont {{Pan}}, \citenamefont {{Pan}}, \citenamefont {{Pankow}},
  \citenamefont {{Pannarale}}, \citenamefont {{Pant}}, \citenamefont
  {{Paoletti}}, \citenamefont {{Paoli}}, \citenamefont {{Papa}}, \citenamefont
  {{Paris}}, \citenamefont {{Parker}}, \citenamefont {{Pascucci}},
  \citenamefont {{Pasqualetti}}, \citenamefont {{Passaquieti}}, \citenamefont
  {{Passuello}}, \citenamefont {{Patricelli}}, \citenamefont {{Patrick}},
  \citenamefont {{Pearlstone}}, \citenamefont {{Pedraza}}, \citenamefont
  {{Pedurand}}, \citenamefont {{Pekowsky}}, \citenamefont {{Pele}},
  \citenamefont {{Penn}}, \citenamefont {{Perreca}}, \citenamefont {{Phelps}},
  \citenamefont {{Piccinni}}, \citenamefont {{Pichot}}, \citenamefont
  {{Piergiovanni}}, \citenamefont {{Pierro}}, \citenamefont {{Pillant}},
  \citenamefont {{Pinard}}, \citenamefont {{Pinto}}, \citenamefont {{Pitkin}},
  \citenamefont {{Poggiani}}, \citenamefont {{Popolizio}}, \citenamefont
  {{Post}}, \citenamefont {{Powell}}, \citenamefont {{Prasad}}, \citenamefont
  {{Predoi}}, \citenamefont {{Premachandra}}, \citenamefont {{Prestegard}},
  \citenamefont {{Price}}, \citenamefont {{Prijatelj}}, \citenamefont
  {{Principe}}, \citenamefont {{Privitera}}, \citenamefont {{Prodi}},
  \citenamefont {{Prokhorov}}, \citenamefont {{Puncken}}, \citenamefont
  {{Punturo}}, \citenamefont {{Puppo}}, \citenamefont {{P{\"u}rrer}},
  \citenamefont {{Qi}}, \citenamefont {{Qin}}, \citenamefont {{Quetschke}},
  \citenamefont {{Quintero}}, \citenamefont {{Quitzow-James}}, \citenamefont
  {{Raab}}, \citenamefont {{Rabeling}}, \citenamefont {{Radkins}},
  \citenamefont {{Raffai}}, \citenamefont {{Raja}}, \citenamefont
  {{Rakhmanov}}, \citenamefont {{Rapagnani}}, \citenamefont {{Raymond}},
  \citenamefont {{Razzano}}, \citenamefont {{Re}}, \citenamefont {{Read}},
  \citenamefont {{Reed}}, \citenamefont {{Regimbau}}, \citenamefont {{Rei}},
  \citenamefont {{Reid}}, \citenamefont {{Reitze}}, \citenamefont {{Rew}},
  \citenamefont {{Reyes}}, \citenamefont {{Ricci}}, \citenamefont {{Riles}},
  \citenamefont {{Robertson}}, \citenamefont {{Robie}}, \citenamefont
  {{Robinet}}, \citenamefont {{Robinson}}, \citenamefont {{Rocchi}},
  \citenamefont {{Rodriguez}}, \citenamefont {{Rolland}}, \citenamefont
  {{Rollins}}, \citenamefont {{Roma}}, \citenamefont {{Romano}}, \citenamefont
  {{Romanov}}, \citenamefont {{Romie}}, \citenamefont {{Rosi{\'n}ska}},
  \citenamefont {{Rowan}}, \citenamefont {{R{\"u}diger}}, \citenamefont
  {{Ruggi}}, \citenamefont {{Ryan}}, \citenamefont {{Sachdev}}, \citenamefont
  {{Sadecki}}, \citenamefont {{Sadeghian}}, \citenamefont {{Salconi}},
  \citenamefont {{Saleem}}, \citenamefont {{Salemi}}, \citenamefont
  {{Samajdar}}, \citenamefont {{Sammut}}, \citenamefont {{Sanchez}},
  \citenamefont {{Sand berg}}, \citenamefont {{Sandeen}}, \citenamefont
  {{Sanders}}, \citenamefont {{Santamar{\'\i}a}}, \citenamefont {{Sassolas}},
  \citenamefont {{Sathyaprakash}}, \citenamefont {{Saulson}}, \citenamefont
  {{Sauter}}, \citenamefont {{Savage}}, \citenamefont {{Sawadsky}},
  \citenamefont {{Schale}}, \citenamefont {{Schilling}}, \citenamefont
  {{Schmidt}}, \citenamefont {{Schmidt}}, \citenamefont {{Schnabel}},
  \citenamefont {{Schofield}}, \citenamefont {{Sch{\"o}nbeck}}, \citenamefont
  {{Schreiber}}, \citenamefont {{Schuette}}, \citenamefont {{Schutz}},
  \citenamefont {{Scott}}, \citenamefont {{Scott}}, \citenamefont {{Sellers}},
  \citenamefont {{Sengupta}}, \citenamefont {{Sentenac}}, \citenamefont
  {{Sequino}}, \citenamefont {{Sergeev}}, \citenamefont {{Serna}},
  \citenamefont {{Setyawati}}, \citenamefont {{Sevigny}}, \citenamefont
  {{Shaddock}}, \citenamefont {{Shah}}, \citenamefont {{Shahriar}},
  \citenamefont {{Shaltev}}, \citenamefont {{Shao}}, \citenamefont {{Shapiro}},
  \citenamefont {{Shawhan}}, \citenamefont {{Sheperd}}, \citenamefont
  {{Shoemaker}}, \citenamefont {{Shoemaker}}, \citenamefont {{Siellez}},
  \citenamefont {{Siemens}}, \citenamefont {{Sigg}}, \citenamefont {{Silva}},
  \citenamefont {{Simakov}}, \citenamefont {{Singer}}, \citenamefont
  {{Singer}}, \citenamefont {{Singh}}, \citenamefont {{Singh}}, \citenamefont
  {{Singhal}}, \citenamefont {{Sintes}}, \citenamefont {{Slagmolen}},
  \citenamefont {{Smith}}, \citenamefont {{Smith}}, \citenamefont {{Smith}},
  \citenamefont {{Son}}, \citenamefont {{Sorazu}}, \citenamefont
  {{Sorrentino}}, \citenamefont {{Souradeep}}, \citenamefont {{Srivastava}},
  \citenamefont {{Staley}}, \citenamefont {{Steinke}}, \citenamefont
  {{Steinlechner}}, \citenamefont {{Steinlechner}}, \citenamefont
  {{Steinmeyer}}, \citenamefont {{Stephens}}, \citenamefont {{Stone}},
  \citenamefont {{Strain}}, \citenamefont {{Straniero}}, \citenamefont
  {{Stratta}}, \citenamefont {{Strauss}}, \citenamefont {{Strigin}},
  \citenamefont {{Sturani}}, \citenamefont {{Stuver}}, \citenamefont
  {{Summerscales}}, \citenamefont {{Sun}}, \citenamefont {{Sutton}},
  \citenamefont {{Swinkels}}, \citenamefont {{Szczepa{\'n}czyk}}, \citenamefont
  {{Tacca}}, \citenamefont {{Talukder}}, \citenamefont {{Tanner}},
  \citenamefont {{T{\'a}pai}}, \citenamefont {{Tarabrin}}, \citenamefont
  {{Taracchini}}, \citenamefont {{Taylor}}, \citenamefont {{Theeg}},
  \citenamefont {{Thirugnanasambandam}}, \citenamefont {{Thomas}},
  \citenamefont {{Thomas}}, \citenamefont {{Thomas}}, \citenamefont {{Thorne}},
  \citenamefont {{Thorne}}, \citenamefont {{Thrane}}, \citenamefont {{Tiwari}},
  \citenamefont {{Tiwari}}, \citenamefont {{Tokmakov}}, \citenamefont
  {{Tomlinson}}, \citenamefont {{Tonelli}}, \citenamefont {{Torres}},
  \citenamefont {{Torrie}}, \citenamefont {{T{\"o}yr{\"a}}}, \citenamefont
  {{Travasso}}, \citenamefont {{Traylor}}, \citenamefont {{Trifir{\`o}}},
  \citenamefont {{Tringali}}, \citenamefont {{Trozzo}}, \citenamefont {{Tse}},
  \citenamefont {{Turconi}}, \citenamefont {{Tuyenbayev}}, \citenamefont
  {{Ugolini}}, \citenamefont {{Unnikrishnan}}, \citenamefont {{Urban}},
  \citenamefont {{Usman}}, \citenamefont {{Vahlbruch}}, \citenamefont
  {{Vajente}}, \citenamefont {{Valdes}}, \citenamefont {{van Bakel}},
  \citenamefont {{van Beuzekom}}, \citenamefont {{van den Brand}},
  \citenamefont {{Van Den Broeck}}, \citenamefont {{Vander-Hyde}},
  \citenamefont {{van der Schaaf}}, \citenamefont {{van Heijningen}},
  \citenamefont {{van Veggel}}, \citenamefont {{Vardaro}}, \citenamefont
  {{Vass}}, \citenamefont {{Vas{\'u}th}}, \citenamefont {{Vaulin}},
  \citenamefont {{Vecchio}}, \citenamefont {{Vedovato}}, \citenamefont
  {{Veitch}}, \citenamefont {{Veitch}}, \citenamefont {{Venkateswara}},
  \citenamefont {{Verkindt}}, \citenamefont {{Vetrano}}, \citenamefont
  {{Vicer{\'e}}}, \citenamefont {{Vinciguerra}}, \citenamefont {{Vine}},
  \citenamefont {{Vinet}}, \citenamefont {{Vitale}}, \citenamefont {{Vo}},
  \citenamefont {{Vocca}}, \citenamefont {{Vorvick}}, \citenamefont {{Voss}},
  \citenamefont {{Vousden}}, \citenamefont {{Vyatchanin}}, \citenamefont
  {{Wade}}, \citenamefont {{Wade}}, \citenamefont {{Wade}}, \citenamefont
  {{Walker}}, \citenamefont {{Wallace}}, \citenamefont {{Walsh}}, \citenamefont
  {{Wang}}, \citenamefont {{Wang}}, \citenamefont {{Wang}}, \citenamefont
  {{Wang}}, \citenamefont {{Wang}}, \citenamefont {{Ward}}, \citenamefont
  {{Warner}}, \citenamefont {{Was}}, \citenamefont {{Weaver}}, \citenamefont
  {{Wei}}, \citenamefont {{Weinert}}, \citenamefont {{Weinstein}},
  \citenamefont {{Weiss}}, \citenamefont {{Welborn}}, \citenamefont {{Wen}},
  \citenamefont {{We{\ss}els}}, \citenamefont {{West}}, \citenamefont
  {{Westphal}}, \citenamefont {{Wette}}, \citenamefont {{Whelan}},
  \citenamefont {{White}}, \citenamefont {{Whiting}}, \citenamefont
  {{Wiesner}}, \citenamefont {{Williams}}, \citenamefont {{Williamson}},
  \citenamefont {{Willis}}, \citenamefont {{Willke}}, \citenamefont {{Wimmer}},
  \citenamefont {{Winkler}}, \citenamefont {{Wipf}}, \citenamefont {{Wiseman}},
  \citenamefont {{Wittel}}, \citenamefont {{Woan}}, \citenamefont {{Worden}},
  \citenamefont {{Wright}}, \citenamefont {{Wu}}, \citenamefont {{Yablon}},
  \citenamefont {{Yam}}, \citenamefont {{Yamamoto}}, \citenamefont {{Yancey}},
  \citenamefont {{Yap}}, \citenamefont {{Yu}}, \citenamefont {{Yvert}},
  \citenamefont {{Zadro{\.Z}ny}}, \citenamefont {{Zangrando}}, \citenamefont
  {{Zanolin}}, \citenamefont {{Zendri}}, \citenamefont {{Zevin}}, \citenamefont
  {{Zhang}}, \citenamefont {{Zhang}}, \citenamefont {{Zhang}}, \citenamefont
  {{Zhang}}, \citenamefont {{Zhao}}, \citenamefont {{Zhou}}, \citenamefont
  {{Zhou}}, \citenamefont {{Zhu}}, \citenamefont {{Zucker}}, \citenamefont
  {{Zuraw}}, \citenamefont {{Zweizig}}, \citenamefont {{LIGO Scientific
  Collaboration}},\ and\ \citenamefont {{Virgo
  Collaboration}}}]{2016PhRvD..93l2003A}%
  \BibitemOpen
  \bibfield  {author} {\bibinfo {author} {\bibfnamefont {B.~P.}\ \bibnamefont
  {{Abbott}}}, \bibinfo {author} {\bibfnamefont {R.}~\bibnamefont {{Abbott}}},
  \bibinfo {author} {\bibfnamefont {T.~D.}\ \bibnamefont {{Abbott}}}, \bibinfo
  {author} {\bibfnamefont {M.~R.}\ \bibnamefont {{Abernathy}}},  \emph
  {et~al.},\ }\href {\doibase 10.1103/PhysRevD.93.122003} {\bibfield  {journal}
  {\bibinfo  {journal} {\prd}\ }\textbf {\bibinfo {volume} {93}},\ \bibinfo
  {eid} {122003} (\bibinfo {year} {2016})},\ \Eprint
  {http://arxiv.org/abs/1602.03839} {arXiv:1602.03839 [gr-qc]} \BibitemShut
  {NoStop}%
\bibitem [{\citenamefont {Jaranowski}\ and\ \citenamefont
  {Kr{\'o}lak}(2012)}]{Jaranowski2012}%
  \BibitemOpen
  \bibfield  {author} {\bibinfo {author} {\bibfnamefont {P.}~\bibnamefont
  {Jaranowski}}\ and\ \bibinfo {author} {\bibfnamefont {A.}~\bibnamefont
  {Kr{\'o}lak}},\ }\href {\doibase 10.12942/lrr-2012-4} {\bibfield  {journal}
  {\bibinfo  {journal} {Living Reviews in Relativity}\ }\textbf {\bibinfo
  {volume} {15}},\ \bibinfo {pages} {4} (\bibinfo {year} {2012})}\BibitemShut
  {NoStop}%
\end{thebibliography}%

\end{document}